\documentclass[11pt]{article}   
\usepackage{graphicx}				
\usepackage{cancel}
\usepackage{amsthm}
\usepackage{tikz}
\usepackage{pgfplots}
\usetikzlibrary{decorations.markings}

\usepackage{xcolor}
\definecolor{mydarkgreen}{RGB}{0,100,0}			

\newtheorem*{theoremA}{Theorem A}

\usepackage{hyperref}
\usepackage{amssymb}
\usepackage{amsmath}
\usepackage{amsfonts}
\usepackage{mathbbol}

\usepackage{color}  
\newcommand{\bcc}{\color{black}}

\newcommand{\rcc}{\color{black}}
\newcommand{\ma}{\color{black}}
\newcommand{\gr}{\color{black}}
\newcommand{\bl}{\color{black}}
\renewcommand{\d}{\mrm{d}}

 \newcommand{\sinc}{\mrm{sinc}}

\newlength{\dinwidth}
\newlength{\dinmargin}

 \usepackage{cancel}       

\newcommand{\one}{\mathbb{1}}
\newcommand{\mrm}{\mathrm}

\newcommand{\Img}{\mrm{Img}}

\newcommand{\wh}{\widehat}

\newcommand{\mcF}{\mathcal{F}}

\newcommand{\De}{\Delta}

\newcommand{\mcL}{\mathcal{L}}

\newcommand{\nin}{\noindent}
\newcommand{\lan}{\langle}
\newcommand{\ran}{\rangle}

\newcommand{\ti}{\tilde}

\newcommand{\pa}{\partial}
\newcommand{\supp}{\mathrm{supp}}

\newcommand{\Om}{\Omega}
\newcommand{\si}{\sigma}

\newcommand{\h}{\fr{1}{2}}
\newcommand{\e}{\mathrm{e}}

\newcommand{\de}{\delta}

\newcommand{\sh}{\mathrm{supp}\: h}

\newcommand{\nat}{\mathbb{N}}
\newcommand{\hil}{\mathcal{H}}

\newcommand{\fr}[2]{\frac{#1}{#2}}
\newcommand{\al}{\alpha}

\newcommand{\real}{\mathbb{R}}

\newcommand{\la}{\lambda}
\newcommand{\ov}{\overline}

\newcommand{\ga}{\gamma}
\newcommand{\non}{\nonumber}
\newcommand{\complex}{\mathbb{C}}

\def\proof{\noindent{\bf Proof. }}
\def\qed{$\Box$\medskip}

\newcommand{\da}{\dagger}

\newcommand{\xintj}{[\fr{x_{\mud}}{L^j\xi}]}

\newcommand{\mud}{\bullet}
\newcommand{\llan}{\lan\!\lan}
\newcommand{\rran}{\ran\!\ran}

\newtheorem{theoreme}{Theorem}[section]
\newtheorem{proposition}[theoreme]{Proposition}
\newtheorem{lemma}[theoreme]{Lemma}
\newtheorem{definition}[theoreme]{Definition}
\newtheorem{corollary}[theoreme]{Corollary}
\newtheorem{remark}[theoreme]{Remark}
\newtheorem{example}[theoreme]{Example}
\newtheorem{criterion}[theoreme]{Criterion}
\newtheorem{assumption}[theoreme]{Assumption}
\newcommand{\bem}{\begin{multline}}
\newcommand{\eem}{\end{multline}}

\newcommand{\beex}{\begin{example}}
\newcommand{\eeex}{\end{example}}
\newcommand{\bea}{\begin{assumption}}
\newcommand{\eea}{\end{assumption}}
\newcommand{\beq}{\begin{equation}}
\newcommand{\eeq}{\end{equation}}
\newcommand\beqa{\begin{eqnarray}}   
\newcommand\eeqa{\end{eqnarray}} 
\newcommand{\ben}{\begin{arabicenumerate}}
\newcommand{\een}{\end{arabicenumerate}}
\newcommand{\bex}{\begin{example}}
\newcommand{\eex}{\end{example}}
\newcommand{\ber}{\begin{remark}}
\newcommand{\eer}{\end{remark}}
\newcommand{\bec}{\begin{corollary}}
\newcommand{\eec}{\end{corollary}}
\newcommand{\bed}{\begin{definition}}
\newcommand{\eed}{\end{definition}}
\newcommand{\bep}{\begin{proposition}}
\newcommand{\eep}{\end{proposition}}
\newcommand{\becr}{\begin{criterion}}
\newcommand{\eecr}{\end{criterion}}
\def\bel{\begin{lemma}}
\def\eel{\end{lemma}}
\def\bet{\begin{theoreme}}
\def\eet{\end{theoreme}}
\def\bed{\begin{definition}}
\def\eed{\end{definition}}

\newcommand{\2}{\!\!\!&}
\newcommand{\kk}{j}
\newcommand{\GG}{O}

\title{{\rcc Lattice Green functions for pedestrians: Exponential decay}}
\author{Wojciech Dybalski\,$^a$, Alexander Stottmeister$^b$ \ and \
Yoh Tanimoto$^c$\,  \\[5mm]
\normalsize ${}^a$ Faculty of Mathematics and Computer Science, 
\normalsize Adam Mickiewicz University, \\
\normalsize ul.~Uniwersytetu Pozna\'nskiego 4, 61-614 Pozna\'n, Poland\\
\normalsize E-mail: {\tt wojciech.dybalski@amu.edu.pl} \\ [2mm]
\normalsize ${}^b$ Institut f\"ur Theoretische Physik, Leibniz Universit\"at Hannover, \\
\normalsize Appelstrasse 2, \ 30167 Hannover, Germany  \\ 
\normalsize E-mail: {\tt alexander.stottmeister@itp.uni-hannover.de} \\ [2mm] 
\normalsize ${}^c$ Dipartimento di Matematica, Universit\`a di Roma ``Tor Vergata''\\ 
\normalsize Via della Ricerca Scientifica, 1 - I--00133 Roma, Italy.\\ 
\normalsize E-mail: {\tt hoyt@mat.uniroma2.it} }

\date{}							

\begin{document}

\maketitle

\begin{abstract} 

{\rcc The exponential decay of lattice Green functions is one of the main technical ingredients of the Bałaban's approach to renormalization. 
We give here a self-contained  proof, whose various ingredients were scattered in the literature. The main sources of exponential decay are the 
Combes-Thomas method and the analyticity of the Fourier transforms. They are combined using a renormalization group equation 
and the method of images.}

\end{abstract}
\newpage

\tableofcontents

\renewcommand{\k}{\mrm{k}}
\newcommand{\eell}{\mrm{l}}
\newcommand{\n}{\mrm{n}}
\newcommand{\m}{\mrm{m}}
\newcommand{\bmu}{\bar{\mu}}
\renewcommand{\v}{f}
\renewcommand{\u}{g}
\newcommand{\w}{h}
\newcommand{\etaq}{q}
\renewcommand{\xi}{\eta}
\renewcommand{\L}{\mathcal{L}}
\newcommand{\ovs}{\overset}
\newcommand{\ovsum}{\ovs{*}{\sum}{}}
\newcommand{\ch}{\mrm{ch}}
\renewcommand{\sh}{\mrm{sh}}
\newcommand{\hmu}{\hat{\mu}} 
\renewcommand{\i}{\mathrm{i}}

\newpage

\section{{\bcc Introduction}}

The approach of Ba{\ma \l }aban, which builds on the Wilsonian renormalization of lattice {\gr quantum field theory} (QFT),   stands out as a charted route
towards a construction of non-trivial just-renormalizable QFT. {\rcc The strategy, outlined e.g. in \cite{BJ85, Ba90}, can be divided into four main steps:
(1)~the exponential decay of  lattice Green functions \cite[II.9,II.10,IV]{BJ85}, (2)~variational problem identifying the background field \cite[pp.~242--243]{BJ85}, (3)~Wilsonian renormalization 
in the small field region by an expansion around the background field \cite[III.1--III.5]{BJ85}, (4)~the large field problem \cite[III.6]{BJ85}\cite{Ba90}}. Unfortunately, most of the literature on this topic was written in a style suitable for a small circle of experts and, after several decades,  is  difficult to comprehend 
for interested readers. The main obstacle, as we see it, is the absence of systematic expositions of the  
methods used and developed in these works. Relevant pieces of information are scattered over many papers 
treating different models and a substantial part of the discussion is left to the reader.  As a valuable exception we would
like to mention  relatively recent papers of Dimock on the UV stability of the  $\phi^4_3$ model \cite{Di13, Di13.1, Di14}, which are largely self-contained and
readable line by line. But even these useful works on a super-renormalizable model do not provide the reader with sufficient background to delve into 
the literature on just-renormalizable theories. One reason is a rather brief discussion of the lattice Green functions in \cite{Di13}, which 
covers only the more elementary part of the subject, leaving the rest to rather cryptic references. As the lattice Green functions
are at the foundation of the entire subject and their detailed control is needed already at the level of the variational problem in
some just-renormalizable models, we decided to write {\gr this} systematic account. {\rcc We recall  that the need to present Bałaban's method
in a more comprehensible manner was stressed already in \cite{Ri00}}.  

Let us describe the content of our article  in an  informal way: Let $\Om$ be a finite square lattice in $d$-dimensions
with lattice spacing $\eta$. In the Wilsonian spirit, we divide this lattice into boxes $B(y)$ of linear size $L$. The points $y$, which
label the boxes, form the coarse lattice $\Om_1:=(L \Om)\cap \Om$. We can repeat this procedure $k$ times, denoting the
resulting boxes $B_k(y)$ and the coarse lattice $\Om_k:=(L^k\Om)\cap \Om$. We set in the following $\eta=L^{-k}$, in which case
$\Om_k$ has a unit spacing.  The operator of $k$-times averaging $Q_{\Om,k}: \mcL^2(\Om)\to \mcL^2(\Om_k)$ is given by
\beqa
(Q_{\Om,k}f)(y)=\fr{1}{L^{kd}}\sum_{x\in B_k(y)} f(x)
\eeqa
and the {\gr relevant} propagator has the form
\beqa
G_k(\Om):=(-\De^{\eta}_{\Om}+\bmu_k+a_kQ_{\Om,k}^*Q_{\Om,k} )^{-1}.  \label{G-k-Om-intro}
\eeqa
Here $\De^{\eta}_{\Om}$ is the discrete Laplacian on $\Om$ with Neumann boundary conditions, $\{a_k\}_{k\in\nat}$ is a concrete bounded  sequence 
which stays away from zero, see formula~(\ref{a-j}), and $\bmu_k=L^{2k}\bmu_0$ are mass parameters.  It turns out that $Q_{\Om,k}^*Q_{\Om,k}$  is a projection operator on the subspace of functions in $\mcL^2(\Om)$ which are constant on the blocks $B_k(y)$. 
The kernel $(x,x')\mapsto G_k(\Om)(x,x')$  is called a Green function.

Leaving  the question of motivation of definition (\ref{G-k-Om-intro}) aside for a moment, let us justify the
existence of the inverse.   Here the 
key observation is that the Laplacian vanishes only on constant functions, whereas the averaging operator leaves
such functions invariant (cf. Lemma~\ref{inverse-lemma}).  Moving towards the exponential decay of Green functions,  one observes that the
denominator in (\ref{G-k-Om-intro})  resembles a simple Schr\"odinger operator consisting of a Laplacian perturbed
by a finite dimensional projection. Using the Combes-Thomas  method \cite{CT74} from the Schr{\gr \"o}dinger operator theory, one easily obtains 
an $\mcL^2$ bound  (cf. Lemma~\ref{Combes-Thomas})
\beqa
|\langle f^{\prime}, G_{k}(\Om) f \rangle| \leq c \e^{-c_1 |y - y^{\prime}| } \| f' \|_{2} \|f \|_{2},  \label{CT-intro}
\eeqa
where $\supp\, f\subset B_k(y)$, $\supp\,f'\subset  B_k(y')$. However, the small field conditions of the Balaban's approach
single out the supremum norm, in particular in the analysis of the variational problem. {\bl This leads us to the main theorem
of this paper:}  
\begin{theoremA} \label{main-theorem-intro}  {\ma There is ${\bl c_1} >0$ s.t. for all  $f\in \mcL^{\infty}(\Om)$} 
\beqa
|(G_k(\Om) f)(x)|\leq  c  L^{2d}  \e^{-{\bl c_1} \d(x, \supp(f)) } \|f\|_{\infty}, \label{main-estimates-intro}
\eeqa
{\bl where $\d(x, \supp(f)):=\mrm{inf}_{y\in \supp(f)}|x-y|$ and   $c,{\bl c_1}$ are constants depending only on dimension $d$ of the lattice. (In particular they are independent of the lattice spacing, the size of the lattice and the parameter $L$).} 

\end{theoremA}
\nin {\bl We note that estimate (\ref{main-estimates-intro}) is non-trivial only because the constants $c, c_1$ are independent of the lattice size. This result is stated  in  many references including \cite{Ba83} and \cite{Di13} but we could not find a  complete and verifiable proof. 
 Theorem~A  will be proven as Theorem~\ref{main-result} below.  We summarize the proof  in the later part of this introduction.}  

We mention as an aside that (\ref{main-estimates-intro}) is not a direct consequence of (\ref{CT-intro}): Setting $f'= \fr{1}{\eta^d} \de_x$ in (\ref{CT-intro})
and then estimating the $\mcL^2$-norm of $f$ by the supremum norm, we immediately obtain a dependence of the constants on the
lattice spacing and the volume of the lattice. Furthermore, (\ref{main-estimates-intro}) does not imply that $(x,x')\mapsto G_k(\Om)(x,x')$ is a bounded
function uniformly in the lattice spacing: by setting $f=\fr{1}{\eta^d} \de_x$ in (\ref{main-estimates-intro}) we obtain again $\eta$-dependence of the constants.
The latter fact  reflects the well{\ma -}known singularity at coinciding points of the free Green function in the continuum,  e.g. for $d\geq 3$  
\beqa
G(x,x'):=\fr{1}{(2\pi)^{d}} \int_{\bl \real^d} \e^{-\i p\cdot (x-x')} \fr{1}{p^2+m^2} dp=\fr{1}{4\pi} \fr{1}{|x-x'|^{d-2}} \e^{-m|x-x'|}.
\eeqa

To explain how the propagators defined in (\ref{G-k-Om-intro})  appear in QFT on the lattice, let us define the following
family of functions on $\mcL^2(\Om)$
\beqa
\rho_0^{J}(\phi):=  \e^{-\h\lan \phi, \De^{(0), \xi}(\Om) \phi\ran} \e^{\i \lan J,\phi\ran}, \quad  \De^{(0),\xi}(\Om)=-\De^{\xi}_{\Om} +\bmu_k,
\eeqa
which,  after integration in $\phi$, gives  the generating functional of the theory in the variable $J\in  \mcL^2(\Om)$.
Denote by $T^{\eta}_{a_j, L^j}$ the renormalization group transformation, acting by averaging $j$-times over boxes, defined precisely in 
(\ref{T-def}). The averaged function $\rho_j^J:=T^{\eta}_{a_j, L^j}[\rho_0^J]$ can be computed via Gaussian integration and has the 
form:
\beqa
\rho_j^J(\psi)=Z^{\xi}_j (\Om) \e^{-\h\lan \psi, \De^{(j), L^j\xi}(\Om) \psi\ran } \e^{\i \lan J, \hil_j^{\eta}(\Om)\psi  \ran}  \e^{-\h \lan J,  G_j^{\eta}(\Om)  J\ran}. \label{rho-j-intro}
\eeqa
By setting $J=0$, we see that the covariance of the averaged theory is the inverse of
\beqa
\De^{(j),L^j\xi}(\Om)\2:=\2 c_{\rcc (j)} - c^2_{\rcc (j)} Q_{\Om,j} G_j^{\xi}(\Om) Q_{\Om,j}^*, \label{Delta-intro}
\eeqa 
where $c_{\rcc (j)}:=a_j(L^j\xi)^{-2}$. The propagator  $G_j^{\xi}(\Om)$ is defined as in (\ref{G-k-Om-intro}), with $a_j$ replaced by $c_{\rcc (j)}$,
and it is related to $G_j(\Om)$ by a simple scaling transformation, cf. (\ref{G-xi-j-Om-lab}). It appears not only in (\ref{Delta-intro}),
but also in another ingredient of (\ref{rho-j-intro})
\beqa
\hil_j^{\eta}(\Om):=c_{\rcc(j)} G_j^{\eta}(\Om) Q_{\Om,j}^*
\eeqa
which will be important below. Thus the Green functions we study in
this paper do not have an immediate meaning as correlations of some physical system. Instead, they are  convenient building blocks to express various
natural  quantities appearing  in the process of renormalization.

The semigroup property of the renormalization group transformation (cf. Lemma~\ref{ren-iteration-x}) turns out to be useful  
for proving Theorem~A.  On the one hand we have $\rho_{j+1}^J:=T^{\eta}_{a_{j+1}, L^{j+1}}[\rho_0^J]$,
on the other hand $\rho_{j+1}^J:=T^{L^j\eta}_{a, L}[\rho_j^J]$. The latter formula, stated explicitly in (\ref{rho-j-final}), can be 
compared with  (\ref{rho-j-intro}), which results in the so called renormalization group formula
\beqa
G^{\xi}_{j+1}(\Om) \2=\2  \hil_j^{\eta}(\Om)   C^{(j),L^j\xi}(\Om)  \hil_j^{\eta}(\Om)^* +G_j^{\xi}(\Om), \quad  C^{(j), L^j\xi}(\Om):=(\De^{(j), L^j\xi}(\Om) +\fr{a}{(L^{j+1}\xi)^2} Q_{\Om_j}^*Q_{\Om_j})^{-1}.\,\,\,\,\,\,\,\, \label{RG-formula-intro}
\eeqa
This formula appears (without proof) already in \cite{Ba82}, the proof sketched above is taken from   \cite{Di04}. 
It turns out that an exponential decay of the kernel of $C^{(j),L^j\xi}(\Om)$ can be obtained by the Combes-Thomas method  {\gr \cite{CT74}}.  The exponential decay of
the kernel of $\hil_j^{\eta}(\Om)$ also holds, but its proof is a separate story, which we will discuss below. The propagator $G_{j=1}^{\xi}(\Om)$ can
be rescaled to $G_{j=1}(L^{k-1}\Om)$, for which the bound (\ref{main-estimates-intro})  is easy to prove, again by the Combes-Thomas method, since the lattice $L^{k-1}\Om$ has spacing $L^{-1}$ and $L$ is fixed in the discussion. It should be clear from these remarks, that the renormalization group formula  (\ref{RG-formula-intro})
allows for an iterative proof of Theorem~A. 

\begin{figure}[t]   
\centering
\begin{tikzpicture}
	\draw[dashed, red] (-7,-0.5) to (8,-0.5);
	\draw[dashed, red] (-0.5,-4) to (-0.5,5);
	\draw[thick, blue] (-7,0) to (8,0);
	\draw[thick, blue] (0,-4) to (0,5);
	
	\foreach \i in {1,4,7,10,13}
	{
	\foreach \j in {1,4,7}
	{
	\draw[thick, blue] (-7+\i,-4+\j) to (-5+\i,-4+\j);
	\draw[thick, blue] (-7+\i,-4+\j) to (-7+\i,-2+\j);
	}
	}
	\foreach \i in {1,4,7,10,13}
	{
	\foreach \j in {3,6,9}
	{
	\draw[thick, blue] (-7+\i,-4+\j) to (-5+\i,-4+\j);
	\draw[thick, blue] (-5+\i,-6+\j) to (-5+\i,-4+\j);
	}
	}	
	
	\filldraw (0.5,1.5) circle(1pt);
	\draw (0.5,1.5) node[below]{$x$};
	\draw[thick,blue] (-1,-0.1) to (-1,0.1);
	\draw[blue] (-1.1,-0.1) node[below]{\tiny$-1$};
	\draw[thick,blue] (-0.5,-0.1) to (-0.5,0.1);
	\draw[blue] (-0.6,-0.1) node[below]{\tiny$-0.5$};
	\draw[thick,blue] (0.5,-0.1) to (0.5,0.1);
	\draw[blue] (0.5,-0.1) node[below]{\tiny$0.5$};
	\draw[thick,blue] (1,-0.1) to (1,0.1);
	\draw[blue] (1,-0.1) node[below]{\tiny$1$};
	\draw[thick,blue] (1.5,-0.1) to (1.5,0.1);
	\draw[blue] (1.5,-0.1) node[below]{\tiny$1.5$};
	\draw[thick,blue] (2,-0.1) to (2,0.1);
	\draw[blue] (2,-0.1) node[below]{\tiny$2$};
	
	\foreach \i in {0,2,6,8,12}
	{
	\foreach \j in {0,2,6}
	{
	\filldraw (-4.5+\i,-1.5+\j) circle(1pt);
	}
	}
	\draw (-4.5+12,-1.5+6) node[below]{$y_{-\!7}$};
	\draw (-4.5+8,-1.5+6) node[below]{$y_{-\!6}$};
	\draw (-4.5+6,-1.5+6) node[below]{$y_{-\!5}$};
	\draw (-4.5+2,-1.5+6) node[below]{$y_{-\!4}$};
	\draw (-4.5,-1.5+6) node[below]{$y_{-\!3}$};
	\draw (-4.5,-1.5+2) node[below]{$y_{-\!2}$};
	\draw (-4.5+2,-1.5+2) node[below]{$y_{-\!1}$};
	\draw (-4.5+6,-1.5+2) node[below]{$y_{0}$};
	\draw (-4.5+8,-1.5+2) node[below]{$y_{1}$};
	\draw (-4.5+12,-1.5+2) node[below]{$y_{2}$};
	\draw (-4.5,-1.5) node[below]{$y_{3}$};
	\draw (-4.5+2,-1.5) node[below]{$y_{4}$};
	\draw (-4.5+6,-1.5) node[below]{$y_{5}$};
	\draw (-4.5+8,-1.5) node[below]{$y_{6}$};
	\draw (-4.5+12,-1.5) node[below]{$y_{7}$};
\end{tikzpicture}
\caption{Set of image points $y_j$ of the argument $y$ in (\ref{Fourier-rep-intro}). The square containing the origin is the set $\Om$.}
\label{picture}
\end{figure}
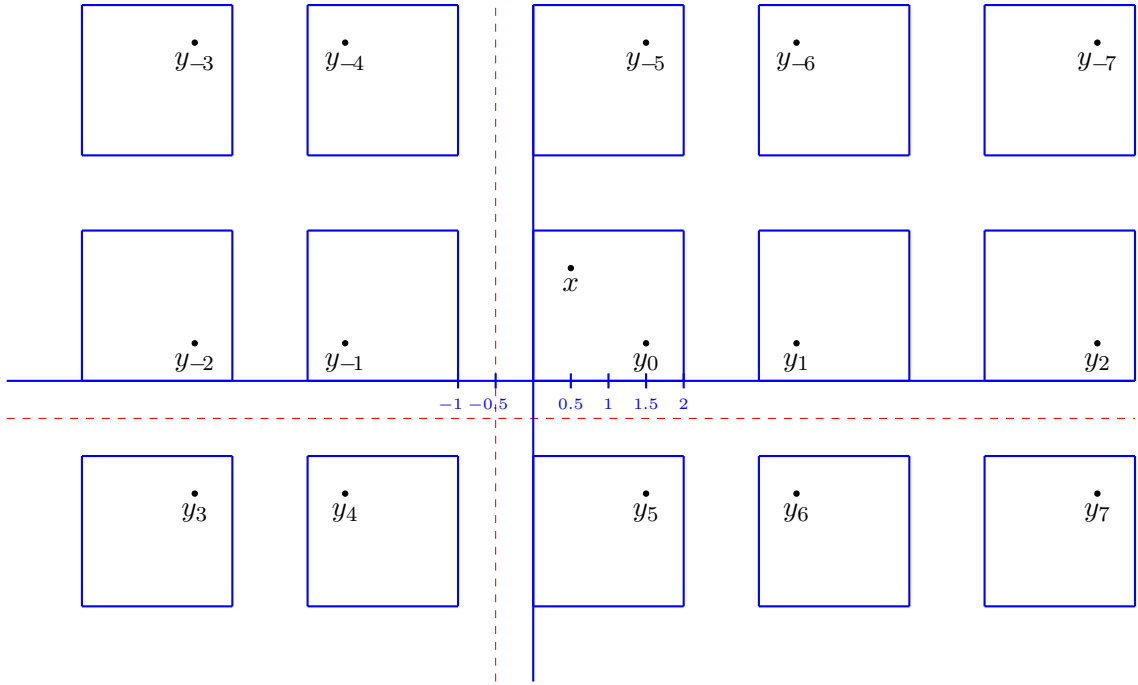

To conclude the discussion of the proof of Theorem~A, let us comment on the exponential decay of the kernel of $\hil_j^{\eta}(\Om)$.
As usual, it suffices to study its rescaled variant $\hil_k(\Om):=a_kG_k(\Om)Q_{\Om,k}^*$. The first step is to express the  Green functions $G_k(\Om)$
 on the finite lattice $\Om$ by their counterparts $G_k$ on an infinite lattice. This is achieved using the
method of images: 
\beqa
G_k(\Om)(x,y)=\sum_{y_j\in \Img}G_k(x, y_j), \label{images-form-intro}
\eeqa
where $\Img $ is the set of images of $y$ as indicated in Figure~\ref{picture}. 
(Think of the problem of determining the electric field of a charge near
infinitely extended conductive planes from basic electrostatic). Once stated, formula~(\ref{images-form-intro}) is not difficult to prove, it suffices
to check that the r.h.s. is indeed the inverse of $(-\De^{\eta}_{\Om}+\bmu_k+a_kQ_{\Om,k}^*Q_{\Om,k} )$. The advantage of translating the problem 
on an infinite lattice is that we can use the Fourier transform to obtain the following representation
\beqa
(G_k Q_k^*)(x,y)=(2\pi)^{-d} \int_{[-\pi/\xi, \pi/\xi[^{\times d}} dp \,\e^{\i p\cdot (x-y)}  h(p).     \label{Fourier-rep-intro}
\eeqa
If $h$ can be analytically continued to a bounded function on a strip in a complex plane, then exponential decay of this expression follows by standard arguments.
The difficult part is in showing that the width of this strip can be chosen uniformly in the lattice spacing and other relevant parameters. This 
step is completely skipped in \cite[p. 586]{Ba83}, a related discussion appears in \cite{BOS89} in a different context, but we did not find a
complete proof of this fact in the literature. We provide such a proof in Appendix~\ref{App-bound-integrand}, which is  the main
technical result of this article. The main idea of the proof is described below Lemma~\ref{bound-integrand}.

The proof of Theorem~A can thus be summarized as follows: there are two main sources of exponential
decay: the Combes-Thomas method (\ref{CT-intro}) and analytic continuation of the Fourier transform~(\ref{Fourier-rep-intro}).
To combine them, two identities are used: the renormalization group equation (\ref{RG-formula-intro}) and the method of images
formula (\ref{images-form-intro}).

To keep this article within reasonable limits, we deliberately omit the topic of random walk expansions. They are useful, e.g., to translate Theorem~A to 
periodic boundary conditions, cf. \cite[Lemma 6]{Di13}, or to control the kernels of operators $(Q_k G_k(\Om) Q^*_k)^{-1}$, which appear at the
level of the variational problem of some just-renormalizable models. We plan to come back to this topic in future publications.

Our paper is organized as follows: In Section~\ref{Neuman-sect} we focus on the Laplacian with  Neumann boundary conditions and provide
a proof of Theorem~A assuming the exponential decay of the kernel of $\hil_k(\Om)$. This part is mostly based on \cite{Di04,Di13}
but our discussion is more detailed. In Section~\ref{Free-section} we 
prove such a decay for the counterpart of $\hil_k(\Om)$ on an infinite lattice, using the Fourier space representation~(\ref{Fourier-rep-intro}). 
As this part is omitted in  \cite{Di13}, we follow mostly \cite{Ba83} and complete the discussion regarding  analytic continuation.
 Finally,
in Section~\ref{Images-section} we use the method of images to prove  formula~(\ref{images-form-intro}) which is only stated in \cite{Ba83}. 
This  implies 
the required decay of the kernel of $\hil_k(\Om)$
and concludes the proof of Theorem~A. More technical steps of the discussion are postponed to the appendices.

\vspace{1cm}

\nin \textbf{Acknowledgements:} W.D. would like to thank J\"urg Fr\"ohlich and Gian Michele Graf for useful discussions. W.D. was supported by the grant `Sonata Bis' 2019/34/E/ST1/00053  of the National Science Centre, Poland. {\ma Y.T. is partially supported by the MIUR Excellence Department Project MatMod@TOV awarded to the Department of  Mathematics, University of Rome ``Tor Vergata'' CUP E83C23000330006, by the University of Rome ``Tor Vergata'' funding OAQM
CUP E83C22001800005 and by GNAMPA–INdAM.}

\vspace{0.5cm} 
\nin\textbf{Notation}

\begin{enumerate}

\item By $c, c', c_1,c_2\ldots$ we denote numerical constants, independent of any parameters except for the dimension $d$.
These constants may change from line to line.

\item We denote by $\xi$ the lattice spacing. {\bl We introduce an odd
positive integer  $L>1$ and set $\xi=L^{-k}$, $k\in \mathbb{N}$}. To change the lattice spacing, we keep $L$
fixed and change $k$. 

\item We set $I=\xi [0,1,\ldots, n-1]$, $n-1=L^m$, so that the parameter $m\geq k$ controls the size of the interval. 
{\bl (We remark that the interesting case is $m$ much larger than $k$, since Theorem~A concerns large-distance properties).}

\item We denote by $\Om\subset \xi \mathbb{Z}^d$  a cube $\Om:=I^{\times d}=\xi [0,1,\ldots, L^m]^{\times d}$.

\item For $j=1,\ldots, m$ we denote by  $\Om_j\subset L^j \xi\mathbb{Z}^d$ the coarse lattices of the form $\Om_j=L^j\xi[0,1,\ldots, L^{m-j}]^{\times d}$.
  
\item We will use $\k,\m, \ldots \in \mathbb{Z}$ to number points of the lattice, as opposed to parameters $k,m$, which control the lattice spacing and size of the box. 

\item The elements of the Hilbert space $\L^2(\xi \mathbb{Z})$ are  complex{\gr -valued} functions denoted $f,f', g,g'$.  The  scalar product has the form
\beqa
 \lan \v, \u\ran=\xi\sum_{\k\in \mathbb{Z}}  \bar{\v}_\k \u_\k=\xi\sum_{x\in \xi\mathbb{Z}} \bar{\v}(x) \u(x),
\eeqa
where we write $f(x)$ or $f_{\k}$ for $x=\xi \k$. We  set $\|f\|^2_2=\lan f, f\ran$. {\gr We use complex (rather than real) $\mcL^2$-spaces for simplicity of Fourier analysis 
in Section~\ref{Free-section}.}

\item The Hilbert spaces $\L^2(I)$, $\L^2(\xi \mathbb{Z}^d)$, $\L^2(\Om_j)$ are defined analogously as $\L^2(\xi \mathbb{Z})$ above and the corresponding 
scalar products and norms will be denoted as for $\L^2(\xi \mathbb{Z})$.  If there is a risk of confusion, we will add the region of summation as subscript, e.g. $\lan\,\cdot\,,\cdot\,\ran_{\Om_j}, \|\,\cdot\,\|_{2,\Om_j}$. {\gr The subspaces of real-valued functions are denoted by $\L^2_{\real}(I)$, $\L^2_{\real}(\xi \mathbb{Z}^d)$, $\L^2_{\real}(\Om_j)$.}

\item For any region $\GG\subset \eta\mathbb{Z}^d$ we will write $\|f\|_{\infty, \GG}=\mrm{sup}_{x\in \GG}|f(x)|$. If there is no risk of confusion, we will
drop $\GG$. {\bcc We denote by $\mcL^{\infty}(\GG)$ the Banach space of functions on $\GG$ equipped with the norm $\|\,\cdot\,\|_{\infty, \GG}$.}

\item For quantities or conditions $X_{\mu}$ depending on the spacetime index $\mu=0,1,\ldots, d-1$ we will often
write $X_{\mud}$ as an abbreviation for $X_{\mu}, \mu=0,1,\ldots, d-1$.

\item We define the boxes in the lattice
\beqa
B_j(y):=\{\, x\in \xi\mathbb{Z}^d   \,|\, y_{\mu}\leq x_{\mu}< y_{\mu}+L^j\xi, \,\, \mu=0,1,\ldots, d-1\}
\eeqa
and call $y$ the \emph{label} of the box. {\bl For $j=k$  we obtain unit boxes $\triangle_y:=B_k(y)$, since $\eta=L^{-k}$. 
We also define $\ti{\triangle}_z:=L^{-1}B_{k+1}(Lz)$, $z\in L^{-1}\Om$, which are unit boxes in the lattice $L^{-1}\Om$.    }

\item We denote by $\De^{\xi}_{\Om}$, resp. $\De^{\xi}$, the Laplacian on $\Om$ with Neumann boundary conditions, resp.
the Laplacian on $\xi \mathbb{Z}^d$ with free boundary conditions. 

\item The mass parameters have the form  
$ \bar{\mu}_k= L^{2k}\bar{\mu}_0$ for $\bar{\mu}_{\rcc 0}\geq 0$. {\gr This corresponds to canonical scaling if $\bar{\mu}_0$ is the mass squared.}  As the mass parameter
will play a minor role in our discussion, we abbreviate
\beqa
-\De_{\Om}^{\xi, \bmu_k}:=-\De_{\Om}^{\xi}+\bmu_k, \quad -\De^{\xi, \bmu_k}:=-\De^{\xi}+\bmu_k.
\eeqa
Furthermore, in Appendix~\ref{App-bound-integrand} we will use the notation $\ovsum_{\mu=0}^{d-1} a_{\mu}= (\sum_{\mu=0}^{d-1} a_{\mu})+\fr{1}{4}\bmu_0$.

\item $\d(x,x'):=|x-x'|=\big(\sum_{\mu=0}^{d-1}(x_{\mu} -x'_{\mu})^2 \big)^{1/2}$, 
 $\d(x,{\bcc \GG}):=\inf_{y\in G } |x-y|$ for $\GG\subset \eta\mathbb{Z}^d$, $|x-x'|_{\infty}:=\sup_{\mu=0,\ldots,d-1}|x_{\mu}-x'_{\mu}|$.
 
\item $\one_{\GG}$ denotes the characteristic function of a set $\GG$.

\end{enumerate}

\section{Green functions  with Neumann boundary conditions} \label{Neuman-sect}
 \setcounter{equation}{0}
 
\subsection{Laplacian on interval with Neumann boundary conditions}
 
We consider the finite dimensional Hilbert  space  $\L^2(I)$, where $I:=\xi[0,1,2,\ldots,n-1]$, with the scalar product
\beqa
 \lan \v, \u\ran=\xi\sum_{\k=0}^{n-1}  \bar{\v}_\k \u_\k. \label{scalar-product-1d}
 \eeqa  
We note that $\k\mapsto \de_{\xi; \m,\k}=\fr{1}{\xi} \de_{\m,\k}$ plays the role of the Dirac
delta, since $\lan \de_{\xi; \m}, \v\ran=\v_\m$.  Let $A : \L^2(I)\to \L^2(I)$ be a linear map. We define its kernel by
\beqa
A_{\m,\m'}=\lan \de_{\xi; \m}, A \de_{\xi; \m'} \ran.
\eeqa 
We have $(A\v)_\k=\lan \de_{\xi;\k}, A\v \ran$. By writing $\v_{\eell}=\xi \sum_{\k'} \de_{\xi;\eell,\k'} \v_{\k'}$, we have 
\beqa
(A\v)_\k=\xi \sum_{\k'\in \mathbb{Z}} A_{\k, \k'} \v_{\k'}.
\eeqa

As specific examples of $A$ we define the discrete derivatives
 \beqa
 (\pa^{\xi}_I \v)_\k:=\fr{1}{\xi} (\v_{\k+1}-\v_\k), \quad (\pa^{\xi, \da}_I\v)_\k:=-\fr{1}{\xi} (\v_{\k}-\v_{\k-1})=-(\pa^{\xi}_I\v)_{\k-1}
  \eeqa 
 with Neumann boundary conditions $\v_{n}:=\v_{n-1}$, $\v_{-1}:=\v_0$, {\gr that is, $(\pa_I^{\xi,\da}\v)_{0}=0=  (\pa_I^{\xi}\v)_{n-1}$} .   By Lemma~\ref{Leibniz} we see that $\pa_I^{\xi,\da}$ is not the adjoint of $\pa_I^{\xi}$, given Neumann boundary 
 conditions. 
 \bel\label{Leibniz} The Leibniz rule holds in the form
 \beqa
 (\pa_I^{\xi} (\v\u))_\k=(\pa^{\xi}_I\v)_\k \u_{\k+1} + \v_\k (\pa_I^{\xi}\u)_\k.  \label{Leibniz-rule}
   \eeqa
 Integration by parts holds in the form
 \beqa
 \lan \v, \pa^{\xi}_I \u\ran\2=\2 -\lan \pa_I^{\xi} \v,  \u_{\cdot+1}\ran-\bar{\v}_0\u_0+\bar{\v}_{n-1}\u_{n-1}\non\\
 \2=\2 \lan \pa^{\xi,\da}_I \v,  \u\ran-\bar{\v}_0\u_0+\bar{\v}_{n-1}\u_{n-1}.
  \eeqa
  \eel
 \proof We compute
 \beqa
 (\pa^{\xi}_I  (\v\u))_\k \2=\2 \fr{1}{\xi}\big( (\v\u)_{\k+1}-(\v\u)_{\k} \big) \non\\
\2=\2  \fr{1}{\xi} \big( (\v_{\k+1} -\v_\k)\u_{\k+1}+\v_\k(\u_{\k+1}-\u_\k) \big),
  \eeqa
 which gives (\ref{Leibniz-rule}). Next, we note that
 \beqa
 \xi\sum_{\k=0}^{  n-1} (\pa^{\xi}_I  (\v\u))_\k=-\v_0\u_0+\v_{n-1}\u_{n-1}.
  \eeqa
 On the other hand
 \beqa
 \xi\sum_{\k=0}^{n-1} (\pa^{\xi}_I (\v\u))_\k=\lan (\pa^{\xi}_I \v), \u_{\cdot+1}\ran + \lan \v, (\pa^{\xi}_I \u)\ran.
\eeqa
Finally, we write
\beqa
\lan (\pa^{\xi}_I \v), \u_{\cdot+1}\ran\2=\2\sum_{\k=0}^{n-1} (\pa^{\xi}_I \v)_\k \u_{\k+1} =\sum_{\k'=1 }^{ n} (\pa^{\xi}_I \v)_{\k'-1} \u_{\k'}\non\\
\2=\2 \sum_{\k'=0}^{n-1} (\pa^{\xi}_I \v)_{\k'-1} \u_{\k'} =-\lan \pa^{\da,\xi}_I\v, \u\ran,  
\eeqa
where in the third step we made use of Neumann boundary conditions.  \qed\\
   The Laplacian with Neumann boundary conditions
 in one dimension on the lattice $I=\xi[0,1,2,\ldots,n-1]$ has the form
 \beqa
(\De^{\xi}_I \v)_\k:=((\pa^\xi_I ) (\pa^{\xi}_I\v)_{\cdot-1} )_\k=-((\pa^\xi_I) (\pa_I^{\xi,\da}\v) )_k= \fr{\v_{\k+1} -2\v_\k+\v_{\k-1}}{\xi^2} \label{second-deriv}
 \eeqa
 with the boundary conditions $\v_{n}:=\v_{n-1}$, $\v_{-1}:=\v_0$.  Although it is not manifest from the definition, $(-\De^{\xi}_I)$ is a positive
 operator:
 \bel\label{Laplacian-s-a} We have
  \beqa
  \lan \v,   (-\De^{\xi}_I)  \v \ran= \lan (-\De^{\xi}_I)\v,     \v \ran= \fr{1}{\xi^2}\sum_{\k=0}^{  n-2} |\v_{\k+1}-\v_{\k}|^2. \label{positivity-form}
  \eeqa
   \eel
 \proof  We compute using  (\ref{Leibniz-rule})
\beqa
  \lan \v,   {\ma (\pa^\xi_I(\pa^{\xi}_I \v)_{\cdot-1} )} \ran= -\lan \pa^\xi_I \v, \pa^\xi_I \v\ran-\bar{\v}_0 (\pa^{\xi}_I\v)_{-1}+\bar{\v}_{n-1} (\pa^{\xi}_I\v)_{n-1}. \label{integration-by-parts}
 \eeqa
Noting that $(\pa^{\xi}_I\v)_{-1}=(\pa^{\xi}_I \v)_{ n-1}=0$ by the Neumann boundary conditions and dropping the last
term in the sum defining the scalar product we obtain the claim.  \qed
 
 Next, let us recall the standard computation of the spectrum of the Laplacian on a finite 
 lattice\footnote{See Wikipedia: \href{https://en.wikipedia.org/wiki/Eigenvalues_and_eigenvectors_of_the_second_derivative}{Eigenvalues and eigenvectors of the second derivative}.}. 
  This will be needed in the proof of existence of lattice Green functions in (\ref{unit-box}) below.
\bel The eigenvalues of $\De^{\xi}_I$ are
\beqa
 \la^{(j)}=-\fr{4}{\xi^2} \sin^2\bigg( \fr{\pi j}{2n} \bigg), \label{Lap-eigenvalue}
\eeqa 
where $j=0,\ldots,n-1$. 
\eel
 \proof The eigenvalue equation reads
 \beqa
 \fr{\v_{\k+1}-2\v_\k+\v_{\k-1}}{\xi^2} = \la \v_\k, \quad \k=0,\ldots, n-1,
 \eeqa
 $\v_{-1}=\v_0$, $\v_{n}=\v_{n-1}$. We immediately note that $\v=\mrm{const}$ and $\la=0$ is a solution, so we can restrict attention to non-constant $\v$. We make a change of variables 
 \beqa
 \w_{\k}:=\v_{\k+1} -\v_\k, \quad \k=-1,\ldots n-1.
 \eeqa
This gives 
\beqa
\w_{\k+1}-\w_\k \2=\2\xi^2 \la \v_\k  \label{recurrence-first-step} \\   
                          \2=\2 \xi^2 \la \w_{\k-1} +  \xi^2 \la \v_{\k-1} \\
                          \2=\2 \xi^2 \la \w_{\k-1} +(\w_\k -\w_{\k-1}), \label{recurrence-third-step} \\
 \w_{\k+1} \2=\2  (2+  \xi^2 \la)\w_\k -\w_{\k-1},
\eeqa
where in (\ref{recurrence-third-step}) we applied (\ref{recurrence-first-step}) with  $\k\to \k-1$. The Neumann
boundary conditions now read $\w_{-1}=0, \w_{n-1}=0$. Setting $2\al= 2+\xi^2 \la$ we have the recurrence
\beqa
\w_{\k+1}=2\al \w_\k -\w_{\k-1}, \quad \w_{-1}=0, \,\, \w_{n-1}=0.
\eeqa 
Assuming $\w_0=1$ (to exclude that $\v=\mrm{const}$) we get
\beqa
\w_{\k}=U_\k(\al), 
\eeqa
where $U_\k$ is the $\k$-th Chebyshev polynomial of the 2nd kind by Lemma~\ref{Chebyshev} below. Now $\w_{n-1}=0$ gives
\beqa
U_{n-1}(\al)=0, 
\eeqa
which holds for  $\al_j=\cos\big( \fr{j}{n}\pi)$, $j=1, \ldots, n-1$ by (\ref{eigenvalues}) below. This, together with $2\al= 2+\xi^2 \la$,
gives the remaining $n-1$ eigenvalues. \qed
\bel\label{Chebyshev}\footnote{\ma See Wikipedia: \href{https://en.wikipedia.org/wiki/Chebyshev_polynomials}{Chebyshev polynomials}.} The Chebyshev polynomials of the 2nd kind, defined by the recurrence relation  
\beqa
U_0(\al)=1, \quad U_1(\al)=2\al, \quad U_{\k+1}(\al)=2xU_\k(\al)-U_{\k-1}(\al),  
\eeqa 
have the following property: Each $U_n$ has $n$ different simple roots in $[-1, 1]$ given by
\beqa
\al_j=\cos\bigg( \fr{j}{n+1}\pi\bigg), \quad j=1, \ldots, n.  \label{eigenvalues}
\eeqa 
 \eel

\subsection{Laplacian on $\Om\subset \xi\mathbb{Z}^d$ with Neumann boundary conditions}

Let $\Om:=I ^{\times d}$ be a {\ma hypercube} in the lattice, where $I=\xi[0,1, 2,\ldots, n-1]$
and write $\k=(\k_0,\ldots, \k_{d-1})=(\k_{\mu})_{\mu=0,\ldots,d-1}$ for integer parameters labelling elements $x:=\eta \k$  of $\Om$. 
The boundary $\pa \Om\subset \Om$ consists of $2d$ faces
\beqa
 \pa \Om\2=\2 \bigcup_{\mu=0}^{d-1}     ( \underbrace{I\times\cdots\times  \{0\}}_{\mu+1} \times\cdots \times I ) \cup 
 \bigcup_{\mu=0}^{d-1}     ( \underbrace{I \times\cdots\times  \{  n-1 \}}_{\mu+1} \times\cdots \times I )\non\\
\2=:\2 \bigcup_{\mu=0}^{d-1}(\pa\Om)_{\mu} \cup \bigcup_{\mu=0}^{d-1}(\pa\Om)^{\mu}.
   \label{boundary}
\eeqa
We consider the Hilbert space $\L^2(\Om)=\L^2(I)^{\otimes d}$  whose scalar product, in accordance with (\ref{scalar-product-1d}), has the form 
\beqa
\lan f,g\ran=\xi^d \sum_{\eta \k \in \Om} \bar{f}_{\k} g_{\k}=\xi^d \sum_{x \in \Om} \bar{f}(x)g(x).
\eeqa
A basis in this space is formed by the delta functions
\beqa
\de_{\xi; \k}:=\de_{\xi;\k_0}\otimes\cdots \otimes \de_{\xi;\k_{d-1}}.\label{delta-definition}
\eeqa
In terms of lattice points $x:=\xi \k$ we denote $\de^{\xi}_x:=\de_{\xi; \k}$.
For a linear map $A: \L^2(\Om)\to \L^2(\Om)$ we define its kernel via
\beqa
A(x,x')=A_{\k,\eell}=\lan \de_{\xi; \k}, A \de_{\xi; \eell}\ran=\lan \de^{\xi}_{x}, A \de^{\xi}_{x'}\ran. \label{2d-kernel}
\eeqa

The Laplacian on $\L^2(\Om)$ with  Neumann boundary conditions is defined by
\beqa
 \De_{\Om}^{\xi}:= \sum_{\mu=0}^{d-1} 1\otimes \cdots\otimes  \De^{\xi}_{I}\otimes \cdots \otimes 1.  \label{d-Laplacian}
\eeqa
This is a trivial generalization of the one-dimensional case, because it is a sum of  commuting operators. 
Next, we note the following consequence of Lemma~\ref{Laplacian-s-a}:
\bel\label{bracketing} Let $\v\in \L^2(\Om)$. Then
\beqa
\lan \v, (-\De^{\xi}_{\Om})\v\ran=\fr{1}{\xi^2} \sum_{b} \lan \pa \v(b), \pa \v(b)\ran, 
\eeqa
where the sum is over all the bonds in the lattice $\Om$ and $(\pa \v)(b)=\v(b_-)-\v(b_+)$.
\eel

Next, we study the behaviour of the Laplacian under scaling. Let  $\GG$ be some subset of $\ti\xi \mathbb{Z}^d$ for $\ti\xi>0$. {\gr (For future  convenience we distinguish $\ti \eta$ from $\eta$ which will be later set equal to $L^{-k}$). }  For $\la>0$ we 
define a scaling transformation $S^{\GG}_{\la} : \L^2(\GG) \to  \L^2(\la \GG)$ by
\beqa
S^{\GG}_{\la}f= \la^{-d/2 } f_{\la}, \quad f_{\la}(x):=f(\la^{-1}x). \label{S-definition}   
\eeqa
It is easy to see that  $S^{\GG}_{\la}$   is a unitary map and $(S^{\GG}_{\la})^*=S^{\la \GG}_{\la^{-1}}$. In particular:
\beqa
\|({S}^{\GG}_{\la} f )\|^2_{2,\la \GG}=(\la \ti\xi)^d  \sum_{x\in \la \GG} |(S^{\GG}_{\la}f)(x)|^2 \2=\2 \ti{\xi}^{d} \sum_{x'\in \GG  } |f(x')|^2=\|f\|_{2,\GG}^2.
\eeqa
{\gr We also note the semigroup property: $S_{\la_2}^{\la_1\GG}S_{\la_1}^{\GG}=S_{\la_2\la_1}^{\GG}$.} For $\GG=\Om$ we will often skip the superscript of $S^{\GG}_{\la}$.

We check the behaviour of the Laplacian under scaling:
\bel\label{De-scaling} We have, for  $\la>0$, 
\beqa
 \De^{\xi}_{\Om} = \la^{2} (S_{\la}^{\Om})^* \De^{\la\xi}_{\la\Om} S^{\Om}_{\la}.
\eeqa
\eel
\proof We note the relation
\beqa
(\De^{\la\xi}_{\la\Om} f_{\la} )(\la x) =\sum_{\mu=0}^{d-1}\fr{f_{\la}(\la x+ e_{\mu} \la \xi ) -2  f_{\la}(\la x) + f_{\la}(\la x-e_{\mu}\la \xi)     }{ (\la \xi)^2}
=  \la^{-2}(\De^{\xi}_{\Om} f)(x),  
\eeqa
where $\{e_{\mu}\}_{\mu=0,1,\ldots, d-1}$,  are basis vectors.
By rewriting this expression using $S_{\la}$ we obtain the formula. Clearly, modifications at the boundaries of the region do not change the result. \qed
\subsection{Averaging operator on $\Om$}\label{finite-lattice-subsection}

Recall that the finite lattice in one dimension has the form $I:=\xi[0,1,2,\ldots,n-1]$. We fix an odd\footnote{The assumption that $L$ is odd is used in Lemma~\ref{Fourier-transform-lemma}. It is consistent with \cite{Di13}.} integer $L>1$, set $\xi=L^{-k}$ 
and  $n-1=L^m$ for some fixed $m\geq k$. Now we define the coarse lattice for $1\leq j\leq k$: 
 \beqa
 I_{j} =  (L^j I)\cap I \2=\2    (L^j \xi)[0,1,2,\ldots,L^m] \cap  \xi[0,1,2,\ldots, L^m] \non\\
\2=\2 (L^j \xi)[0,1,2,\ldots, L^{m-j}].
\eeqa
Next, consider the $d$-dimensional case with $\Om=I^{\times d}$. The coarse finite lattice
has the form
\beqa
\Om_{j}= L^j\Om\cap \Om=(L^j \xi)[0,1,2,\ldots, L^{m-j}]^{\times d}.
\eeqa
For future reference, we note the following lemma, which checks that scaling is compatible with the procedure of making the lattice
coarser. 
\bel\label{Scaling-consistency} With $\Om=L^{-k}[0,1,2,\ldots, L^{m}]^{\times d}$, we have  for any $\ell\in \mathbb{Z}$,
\beqa
(L^{\ell}\Om)_j=L^{\ell}(\Om_j).
\eeqa
\eel
\proof We note that $L^{\ell}\Om$ is obtained from $\Om$ by changing the lattice spacing $\xi$ to $L^{\ell} \xi$. Thus
\beqa
(L^{\ell}\Om)_j=L^j (L^{\ell}\Om)\cap (L^{\ell}\Om)=(L^j L^{\ell}\xi)[0,1,2,\ldots, L^{m-j}]^{\times d}=L^{\ell}(\Om_j)
\eeqa
as claimed. \qed 
 
 We define the  averaging operator  $Q: \L^2(\Om)\to  \L^2(\Om_1)$ by
\beqa
(Q\v)(y)\2=\2\fr{1}{L^d} \sum_{y_{\mud}\leq x_{\mud}'< y_{\mud}+L\xi} \v(x'), \label{averaging-start}
\eeqa
where the condition under the sum is an abbreviation for $y_{\mu}\leq x_{\mu}'< y_{\mu}+L\xi, \,\, \mu=0,1,\ldots, d-1$. 
We can iterate this procedure, remembering that upon second application $Q$  acts between different spaces, namely   $Q: \L^2(\Om_1)\to \L^2(\Om_2)$. We obtain
\beqa
(Q_2\v)(z)\2=\2(Q^2\v)(z)= \fr{1}{L^d} \sum_{   z_{\mud} \leq y_{\mud}'< z_{\mud}+L^2\xi  }   (Q\v)(y')  \non\\
           \2=\2 \fr{1}{L^{2d}} \sum_{   z_{\mud} \leq y'_{\mud}< z_{\mud}+L^2\xi  } \,\,  \sum_{y'_{\mud}\leq x'_{\mud}< y'_{\mud}+L\xi} \v(x')= \fr{1}{L^{2d}}  \sum_{   z_{\mud} \leq x'_{\mud}< z_{\mud}+L^2\xi  } \v(x').
  \eeqa
 By iterating, we obtain $Q_j: \L^2(\Om)\to \L^2(\Om_j)$: 
 \beqa
 (Q_j\v)(z)= \fr{1}{L^{jd}}  \sum_{   z_{\mud} \leq x'_{\mud}< z_{\mud}+L^j\xi  } \v(x'). \label{Q-k-x}
  \eeqa
We denote the region of summation in (\ref{Q-k-x}) $B_j(z)$ and call $z$ the label of this box. For $j=k$ we have $L^k\xi=1$, hence $\triangle_z:=B_k(z)$
is a unit box. If there is a risk of confusion, we will denote $Q_j$ by $Q_{\Om,j}$.

It is easy to see that the adjoint    $Q_j^*: \L^2(\Om_j)\to \L^2(\Om)$ has the form
\beqa
(Q^*_j \w)(x)=\w(y_x),  \label{y-x}
\eeqa
where $y_{x,\mu}:=\big[ \fr{x_{\mu}}{L^j\xi} \big] (L^j\xi)$ is an element of the coarser lattice and $[a]$ denotes the integer part of $a\in\real$.
(Equivalently, one can say that $y_x$ is the unique element of the coarser lattice $\Om_j$ s.t. $x\in B_j(y_x)$. That is  $y_x$ is the 
label of the box to which $x$ belongs).  To verify that (\ref{y-x}) defines the  adjoint of $Q_j$ we compute
\beqa
\lan \w, Q_j \v\ran_{\Om_j} \2=\2(L^j\xi)^d \sum_{y\in (L^j\xi) \mathbb{Z}^d} \ov{\w}(y) \fr{1}{L^{jd}} \sum_{y_{\mud}\leq x_{\mud}'< y_{\mud}+L^j\xi} \v(x')
= \xi^d \sum_{y\in (L^j\xi) \mathbb{Z}^d} \,\,  \sum_{y_{\mud}\leq x_{\mud}'< y_{\mud}+L^j\xi}\ov{\w}(y) \v(x') \non\\
\2=\2  \xi^d \sum_{x' \in \xi \mathbb{Z}^d}  \ov{\w}(y_{x'}) \v(x')=\lan Q^*_j \w, \v\ran_{\Om}.
\eeqa

Thus we get for $Q_j^*Q_j: \L^2(\Om)\to  \L^2(\Om)$
\beqa
(Q_j^*Q_j \v)(x) \2=\2\fr{1}{L^{jd}} \sum_{y_{x,\mud}\leq x'_{\mud}< y_{x,\mud}+L^j\xi} \v(x')= \fr{1}{L^{jd}} \sum_{  \xintj (L^j\xi) \leq x_{\mud}'<  \xintj (L^j\xi)+L^j\xi} \v(x'). \label{QQ}
\eeqa
This operator is the main building block of the lattice Green functions in Subsection~\ref{Green-subsection}.
\begin{remark} \label{remark-block-constant}
Since $Q_jQ_j^*=1$, the operator $Q_j^*Q_j$ is an orthogonal projection, {\gr hence $\|Q_j\|=\|Q_j^*\|=1$}. Its range is the subspace of `block-constant functions', i.e.,  step functions which are constant on $B_j(y)$, $y\in \Om_j$.   The statement remains true \emph{mutatis mutandis} in the case of the infinite lattice studied in Subsection~\ref{averaging-infinite-lattice}. 
\end{remark}

Now we note the behaviour of the averaging operators under the scaling transformations (\ref{S-definition}).
\bel\label{Q-scaling-x} We have, for $\la=L^{\ell}$, $\ell\in \mathbb{Z}$,
\beqa
Q_{\la\Om,j} S^{\Om}_{\la}=S^{\Om_j}_{\la}Q_{\Om,j}. \label{scaling-of-Q}
\eeqa 
\eel
\proof We compute the l.h.s. on $f\in \L^2(\Om)$
\beqa
(Q_{\la\Om} S^{\Om}_{\la}f)(y)=\fr{1}{L^{jd}}\sum_{y_{\mud}\leq x_{\mud} < y_{\mud}+\la L^j\xi} (S^{\Om}_{\la}f)(x)  
=\fr{1}{L^{jd}}\sum_{ \la^{-1}y_{\mud}\leq \la^{-1}x_{\mud} < \la^{-1}y_{\mud}+L^j\xi}  \la^{-d/2} f( \la^{-1}x). 
\eeqa
Now the r.h.s. gives
\beqa
(S^{\Om_j}_{\la}Q_{\Om,j}f)(y)\2 =\2  \la^{-d/2}  (Q_{\Om,j}f)(\la^{-1}y)= \la^{-d/2} \fr{1}{L^{jd}} \sum_{ \la^{-1}y_{\mud}\leq x_{\mud} < \la^{-1}y_{\mud}+ L\xi } f(x) \non\\
\2=\2   \fr{1}{L^{\gr jd}} \sum_{ \la^{-1}y_{\mud}\leq \la^{-1}x'_{\mud} < \la^{-1}y_{\mud}+ L^j\xi }  \la^{-d/2} f(\la^{-1}x'). 
\eeqa
This concludes the proof. \qed

\subsection{Green functions  with Neumann boundary conditions}\label{Green-subsection}
Now we define the propagator with Neumann boundary conditions as a map on $\L^2(\Om)$:
\beqa
G_k(\Om):=(-\De_{\Om}^{\xi}+  \bar{\mu}_k+ a_k Q_{\Om,k}^*Q_{\Om,k})^{-1}=:\big[-\De^{\xi}+ \bar{\mu}_k+a_k Q_{k}^*Q_{k}\big]_{\Om}^{-1}, \label{bracket-notation}
\eeqa
where  $0< c \leq a_k \leq c'<\infty$ will be {\ma specified} in (\ref{a-j}) below. The mass parameters have the form  
$\bar{\mu}_k=L^{2k}\bar{\mu}_0$ for $\bar{\mu}_0\geq 0$.  As the mass parameter will play a minor role in our discussion, we abbreviate
\beqa
-\De_{\Om}^{\xi, \bmu_k}:=-\De_{\Om}^{\xi}+\bmu_k.
\eeqa
 We first have to show that the inverse exists:
\bel \emph{\cite{Di13}} \label{inverse-lemma} The following holds: 
\begin{enumerate}
  \item For a unit cube $\triangle$, as operators on $\L^{2}(\triangle)$
\beqa 
\label{unit-box}
-\Delta^{\xi,\bmu_k}_{\triangle}  + a_{k}Q_{\triangle,k}^{*}Q_{\triangle, k} \geq c (-\Delta^{\xi}_{\triangle} + 1),
\eeqa
where $\Delta^{\xi}_{\triangle}$ has Neumann boundary conditions on $\triangle$.

\item For $\Omega$ a union of disjoint unit cubes the following inequality holds as operators on $\mcL^{2}(\Omega)$
\beqa
\label{many-boxes} 
-\Delta^{\xi,\bmu_k}_{\Omega} + a_{k}Q_{\Om, k}^{*}Q_{\Om,k} \geq c (-\Delta^{\xi}_{\Om} + 1), 
\eeqa 
\end{enumerate}
where $c>0$ is independent of $\xi$ and of the size of the box.
\eel
\begin{remark} The choice of Neumann boundary conditions {\bcc is used in the proof of the} independence of the 
constant in (\ref{many-boxes}) of the size of the box $\Om$.
  \end{remark}
\proof  If $f \in \L^{2}(\triangle)$ is constant then $-\Delta^{\xi}_{\triangle} f = 0$ and
\beqa
\langle f, a_{k} Q_{\triangle, k}^{*}Q_{\triangle, k} f \rangle = a_{k}\|f\|^{2}_2 \geq c \|f\|^{2}_2.
\eeqa
If $f \in \L^{2}(\triangle)$ is orthogonal to constant functions, then $-\Delta^{\xi}_{\triangle}$ is strictly positive with lowest eigenvalue
given by (\ref{Lap-eigenvalue}) and (\ref{d-Laplacian}): 
\beqa
(-\la^{(1)})=\fr{4}{\xi^2} \sin^2\bigg( \fr{\pi }{2n} \bigg)=\fr{4}{\xi^2} \sin^2\bigg( \fr{\pi\xi }{2(\xi+1)} \bigg),
\eeqa
where we used that we are in a unit box, so $(n-1)\xi=1$. We have $(-\la^{(1)})\geq c>0$ uniformly in $\xi\leq 1$.
Therefore,
\beqa
\langle f, (- \Delta^{\xi}_{\triangle}) f\rangle \geq c \|f \|^{2}_2
\eeqa
and since $Q_{\triangle,k}^{*}Q_{\triangle,k}$ is also positive and $\bmu_k\geq 0$
\beqa
\langle f, (-\Delta^{\xi,\bmu_k}_{\triangle} + a_{k}Q_{\triangle, k}^{*}Q_{\triangle, k}\big)f \rangle \geq c \|f\|^{2}_2.
\eeqa
This proves (\ref{unit-box}).

Now let $f \in \L^{2}(\Omega)$ {\bcc and set $f_{\triangle}=f|_{\triangle}$}. {\bcc We choose as $\triangle$ the boxes  $B_k(y), y\in \Om_k$, to ensure that $Q^*_kQ_k f_{\triangle}$ is supported in  $\triangle$}. Since $\triangle \subset \Omega$ is a unit box, we have
\beqa
\lan f, (-\Delta^{\xi, \bmu_k}_{\Om} + a_{k}Q_{\Om, k}^{*}Q_{\Om,k}) f \ran \2\geq\2 \sum_{\triangle \subset \Omega} 
\big\langle f_{\triangle}, \big(-\Delta^{\xi, \bmu_k}_{\triangle} + a_{k}Q_{\triangle, k}^{*}Q_{\triangle, k}\big) f_{\triangle} \big\rangle \\
\2\geq\2 {\ma c} \sum_{\triangle \subset \Omega} \|f_{\triangle}\|^{2}_{2} = {\ma c} \|f\|^{2}_{2}.
\eeqa
Here in the first inequality we used Lemma~\ref{bracketing} {\ma and the Neumann boundary conditions} to justify that we can drop the bonds linking different 
unit boxes $\triangle$. \qed\\ 
Now it is easy to obtain exponential decay of $G_k(\Om)$ in the $\L^2$ sense by a Combes-Thomas  \cite{CT74} argument.
Our efforts in later sections will aim at improving this decay from $\L^2$ to $L^{\infty}$.  We recall that $\triangle_y$
denotes the unit box with label $y$.
\bel\label{Combes-Thomas}\emph{\cite{Di13}} Let $\supp(f) \subset \triangle_{y}$, $\supp(f^{\prime})  \subset \triangle_{y^{\prime}}$
with $\triangle_{y}, \triangle_{y^{\prime}} \subset \Omega$  unit boxes with labels $y,y'$. Then 
\beqa
|\langle f, G_{k}(\Omega) f^{\prime} \rangle| \leq c \e^{-c_1 |y - y^{\prime}| } \| f \|_{2} \|f' \|_{2}
\eeqa
for some numerical constants $c$ and $c_1>0$. {\rcc These constants are, in particular, independent of the size of $\Om$.}
\eel
\proof See Appendix~\ref{Combes-Thomas-App}. \qed\\ 
{\bcc As this proof is rather technical, let us explain the basics of the Combes-Thomas method in
a simple example. Consider a Schr\"odinger operator $H=-\De+V(x)$ on $\real^d$, where $V$ is
a  measurable function s.t. $V\geq 1$. Let ${\rcc \triangle}_{y}$, $y\in \mathbb{Z}^d$, be unit boxes of a unit lattice, which we draw on $\real^d$. 
{\rcc Then there exists $\de>0$ s.t. 
\beqa
|\lan f_1, H^{-1} f_2\ran| \leq c \e^{-\de|y_1-y_2|} \|f_1\|_2 \|f_2\|_2,  \label{CT-expl}
\eeqa
for all square-integrable  $f_1,f_2$ s.t. $\supp(f_1)\subset \triangle_{y_1}$, $\supp(f_2)\subset \triangle_{y_2}$. }

First, set $H_q=\e^{qx}H\e^{-qx}$ and assume that $\|H_q^{-1} h\|_{\rcc 2}\leq c\|h\|_2$ uniformly in $|q|\leq {\rcc \de}{\bl \leq 1}$.
Then, we immediately get (\ref{CT-expl}) {\rcc by computing} 
\beqa
|\lan f_1, H^{-1} f_2\ran|=|\lan \e^{-q x} f_1, H^{-1}_{q} \e^{q x} f_2\ran|\leq c\|\e^{- qx}f_1\|_2 \|\e^{qx}f_2\|_2\leq c' \e^{ - q(y_1-y_2)}\|f_1\|_2\|f_2\|_2.  
\eeqa
{\rcc and setting $q=\de \fr{(y_1-y_2)}{|y_1-y_2|}$.}

In order to justify our assumption, we note that $H_q=(p+\i q)^2+V(x)$, where $p_{\mu}=-\i\partial_{x_\mu}$. Hence $H_q-H= \i 2p\cdot q-q^2$ and
 \beqa
\|(-\De+1)^{-1/2}(H_q-H)(-\De+1)^{-1/2}\|\leq c q,
\eeqa
{\ma since  $(-\De+1)^{-1/2}p_{\mu} (-\De+1)^{-1/2}$ is a bounded operator.}
Consequently, 
\beqa
|\lan f,  H_q f\ran|\geq |\lan f, H f\ran|-  | \lan f, (H_q-H)f\ran |\geq \lan f, (-\De+1) f\ran - c q\lan f,  (-\De+1) f\ran \geq c'\|f\|^2_2.
\eeqa
Setting $q$ small we get $c'>0$. Now for $f=H_q^{-1}h$
\beqa
{\bcc \|h\|_2\|  H^{-1}_q h  \|_2  }\geq |\lan h,  H^{-1}_q h\ran| \geq  c_1\| H_q^{-1}h  \|^2_2
\eeqa 
which gives $\|H_q^{-1} h\|_{\rcc 2} \leq c\|h\|_2$. \vspace{0.2cm}
}

Coming back to the main course of our discussion, we note the following lemma  for future reference in (\ref{G-k-xi-sec4}):
\bel\label{G-scaling}  For $\la_j=L^{k-j}$,  $k\geq j\geq 1$,
\beqa
G^{\xi}_j(\Om)\2:=\2\la_j^{-2}  (S^{\Om}_{\la_j})^*G_j(\la_j \Om)  S^{\Om}_{\la_j}=[-\De^{\xi, \bmu_k}_{\Om}+a_j(L^j\xi)^{-2}Q^*_{\Om,j} Q_{\Om,j}]^{-1}, 
\label{G-xi-j-Om-lab} \\
  \ti{G}_{j+1}(\la_j \Om)\2:=\2\la_j^{2}(S^{\Om}_{\la_j}) G^{\xi}_{j+1}(\Om)(S^{\Om}_{\la_j})^*=  [- \De^{\la_j\xi, \bmu_j}_{\la_j\Om}+\fr{a_{j+1}}{L^2} Q^*_{\la_j\Om,j+1} Q_{\la_j\Om,j+1}]^{-1}.
    \label{j+1-term}
\eeqa
\eel
\begin{remark} The notation $\ti{G}_{j+1}$ corresponds to $G^0_{j+1}$ in   \emph{\cite{Di13}}.
\end{remark}
\proof  Formula~(\ref{G-xi-j-Om-lab})  is obtained from  Lemmas~\ref{Q-scaling-x}, \ref{De-scaling} and the following computation
 \beqa
G^{\xi}_{j}(\Om)^{-1}\2=\2 (-\De^{\xi,\bmu_k}_{\Om}+a_j (L^j\xi)^{-2}Q_{\Om,j}^* Q_{\Om,j})\non\\
 \2 =\2 S_{\la_j}^* (- \la_j^2 \De^{\la_j \xi, \la_j^{-2}\bmu_k}_{\la_j\Om} +a_j (L^j\xi)^{-2} Q_{\la_j\Om, j}^* Q_{\la_j\Om, j})S_{\la_j} \non\\
\2=\2  \la_j^2 S_{\la_j}^*( - \De^{\la_j\xi, \la_j^{-2}\bmu_k}_{\la_j\Om}+a_j  Q_{\la_j \Om, j}^* Q_{\la_j\Om, j}) S_{\la_j}=\la_j^2 S_{\la_j}^* G_j(\la_j \Om)^{-1} S_{\la_j},  \label{G-inverting}
\eeqa
where we used $L^j\xi=\la_j^{-1}$ and $\la_j^{-2}\bmu_k=\bmu_j$. 

Now we consider~(\ref{j+1-term}): Since $Q_{\Om,j+1} (S^{\Om}_{\la_j})^*= (S^{\Om_{j+1}}_{\la_j})^* Q_{\la_j\Om,j+1}$
\beqa
S^{\Om}_{\la_j} G^{\xi}_{j+1}(\Om)(S^{\Om}_{\la_j})^*\2=\2 S^{\Om}_{\la_j} 
[ -\De^{\xi,\bmu_k}_{\Om}+a_{j+1}(L^{j+1} \xi)^{-2}Q^*_{\Om,j+1} Q_{\Om,j+1}]^{-1}(S^{\Om}_{\la_j})^* \non\\
\2=\2 [ - \la_j^2\De^{\la_j\xi, \la_j^{-2}\bmu_k}_{\la_j\Om}+a_{j+1}(L^{j+1} \xi)^{-2}Q^*_{\la_j\Om,j+1} Q_{\la_j\Om,j+1}]^{-1} \non\\
\2=\2\la_j^{-2}[- \De^{\la_j\xi,\la_j^{-2}\bmu_k}_{\la_j\Om}+\fr{a_{j+1}}{L^2} Q^*_{\la_j\Om,j+1} Q_{\la_j\Om,j+1}]^{-1}.
\eeqa
This concludes the proof.  \qed\\
Now we note the following corollary of Lemma~\ref{Combes-Thomas}. A variant of  estimate (\ref{G-zero-bound})
 is stated without detailed justification in \cite[formula (364)]{Di13}. 
\bel\label{decay-green} Let $f,f'\in \L^2(\Om )$ and ${\bl \ti{\triangle}}_y, {\bl\ti{\triangle}}_{y'}$ unit boxes in $L^{-1}\Om$. For $\supp(f_{L^{-1}})\subset {\bl \ti\triangle}_y$ and $\supp(f'_{L^{-1}})\subset {\bl \ti\triangle}_{y'}$ we have
\beqa
|\langle f, \ti{G}_{k+1}(\Omega) f^{\prime} \rangle| \leq c L^2 \e^{-c_1  |y- y^{\prime}| } \|f \|_{2} \|f' \|_{2} \label{G-zero-bound}
\eeqa
{\rcc for some numerical constants $c$ and $c_1>0$.  These constants are, in particular, independent of the size of $\Om$.}
\eel
\proof By Lemma~\ref{G-scaling}, we can write $\ti{G}_{j+1}(\la\Om)$ as
\beqa
\ti{G}_{j+1}(\la_j\Om)\2=\2 (\la_j/\la_{j+1})^2 S^{\Om}_{\la_j} (S^{\Om}_{\la_{j+1}})^* G_{j+1}(\la_{j+1}\Om) S^{\Om}_{\la_{j+1}} (S^{\Om}_{\la_j})^*\non\\
\2=\2 L^2 (S^{\la_j\Om}_{L^{-1}})^*  G_{j+1}(\la_{j+1}\Om) S^{\la_j\Om}_{L^{-1}},
\eeqa
where $\la_j:=L^{k-j}$. By setting $j=k$, we get
\beqa
\ti{G}_{k+1}(\Om)=L^2(S^{\Om}_{L^{-1}})^*  G_{k+1}(L^{-1}\Om) S^{\Om}_{L^{-1}}.
\eeqa
Thus we can write
\beqa
|\langle f, \ti{G}_{k+1}(\Omega) f^{\prime} \rangle |\2=\2 L^2|\lan (S^{\Om}_{L^{-1}})f, G_{k+1}(L^{-1}\Om) S^{\Om}_{L^{-1}}f'\ran |\non\\
\2=\2  L^{d+2} |\lan f_{L^{-1}} , G_{k+1}(L^{-1}\Om)  f'_{L^{-1}} \ran| \non\\
\2 \leq \2 c L^{d+2} \e^{-c_1 | y -   y^{\prime}| } \| f_{L^{-1}} \|_{2} \|f'_{L^{-1}} \|_{2} \non\\
\2 \leq \2 c L^{2} \e^{-c_1 |y - y^{\prime}| } \| f \|_{2} \|f' \|_{2},
\eeqa
{\bl where in the third step we applied Theorem~\ref{Combes-Thomas} with  $(\Omega, \eta, k, \triangle_y)$ replaced with $(L^{-1}\Om, L^{-1}\eta, k+1, \tilde\triangle_y)$.
This  is legitimate, because $(L^{-1}\eta)=L^{-(k+1)}$,  $\supp(f_{L^{-1}}) \subset    \ti\triangle_y $ and constants
in Theorem~\ref{Combes-Thomas} are independent of the lattice spacing and size of the lattice}. In the last step
we also exploited
\beqa
\| f_{L^{-1}} \|_{2, L^{-1}\Om}^2= L^{(-1-k)d}\sum_{x\in L^{-1}\Om}  |f(Lx)|^2=L^{-d}  \|f\|_{2,\Om}^2,
\eeqa
which concludes the proof.  \qed
\subsection{Renormalization group formula}

In this section we derive a key formula, stated in (\ref{recursive-ident}) below, which allows us to conclude exponential decay of
the kernels of Green functions. We follow the discussion from~\cite{Ba82}, where, however, the actual proof is left to the reader. {\bcc The proof 
below is adapted from \cite[Section 1.2.3]{Di04}.}
As before, we work on a finite lattice $\Om$ with spacing $\xi=L^{-k}$ and Neumann boundary conditions and denote by $\Om_{j}$ the coarse
lattices. 

We define the renormalization transformation which maps a {\ma measurable  function\footnote{\ma We identify here $\L^2_{\gr \real}(\Om)$ with ${\gr \real}^{|\Om|}$ and assume that $\rho_0$ is Borel measurable.}} $\rho_0: \L^2_{\gr \real}(\Om)\to {\bcc \complex}$ to a new {\bcc function} 
$\rho_{j}: \L^2_{\gr \real}(\Om_{j} )\to \complex$. 
We set $\rho_0$ unspecified, only require that $\int d\phi\, {\ma  |}\rho_0(\phi){\ma |}:= \int \prod_{x\in \Om} d\phi(x) {\ma |}\rho_0(\phi(\{x\}_{x\in \Om}){\ma |}$ is finite. 
We set {\gr for $j=1,2,3,\ldots$}
\beqa
T^{\xi}_{a_j,L^j}[\Om,\rho_0](\psi):=    \rho_{j}(\psi)
=\bigg(  \fr{b_j^{\xi} }{2\pi}  \bigg)^{ \fr{|\Om_{j}|}{2} } \int d\phi\,   \e^{-\h b_j^{\xi}  \sum_{y\in \Om_{j}} |\psi(y)-(Q_{\Om,j}\phi)(y)|^2    }  \rho_0(\phi), \label{T-def}
\eeqa
where $b_j^{\xi}:=a_j (L^j\xi)^{d-2}$. {\gr (We note that for $j=0$ the r.h.s. of (\ref{T-def}) is undefined, in particular it does not reproduce $\rho_0$)}.  In the notation $T^{\xi}_{a,L^j}$ the superscript $\xi$ indicates  the lattice spacing of $\Om$, the subscript $L^j$ defines the size of the averaging box and we set for some $a>0$
\beqa
a_1=a, \quad a_{j+1}=\fr{aa_j}{aL^{-2}+a_j} \quad \Rightarrow \quad a_j=a\fr{1-L^{-2}}{1-L^{-2j}}. \label{a-j}
\eeqa
The normalization constant  $\big(   b_j^{\xi}  /(2\pi)  \big)^{ \fr{|\Om_{j}|}{2} } $  ensures that
\beqa
\int d\psi \rho_{j}(\psi)=\int d\phi\, \rho_0(\phi). \label{equality-of-density-integrals}
\eeqa
We note the following:
\bel\label{ren-iteration-x} We have $T^{L^j\xi}_{a,L} T^{\xi}_{a_j, L^j}=T^{\xi}_{a_{j+1}, L^{j+1}}$ and similarly
\beqa
T^{L^{j-1} \xi}_{a,L}\ldots T^{L\xi}_{a,L} T^{\xi}_{a,L} =T^{\xi}_{a_j, L^j}.  \label{T-product}
\eeqa
\eel
\proof See Appendix \ref{ren-iteration}. \qed

Next, we choose $\rho_0$ as  {\bcc a function whose integral gives the generating functional of} the free field theory:
\beqa
\rho_0^{\bcc J}(\phi):=  \e^{-\h\lan \phi, \De^{(0), \xi}(\Om) \phi\ran} {\bcc \e^{\i \lan J,\phi\ran}}, \quad  \De^{(0),\xi}(\Om)=-\De^{\xi}_{\Om} +\bmu_k,
\label{De+bmu}
\eeqa 
{\bcc where $J\in \mcL^2_{\gr \real}(\Om)$. }
Here we set $\bmu_k>0$ to ensure that the integral of $\rho^J_0$ is finite. This assumption will be relaxed to 
$\bmu_k\geq 0$ in Theorem~\ref{Key-formula-thm} below.  

We define $\rho^J_j$, $j=1,2,3\ldots$, iteratively, using the renormalization transformation
\beqa
\rho^J_{j+1}(\psi):=T^{L^j\xi}_{a,L}[{\bcc \Om_j}, \rho^J_j] (\psi)= T^{\xi}_{a_{j+1},L^{j+1}} [\Om, \rho^J_0] (\psi), \label{rho-k-steps}
\eeqa
where in the second step we made use of (\ref{T-product}). 
\bel\label{G-lemma} The following identity holds for $j=1,2,3\ldots$
\beqa
\rho^J_{j}(\psi)= Z^{\xi}_j (\Om) \e^{-\h\lan \psi, \De^{(j), L^j\xi}(\Om) \psi\ran }{\bcc \e^{\i \lan J, \hil_j^{\eta}(\Om)\psi  \ran}} {\bcc \e^{-\h \lan J,  G_j^{\eta}(\Om)  J\ran}},
\label{rho-J-j-psi}
\eeqa
where
\beqa
\De^{(j),L^j\xi}(\Om)\2:=\2 a_j(L^j\xi)^{-2} -a_j^2 (L^j\xi)^{-4} Q_{\Om,j} G_j^{\xi}(\Om) Q_{\Om,j}^*, \label{2.81} \\
G_j^{\xi}(\Om)\2:=\2(-\De^{\xi}_{\Om}+\bmu_k+a_j(L^j\xi)^{-2}Q^*_{\Om,j} Q_{\Om,j})^{-1}, \label{G-k-xi-sec4} \\
\hil_j^{\eta}({\bcc\Om})\2:=\2 \fr{a_j}{(L^j\xi)^2 } G_j^{\xi}(\Om) Q_{\Om,j}^*
\eeqa
and $Z^{\xi}_j(\Om){\gr=\sqrt{(2\pi)^{|\Om|}\det(G_j^{\eta}(\Om))} }$ is determined by (\ref{equality-of-density-integrals}).
\eel
\proof See Appendix~\ref{G-appendix}. \qed

Finally, we define the following map on $\mcL^2(\Om_j)$
\beqa
C^{(j), L^j\xi}(\Om):=(\De^{(j), L^j\xi}(\Om)+\fr{a}{(L^{j+1}\xi)^2} Q_{\Om_j}^*Q_{\Om_j})^{-1}. \label{2.84}
\eeqa
It governs the step from $\rho^J_j$ to $\rho^J_{j+1}$ as we will see in the proof of the following theorem.
\bet\label{Key-formula-thm} The following relations hold for $\bmu_k\geq 0$  
\beqa
G^{\xi}_{j+1}(\Om)\2=\2 a_j^2(L^j\xi)^{-4} G_j^{\xi}(\Om) Q_{\Om,j}^* C^{(j),L^j\xi}({\bcc \Om}) Q_{\Om,j} G_j^{\xi}(\Om) +G_j^{\xi}(\Om), \label{main-property}\\
G^{\xi}_{k}(\Om) \2 =\2 \sum_{j=1}^{k-1} a_j^2(L^j\xi)^{-4} G_j^{\xi}(\Om) Q_{\Om,j}^* C^{(j), L^j\xi}(\Om)  Q_{\Om,j} G_j^{\xi}(\Om) + G^{\xi}_{1}(\Om),  \label{recursive-ident}
\eeqa
 and the sum in (\ref{recursive-ident}) should be skipped for $k=1$.
\eet
\newcommand{\G}{G^{\xi}}
\proof {\bcc We note that~(\ref{recursive-ident}) follows by iteration from (\ref{main-property}), so it suffices to prove the latter equation.
By Lemma~\ref{G-lemma} we can write
\beqa
\rho^J_{j+1}(\psi)= Z^{\xi}_{j+1}(\Om) \e^{-\h\lan \psi, \De^{(j+1), L^{j+1}\xi}(\Om) \psi\ran }   \e^{ \i \lan J, \hil_{j+1}^{\eta}(\Om)\psi  \ran}   \e^{-\h \lan J,  G_{j+1}^{\eta}(\Om)  J\ran}.
\label{rho+j+1}
\eeqa
On the other hand, Lemma~\ref{ren-iteration-x} gives
\beqa
\rho^J_{j+1}(\psi)=T^{L^j\xi}_{a,L} [\Om_j,\rho^J_j](\psi). \label{rho-J-j+1}
\eeqa
Thus using {\ma (\ref{rho-J-j+1}), (\ref{rho-J-j-psi})}  and definition~(\ref{T-def}), we obtain
\beqa
\rho^J_{j+1}(\psi)\2=\2 \bigg(  \fr{b_1^{L^j\xi} }{2\pi}  \bigg)^{ \fr{|\Om_{j+1}|}{2} }  \int d\ti\psi \,   \e^{-\h b_1^{L^j\xi}  \sum_{y\in \Om_{j+1}} |\psi(y)-(Q_{\Om_j}\ti\psi)(y)|^2 }
\rho^J_j(\ti\psi) \non\\
\2=\2 {\gr Z_j}  \e^{-\h \lan J,  G_{j}^{\eta}(\Om)  J\ran} \int d\ti\psi \,  \e^{-\h b_1^{L^j\xi}  \sum_{y\in \Om_{j+1}} |\psi(y)-(Q_{\Om_j}\ti\psi)(y)|^2 } 
\e^{-\h\lan \ti\psi, \De^{(j), L^{j}\xi}(\Om) \ti\psi\ran } \e^{ \i\lan J, \hil_{j}^{\eta}(\Om)\ti\psi  \ran}, \label{rho-j+1}
\eeqa
where ${\gr Z_j}:=\big(  \fr{b_1^{L^j\xi} }{2\pi}  \big)^{ \fr{|\Om_{j+1}|}{2} }Z^{\xi}_{j}(\Om)$ is an inessential constant. 
Let us now diagonalize the quadratic form in the exponent above. We define the function
\beqa
F(\ti\psi):= \h b_1^{L^j\xi}  \sum_{y\in \Om_{j+1}} |\psi(y)-(Q_{\Om_j}\ti\psi)(y)|^2 +\h\lan \ti\psi, \De^{(j), L^{j}\xi}(\Om) \ti\psi\ran
\eeqa
and compute its derivatives as in (\ref{Q-derivative})
\beqa
\fr{\pa}{\pa \ti\psi(\ti y)}F(\ti\psi)\2=\2 - \fr{b_1^{L^j\xi}}{L^d}  \big( \psi(y_{\ti y})-(Q_{\Om_j}\ti \psi)(y_{\ti y})\big)+ (L^j\eta)^d (\De^{(j), L^{j}\xi}(\Om)\ti\psi)(\ti y)\non\\
\2=\2 - \fr{b_1^{L^j\xi}}{L^d}  \big(  (Q^*_{\Om_j} \psi) (\ti y )-( Q^*_{\Om_j} Q_{\Om_j}\ti \psi)(\ti y)\big)+ (L^j\eta)^d (\De^{(j), L^{j}\xi}(\Om)\ti\psi)(\ti y), \label{first-der-X}\\
\fr{\pa}{\pa \ti\psi(\ti y')}\fr{\pa}{\pa \ti\psi(\ti y)}F(\ti\psi)\2=\2  [  \fr{b_1^{L^j\xi}}{L^d} {\rcc (L^j\eta)^d} Q^*_{\Om_j} Q_{\Om_j}+ 
(L^j\eta)^{{\rcc 2}d} \De^{(j), L^{j}\xi}(\Om)]{\ma (\ti y',\ti y)} \non\\
\2=\2(L^j\eta)^{{\rcc 2}d} (C^{(j), L^j\xi}(\Om))^{-1} {\ma (\ti y',\ti y)}.
\eeqa
We compute $\ti\psi_0$ at which the first derivative (\ref{first-der-X}) vanishes:
\beqa
- \fr{a}{ (L^{j+1}\eta)^2 }  \big(  Q^*_{\Om_j} \psi -  Q^*_{\Om_j} Q_{\Om_j}\ti \psi_0 \big)+  \De^{(j), L^{j}\xi}(\Om)\ti\psi_0=0 \quad\Rightarrow\quad
 \ti\psi_0=   \fr{a}{ (L^{j+1}\eta)^2 } C^{(j), L^j\xi}(\Om)  Q^*_{\Om_j} \psi.
\eeqa
Altogether,  we can write for $\ti\psi=\ti\psi_0+\ti\psi_1$
\beqa
F(\ti\psi)=F(\ti \psi_0)+\h \lan \ti \psi_1, C^{(j), L^j\xi}(\Om)^{-1}  \ti \psi_1\ran,
\eeqa
{\ma where the part linear in $\tilde \psi_1$ vanishes as $F(\ti \psi_0)$  is a minimum.}
By substituting this to (\ref{rho-j+1}), we obtain, {\ma referring to (\ref{Gaussian})},
\beqa
\rho^J_{j+1}(\psi)\2=\2{\gr Z_j}  \e^{-\h \lan J,  G_{j}^{\eta}(\Om)  J\ran} \e^{-F(\ti \psi_0)} \e^{\i \lan (\hil_{j}^{\eta}(\Om))^*J, \ti\psi_0  \ran} \int d\ti\psi_1\, \e^{-\h \lan \ti \psi_1, C^{(j), L^j\xi}(\Om)^{-1}  \ti \psi_1\ran}    \e^{\i \lan (\hil_{j}^{\eta}(\Om) )^*J, \ti\psi_1  \ran}\non\\
\2=\2 {\gr Z_j'(\psi)}  \e^{\i \lan (\hil_{j}^{\eta}(\Om) )^*J, \ti\psi_0  \ran} \e^{-\h \lan J,  G_{j}^{\eta}(\Om)  J\ran} \e^{-\h \lan J, \hil_{j}^{\eta}(\Om)  C^{(j), L^j\xi}(\Om)(\hil_{j}^{\eta}(\Om) )^* J\ran}, 
\label{rho-j-final}
\eeqa
where ${\gr Z_j'(\psi)}$ is independent of $J$. By comparing the expressions quadratic in $J$ in (\ref{rho-j-final}) and (\ref{rho+j+1}), we arrive at (\ref{main-property}) 
using the following fact: {\gr Suppose that $\mrm{j}$ is the complex conjugation on $\mcL^2(\Om)$ and $A$ a bounded operator on $\mcL^2(\Om)$ s.t.  $\mrm{j}A\mrm{j}=A$ and  $\lan J,AJ\ran=0$ for all $J\in  \mcL^2_{\real}(\Om)$. Then $A=0$.  }

}

 To check that this relation is valid also for $\bmu_k=0$, we note that
\beqa
G_j^{\xi,\bmu_k} - G_j^{\xi,\bmu'_k}= G_j^{\xi,\bmu_k}G_j^{\xi,\bmu'_k}(\bmu_k'-\bmu_k)
\eeqa
{\ma is a Cauchy sequence in the operator norm by Lemma~\ref{inverse-lemma}}, thus $G_j^{\xi,\bmu_k=0}=\lim_{\bmu_k'\to 0} G_j^{\xi,\bmu'_k}$ exists in norm. \qed\\
Let us recall definitions {\ma (\ref{2.81}), (\ref{2.84}) } of the following mappings on $\L^2(\Om_j)$:
\beqa
\De^{(j),L^j\xi}(\Om) \2:=\2 a_j(L^j\xi)^{-2} -a_j^2 (L^j\xi)^{-4} Q_{\Om,j} G_j^{\xi}(\Om) Q_{\Om,j}^*,\non\\
C^{(j), L^j\xi}(\Om) \2:=\2(\De^{(j), L^j\xi}(\Om) +  \fr{ a(L^{j}\xi)^{-2} } {L^2}  Q^*_{\Om_j}Q_{\Om_j} )^{-1}. \non
\eeqa
We have the following scaling properties:
\bel\label{D-G-C} For $\xi=L^{-k}$, $\la_j=L^{k-j}$ the following equalities hold, 
\beqa
\De_j(\la_j\Om)\2:=\2  \la_j^{-2} S^{\Om_j}_{\la_j} \De^{(j), L^j\xi}(\Om) (S^{\Om_j}_{\la_j})^*= (a_j-a_j^2 Q_{\la_j\Om,j} G_j(\la_j\Om)Q_{\la_j\Om, j}^*), \label{De-transf}\\
C_j(\la_j\Om)\2 := \2\la_j^{2} S_{\la_j}^{\Om_j} C^{(j), L^j\xi}(\Om) (S^{\Om_j}_{\la_j})^* = [\De_j(\la_j\Om) +\fr{a}{L^2} Q_{\la_j\Om_j}^* Q_{\la_j\Om_j} ]^{-1}, \label{C-transf}
\eeqa
where the maps $\De_j(\la_j\Om), C_j(\la_j\Om)$ act on  $\L^2(\la_j \Om_j)$.
\eel
\proof We recall from~(\ref{G-xi-j-Om-lab}) and (\ref{scaling-of-Q}) that
\beqa
Q_{\la_j\Om,j'} S^{\Om}_{\la_j} \2=\2 S^{\Om_{j'}}_{\la_j}Q_{\Om,j'}, \label{S-Q-swap} \\
 S^{\Om}_{\la_j} G^{\xi}_j(\Om)  (S^{\Om}_{\la_j})^*\2=\2  \la_j^{-2}  G_j(\la_j \Om). 
 \eeqa
Hence
\beqa
 S^{\Om_j}_{\la_j} \De^{(j),L^j\xi}(\Om) (S^{\Om_j}_{\la_j})^*\2=\2 (L^j\xi)^{-2} \big(a_j -a_j^2 (L^j\xi)^{-2} S^{\Om_j}_{\la_j}Q_{\Om,j} G_j^{\xi}(\Om) Q_{\Om,j}^*(S^{\Om_j}_{\la_j})^*\big) \non\\
\2=\2 (L^j\xi)^{-2}\big(a_j -a_j^2 (L^j\xi)^{-2} Q_{\la_j\Om,j} S^{\Om}_{\la_j}G_j^{\xi}(\Om) (S^{\Om}_{\la_j})^* Q_{\la_j\Om,j}^*\big) \non\\
\2=\2 (L^j\xi)^{-2} \big(a_j -a_j^2 (L^j\xi)^{-2}\la_j^{-2} Q_{\la_j\Om,j}  G_j(\la_j\Om) Q_{\la_j\Om,j}^*\big) \non\\
\2=\2 \la_j^2 \big(a_j -a_j^2  Q_{\la_j\Om,j}  G_j(\la_j\Om) Q_{\la_j\Om,j}^*\big).
\eeqa
Now we compute
\beqa
\la_j^{2} S_{\la_j}^{\Om_j} C^{(j), L^j\xi}(\Om) (S^{\Om_j}_{\la_j})^*
\2=\2 \la_j^{2} S_{\la_j}^{\Om_j}  (\De^{(j), L^j\xi}(\Om) +  \fr{ a(L^{j}\xi)^{-2} } {L^2}  Q^*_{\Om_j}Q_{\Om_j} )^{-1}   (S^{\Om_j}_{\la_j})^*\non\\
\2=\2 (\De_j(\la_j\Om) +  \fr{ a } {L^2}  Q^*_{\la_j\Om_j}Q_{\la_j\Om_j} )^{-1},
\eeqa
where we used that by (\ref{S-Q-swap}) and $(S_{\la}^{\gr \GG})^*=S_{\la}^{\la^{-1}{\gr \GG}}$
\beqa
Q_{\Om_j,1} (S^{\Om_j}_{\la_j})^*=Q_{\la_j^{-1} \la_j\Om_j,1} S^{\la_j\Om_j}_{\la_j^{-1}}= S^{\la_j\Om_{j+1}}_{\la_j^{-1}} Q_{\la_j\Om_j,1}.
\eeqa
This concludes the proof. \qed\\
We define the following objects:
\beqa
\hil_j(\Om)\2:=\2 a_j G_j(\Om)Q_{{\ma \Om},j}^*, \quad C_j'(\Om) :=\hil_j(\Om) C_j(\Om) \hil_j(\Om)^* \label{H-C-prime}
\eeqa
and prove a rescaled variant of the renormalization group formula~(\ref{recursive-ident}).  Relation~(\ref{ren-group-formula}) is
stated without proof in \cite[formula (373)]{Di13}, with unspecified $j=0$ term.  A proof is sketched in \cite{Di04}.
\bet For $\xi=L^{-k}$ and $\la_j=L^{k-j}$,  we have the following relation 
\beqa
(G_k(\Om)f)(x)=\sum_{j=1}^{k-1}   \la_j^{-2} \big(C'_j(\la_j\Om) f_{\la_j}\big)(\la_j x)+
\la_1^{-2}(G_1(\la_1\Om)f_{\la_1}) (\la_1 x).  \label{ren-group-formula}
\eeqa
\eet
\begin{remark} {\ma  If we tried to estimate the kernel of $G_k(\Om)$ by substituting $f=\delta^{\eta}_{x'}$ the first  term on the r.h.s. of 
(\ref{ren-group-formula}) would stay regular, while the last term would acquire $\eta^{-d}$.}
\end{remark}
\proof As a preparation, we note that, by taking adjoints in (\ref{scaling-of-Q}), using $(S^{\GG}_{\la})^*=S^{\la \GG}_{\la^{-1}}$ and rescaling using Lemma~\ref{Scaling-consistency},  we get
\beqa
Q_{\la\Om,j} S^{\Om}_{\la} \2=\2 S^{\Om_j}_{\la}Q_{\Om,j}\quad \Rightarrow\quad  S_{\la}^{\Om} Q^*_{\Om,j} =Q^*_{{\la}\Om,j} S_{\la}^{\Om_j}. \label{recall-SQ}
\eeqa
We also recall the defining relations (\ref{G-xi-j-Om-lab}), (\ref{C-transf}) 
\beqa
G^{\xi}_j(\Om)\2:=\2  \la_j^{-2} (S^{\Om}_{\la_j})^* G_j(\la_j \Om) S^{\Om}_{\la_j}, \label{G-rescaling}\\
C_j(\la_j\Om)\2 := \2\la^{2}_j S_{\la}^{\Om_j} C^{(j), L^j\xi}(\Om) (S^{\Om_j}_{\la})^*. \label{C-rescaling}
\eeqa
Now we restate formula~(\ref{recursive-ident})
\beqa
G^{\xi}_{k}(\Om) \2 =\2 \sum_{j=1}^{k-1} a_j^2 \la^{4}_j G_j^{\xi}(\Om) Q_{\Om,j}^* C^{(j), L^j\xi}(\Om)  Q_{\Om,j} G_j^{\xi}(\Om) +  G^{\xi}_{1}(\Om)
\eeqa
and note that  $G^{\xi}_{k}(\Om)=G_k(\Om)$. Thus, by (\ref{G-rescaling}), (\ref{C-rescaling}), (\ref{recall-SQ}),
\beqa
G_{k}(\Om) \2 =\2 \sum_{j=1}^{k-1} a_j^2     (S^{\Om}_{\la_j})^* G_j(\la_j \Om) S^{\Om}_{\la_j}  Q_{\Om,j}^* C^{(j), L^j\xi}(\Om)  Q_{\Om,j}
 (S^{\Om}_{\la_j})^* G_j(\la_j \Om) S^{\Om}_{\la_j}+  G^{\xi}_{1}(\Om) \\
           \2 =\2 \sum_{j=1}^{k-1} a_j^2     (S^{\Om}_{\la_j})^* G_j(\la_j \Om)   Q_{\la_j\Om,j}^* S^{\Om_j}_{\la_j}C^{(j), L^j\xi}(\Om) (S^{\Om_j}_{\la_j})^* Q_{\la_j\Om,j}
  G_j(\la_j \Om) S^{\Om}_{\la_j}+  G^{\xi}_{1}(\Om) \\
      \2 =\2 \sum_{j=1}^{k-1} a_j^2  \la_j^{-2}   (S^{\Om}_{\la_j})^* G_j(\la_j \Om)   Q_{\la_j\Om,j}^*  C_j(\la_j\Om)  Q_{\la_j\Om,j}
  G_j(\la_j \Om) S^{\Om}_{\la_j}+ \la_1^{-2} (S^{\Om}_{\la_1})^* G_1(\la_1 \Om) S^{\Om}_{\la_1}. \quad
   \eeqa 
By evaluation on a function we obtain (\ref{ren-group-formula}). \qed\\
{\gr In the following lemma we use notation $[\ldots]_{\GG}$ which was introduced in (\ref{bracket-notation}).}
\bel Let $\xi=L^{-k}$, $\la_j=L^{k-j}$. The following relation holds
\beqa
C_j(\la_j\Om)
= [ \la_j^{2} A_{j}+\ti{a}_{j,j}^2 A_{j}Q_{j} \ti{G}_{j+1}(\la_j\Om)  Q_{j}^* A_{j}]_{\la_j\Om_j }, \label{C-formula}
\eeqa
where, {\bcc $\ti{a}_{j,\ell}:=a_j(L^\ell\xi)^{-2}$ and,   as in (\ref{A-notation-app})},
\beqa
A_{j,\Om_j}=\fr{1}{\ti{a}_{j,j}} +\bigg(\fr{1}{\ti{a}_{j,j}+\ti{a}_{1,j}L^{-2}} -\fr{1}{\ti{a}_{j,j}}      \bigg) Q^*_{\Om_j}Q_{\Om_j}, \quad \ti{a}_{j,\ell}=a_j(L^\ell\xi)^{-2}.
\label{A-j-form}
\eeqa
\eel
\proof We have, according to Lemma~\ref{D-G-C},
\beqa
C_j(\la_j\Om)\2 := \2 \la_j^{2} S_{\la_j}^{\Om_j} C^{(j), L^j\xi}(\Om) (S^{\Om_j}_{\la_j})^*. \label{C-transf-x}
\eeqa
Now Lemma~\ref{App-C-lemma} gives
\beqa
C^{(j), L^j\xi}(\Om)\2=\2[A_{j}+\ti{a}_{j,j}^2 A_{j}Q_j G_{j+1}^{\xi} Q_j^* A_{j}]_{\Om_j}.
\eeqa
Furthermore, in Lemma~\ref{G-scaling} we defined
\beqa
\ti{G}_{j+1}(\la_j \Om)\2:=\2\la_j^{2} S^{\Om}_{\la_j} G^{\xi}_{j+1}(\Om)(S^{\Om}_{\la_j})^*.
 \eeqa  
Moreover, we have by $(S^{\GG}_{\la})^*=S^{\la \GG}_{\la^{-1}}$ and  $Q_{\la\Om,j'} S^{\Om}_{\la}=S^{\Om_{j'}}_{\la}Q_{\Om,j'}$ (cf. (\ref{scaling-of-Q}))  
\beqa
Q_{\Om_j,1} (S^{\Om_j}_{\la_j})^*\2=\2(S^{\Om_{j+1}}_{\la_j})^* Q_{\la_j\Om_j,1}, \\
S_{\la_j}^{\Om_j}Q_{\Om_j,1}^*Q_{\Om_j,1}  (S^{\Om_j}_{\la_j})^*\2=\2Q_{\la_j\Om_j,1}^*Q_{\la_j\Om_j,1}, \\
Q^*_{\Om,j} (S^{\Om_{j}}_{\la_j})^*\2=\2(S_{\la_j}^{\Om})^* Q^*_{\la_j\Om,{j}}. 
\eeqa
Now we have all the information to compute (\ref{C-transf-x}):
\beqa
C_j(\la_j\Om)  \2=\2 \la_j^{2} S_{\la_j}^{\Om_j}  [A_{j,\Om_j}+\ti{a}_{j,j}^2 A_{j,\Om_j}Q_{\Om,j} G_{j+1}^{\xi}(\Om)  Q_{\Om,j}^* A_{j, \Om_j}] 
(S^{\Om_j}_{\la_j})^* \non\\
\2=\2 \la_j^{2}   [A_{j,\la_j\Om_j}+\ti{a}_{j,j}^2 A_{j,\la_j\Om}Q_{\la_j\Om,j} S^{\Om}_{\la_j} G_{j+1}^{\xi}(\Om)(S^{\Om}_{\la_j})^*  Q_{\la_j\Om,j}^* 
A_{j, \la_j\Om_j}] \non\\
\2=\2    [ \la_j^{2} A_{j,\la_j\Om_j}+\ti{a}_{j,j}^2 A_{j,\la_j\Om}Q_{\la_j\Om,j} \ti{G}_{j+1}(\la_j\Om)  Q_{\la_j\Om,j}^* A_{j, \la_j\Om_j}].
\eeqa
This concludes the proof.  \qed
\subsection{{\rcc Exponential decay of lattice Green functions}}

In  Theorem \ref{main-result} below we give a proof  {\ma of} exponential decay of Green functions.
The main ingredients of the proof are Lemmas \ref{Combes-Thomas}, \ref{decay-green}, formula (\ref{ren-group-formula}) 
and the  estimate
\beqa
|\hil_k(\Om)(x,y)|\leq c \e^{-c_1|x-y|}, \quad x\in \Om, \quad y\in \Om_k, \quad c_1>0, \label{images-estimate}
\eeqa
on the expression $\hil_k(\Om):=a_k G_k(\Om)Q_k^*$ which appeared in (\ref{H-C-prime}).
It is the goal of Sections~\ref{Free-section} and \ref{Images-section} to prove  (\ref{images-estimate}). 

First, using Lemma~\ref{decay-green}, we obtain: 
\bel The following bound holds for $y,y'\in \Om_k$
\beqa
|C_k(\Om)(y,y')|\leq c  L^2 \e^{-c_1 {\bl L^{-1}} |y-y'| }, \quad y,y'\in \Om_k. \label{C-bound}
\eeqa
\eel
\proof By evaluating~(\ref{C-formula}) for $j=k$, we get
\beqa
C_k(\Om)
= [  A_{k}+\ti{a}_{k,k}^2 A_{k}Q_{k} \ti{G}_{k+1}(\Om)  Q_{k}^* A_{k}]_{\Om_k} \label{C-k-formula}
\eeqa 
and we observe that $\Om_k$ is a unit lattice. Writing $\de_y:=\de^{1}_y$, {\rcc (cf. (\ref{delta-definition}) and the line below)} 
we want to compute  $ \lan \de_y, C_k(\Om)\de_{y'} \ran$.  First, we recall {\ma from~(\ref{A-k-formula-new})}
\beqa
A_k=[\ti {a}_{k,k}+\fr{\ti{a}_{1,k}}{L^2} Q^*Q]_{\Om_k}^{-1}=[\fr{1}{\ti{a}_{k,k}} -\fr{\ti{a}_{k+1,k}  }{\ti{a}_{k,k}^2 L^{2}  } Q^*Q]_{\Om_k}
=[\fr{1}{a_k} -\fr{a_{k+1} }{a_{k}^2  L^{2} } Q^*Q]_{\Om_k},  \label{A-k-repeated}
\eeqa
since $Q^*Q$ is a projection and $ \ti{a}_{j,\ell}:=a_j(L^\ell\xi)^{-2}$, $L^k\xi=1$. Clearly, the kernel $A_k(y,y')$ vanishes for $|y-y'|_{\infty}>L$
so we only have to consider the second term in (\ref{C-k-formula}).
 
 We note that $y\mapsto ({\ma A_k} \de_{y'})(y) $  is supported in an $L$-box $B_{k+1}(z){\bl \cap \Om_k}$ with label $z$ s.t. $|z-y'|_{\infty}\leq L$.
  Now $g(x):=(Q_k^* A_k \de_{y'})(x)$ is a function in $\L^2(\Om)$ {\bl supported in $B_{k+1}(z)\subset \Om$. It has the    property that $g_{L^{-1}}\in \L^2(L^{-1}\Om) $ } is supported in ${\ma L^{-1}B_{k+1}(z)=: }{\bl \ti\triangle_{L^{-1}z}\subset L^{-1}\Om}$
  and $ |L^{-1}z-L^{-1}y'|_{\infty}\leq 1$. 
 Thus Lemma~\ref{decay-green} and the {\gr triangle inequality} give
 \beqa
 |\lan \de_y,   A_k Q_k {\ma \ti{G}}_{k+1}(\Om) Q_k^* A_k \de_{y'} \ran|\2 \leq\2 c  L^2 \e^{-c_1 |L^{-1}z-L^{-1}z'|} \| Q_k^* A_k \de_{y'}\|_{2,\Om}
 \| Q_k^* A_k \de_{y}\|_{2,\Om}\non\\
  \2\leq \2 c  L^2  \e^{2c_1  } \e^{-c_1{\bl L^{-1}} |y-y'|} \| Q_k^* A_k \de_{y'}\|_{2,\Om}  \| Q_k^* A_k \de_{y}\|_{2,\Om}\non\\
 \2\leq \2 c  L^2  \e^{2c_1} \e^{-c_1{\bl L^{-1}} |y-y'|} \| Q_k^* A_k \de_{y'}\|_{2,\Om}  \| Q_k^* A_k \de_{y}\|_{2,\Om}. \label{Estimate-A-k-g}
   \eeqa

Now we estimate the norms on the r.h.s. of (\ref{Estimate-A-k-g}). Let us first consider the $1/a_k$-contribution from $A_k$ in (\ref{A-k-repeated}).  Then $Q_k^* \de_{\ma y}$
is a characteristic function of a  unit block in $\Om$, hence,
\beqa
 \| Q_k^*  \de_{\ma y} \|_{2,\Om}=1.
 \eeqa 
Now we consider the $Q^*Q$ contribution from $A_k$. Then $Q^*Q \de_y$ is a function of 
value $1/L^d$ on one $L$-box  ${B_{k+1}(z){\bl \cap \Om_k} }$ of the unit lattice and zero everywhere else. The action of  $Q_k^*$ creates a 
block-constant function on $\Om$. We have
\beqa
 \| Q_k^*  Q^*Q\de_{y} \|_{2,\Om}=\|Q\de_{y}\|_{2,\Om_{k+1}}{\ma = L^{-d/2} \leq 1}. 
 \eeqa
 {\gr The last inequality also follows from the fact that $\|Q\|=\|Q^*\|=1$, cf. Remark~\ref{remark-block-constant}.} This concludes the proof. \qed\\
Now we  recall from (\ref{H-C-prime}) that   $C_k'(\Om):=\hil_k(\Om) C_k(\Om) \hil_k^*(\Om)$
and prove the following lemma.
\bel\label{C-k-lemma}  For $\ga_0:=\fr{1}{4} c_1 {\bl L^{-1}}$, $c_1>0$,
\beqa
|(C_k'(\Om)f)(x)|\leq c  L^{2d+2} \e^{-\ga_0 \d(x, \supp(f) )} \|f\|_{\infty},
\eeqa
where $\d(x,\supp(f)):=\inf_{y\in \supp(f)} |x-y|$. 
\eel
\proof We have 
\beqa
(C_k'(\Om)f)(x)=\sum_{y,y'\in\Om_k} \hil_k(\Om)(x,y) C_k(\Om)(y,y') (\hil_k^*(\Om) f)(y').
\eeqa
By (\ref{images-estimate}) we have $\big|\hil_k(\Om)(x,y) \big| \leq c \e^{-c_1|x-y|}$. Hence
\beqa
|(\hil_k^*(\Om) f)(y')|  \leq {\bcc \eta^d} \sum_{x\in \Om} |\lan  \hil_k(\Om) \de_{y'},  \de^{\xi}_x\ran f(x)|
\2\leq\2  c{\bcc \eta^d} \|f\|_{\infty} \e^{-\fr{c_1}{2} \d(y',\supp(f))} \sum_{ z\in \mathbb{Z}^d }   \e^{-{\bcc \eta}\fr{c_1}{2} |z| } \non\\ 
\2 \leq \2 c'  \|f\|_{\infty} \e^{-\fr{c_1}{2} \d(y',\supp(f))}.
\eeqa
Recalling from (\ref{C-bound}) that $|C_k(\Om)(y,y')|\leq c  L^2 \e^{-c_1{\bl L^{-1}}\d(y,y')}$, we obtain
\beqa
|(C_k'(\Om)f)(x)|\leq  c  L^2 \sum_{y,y'\in\Om_k  } \e^{-c_1 \d(x,y)}   \e^{-c_1{\bl L^{-1}}\d(y,y') }  \e^{-\fr{c_1}{2} \d(y',\supp(f))}   \|f \|_{\infty}.
\eeqa
By the triangle inequality 
\beqa
\fr{1}{4} c_1 {\bl L^{-1} } \d(x, \supp(f)) \2\leq\2 \fr{1}{4} c_1 {\bl L^{-1}} [ \d(x, y)+  \d(y,y')+  \d(y', \supp(f)) ] \non\\ 
\2 \leq \2  \h [c_1 \d(x, y)+  c_1 {\bl L^{-1}} \d(y,y') + \h c_1 \d(y', \supp(f))].
\eeqa
Thus we obtain
\beqa
|(C_k'(\Om)f)(x)|\2\leq\2  c  L^2  \e^{- \fr{1}{4} c_1 L^{-2}\d(x, \supp(f))} \sum_{y,y'\in \mathbb{Z}^d} \e^{-\h c_1 \d(x,y)} \e^{-\h c_1 L^{-2}\d(y,y')} \|f\|_{\infty}  \non\\
\2\leq \2 c  L^{2d+2} \e^{-\ga_0 \d(x,\supp(f))} \|f\|_{\infty},
\eeqa
where the $L$-dependence is determined by summing up the geometric series:
\beqa
\sum_{y'\in \mathbb{Z}^d} \e^{-\h c_1 L^{-2}|y'|} \leq c\bigg(\fr{1}{1-\e^{-\h c_1 L^{-2}}} \bigg)^d\leq cL^{2d}. \label{sum-geom}
\eeqa
This completes the proof. \qed
\bet\label{main-result}  {\ma There is $c_1'>0$ s.t. for all  $f\in \mcL^{\infty}(\Om)$,}
\beqa
|(G_k(\Om)f)(x)|\leq  c  L^{2d}  \e^{-c_1'  \d(x, \supp(f)) } \|f\|_{\infty}. 
\eeqa
\eet
\proof  The proof is based on formula~(\ref{ren-group-formula}), which has the form 
\beqa
(G_k(\Om)f)(x)=\sum_{j=1}^{k-1}   \la_j^{-2} \big(C'_j(\la_j\Om) f_{\la_j}\big)(\la_j x)+
\la_1^{-2}(G_1(\la_1\Om)f_{\la_1}) (\la_1 x), \label{ren-group-formula-x}
\eeqa
where $\la_j:=L^{k-j}$. .  Regarding the last term on the r.h.s. of (\ref{ren-group-formula-x}), we apply Lemma~\ref{Combes-Thomas}:
\beqa
|\la_1^{-2}(G_1(\la_1\Om)f_{\la_1}) (\la_1 x)|\2=\2\la_1^{-2}|\sum_{y'\in (\la_1\Om)_1} \lan \de^{L^{-1}}_{\la_1x},   
G_1(\la_1\Om)\mathbb{1}_{\triangle_{y'}} f_{\la_1} \ran | \non\\
\2\leq \2 \la_1^{-2} \sum_{y'\in (\la_1\Om)_1}  \e^{-c_1|y-y'|} \| \de^{L^{-1}}_{\la_1x}\|_{2,\la_1\Om}  \|\mathbb{1}_{\triangle_{y'}} f_{\la_1}\|_{2,\la_1\Om}.
\label{G-1-estimate}
\eeqa
We note that $\de^{L^{-1}}_{\la_1x}$ is supported in $\triangle_y$  s.t. $|y-\la_1x|_{\infty} \leq 1$. Furthermore, if 
$\mathbb{1}_{\triangle_{y'}} f_{\la_1}\neq 0 $ then $|y'-\supp(f_{\la_1})|_{\infty}\leq 1$. Consequently, by the inverse triangle
inequality,
\beqa
|y-y'|\geq |\la_1x - \supp(f_{\la_1})|-c=\la_1|x-\supp(f)|-c. \label{triangle-G-decay}
\eeqa
Now we estimate the norms on the r.h.s. of (\ref{G-1-estimate}):
\beqa
 \| \de^{L^{-1}}_{\la_1x}\|^2_{2,\la_1\Om}\2 = \2  L^{-d} \sum_{x'\in \la_1\Om} {\ma \fr{1}{L^{-2d}} }\de_{\la_1x,x'}  =L^d, \label{norm-L-d}\\
\|\mathbb{1}_{\triangle_{y'}} f_{\la_1}\|^2_{2,\la_1\Om}\2=\2 L^{-d} \sum_{x'\in \la_1\Om} \mathbb{1}_{\triangle_{y'}}(x')| f_{\la_1}(x')|^2\non\\
\2=\2 L^{-d} \sum_{x''\in \Om} \mathbb{1}_{\la_1^{-1}\triangle_{y'}}(x'')| f(x'')|^2\non\\
\2\leq \2 L^{-d} \|f\|^2_{\infty}\, \#(\la_1^{-1}\triangle_{y'})\leq c\|f\|^2_{\infty}, \label{norm-f-infty}
\eeqa
where we noticed that  if $x''\in \la_1^{-1}\triangle_{y'}$ then $|\la_1x''- y'|_{\infty}\leq 1$, i.e., $|x''-\la_1^{-1}y' |_{\infty} \leq \la_1^{-1}$. The
latter set of $x''\in \Om$ contains {\rcc $(2L+1)^d\leq cL^d$ elements. (Here we used that $\la_1^{-1}=\eta L$ and the ball in supremum metric is a cube)}. Making use of (\ref{G-1-estimate}), (\ref{triangle-G-decay}), (\ref{norm-L-d}), (\ref{norm-f-infty}) we get
\beqa
|\la_1^{-2}(G_1(\la_1\Om)f_{\la_1}) (\la_1 x)| \2\leq\2  c \la_1^{-2} \e^{-\h c_1\la_1|x-\supp(f)|} \sum_{y'\in L^{-1}\mathbb{Z}^d } \e^{-\h c_1|y-y'|}  {\ma L^{d/2} } 
\|f\|_{\infty} \non\\
\2\leq \2 c \la_1^{-2}{\ma L^{d/2}} \|f\|_{\infty} \e^{-\h c_1\la_1|x-\supp(f)|} \sum_{y''\in\mathbb{Z}^d } \e^{-\h c_1L^{-1}|y''|} \non\\
\2\leq \2 c \la_1^{-2} {\ma L^{3d/2} }   \e^{-\h c_1\la_1|x-\supp(f)|}  \|f\|_{\infty}\label{sum-geom-appl}\\
\2\leq\2  c {\ma L^{3d/2-2k+2 } }\e^{-\h c_1 |x-\supp(f)|}  \|f\|_{\infty},
\eeqa
where in (\ref{sum-geom-appl}) we argued as in (\ref{sum-geom}). We also note that $L^{-2k+2}\leq 1$ since $k\in \nat$.

Now we consider the sum in (\ref{ren-group-formula-x}). Recall from Lemma~\ref{C-k-lemma} that 
\beqa
|(C_k'(\Om)f)(x)|\leq  c  L^{2d+2} \e^{-\ga_0 \d(x, \supp(f) )} \|f\|_{\infty}.
\eeqa
Hence,
\beqa
|(C_j'(\la_j\Om)f_{\la_j})(\la_j x)|  \leq  c   L^{2d+2}  \e^{-\ga_0 \d(\la_j x,\, \supp( f_{\la_j}) )} \|f\|_{\infty}
\2 \leq \2 c  L^{2d+2}   \e^{-\ga_0 \la_j \d( x,\, \supp( f ) )} \|f\|_{\infty} \non\\
\2\leq\2  c    L^{2d+2}  \e^{-\ga_0 L \d( x,\, \supp( f ) )} \|f\|_{\infty},
\eeqa
where we used that
\beqa
\d(\la_j x,\, \supp( f_{\la_j}) )=\inf_{y\in \supp(f_{\la_j}) } |\la_j x- y|=\inf_{y'\in \supp(f)} \la_j|x-y'|=\la_j \d(x, \supp(f)).
\eeqa
Thus we obtain
\beqa
\big|\sum_{j=1}^{k-1}   \la_j^{-2} \big(C'_j(\la_j\Om) f_{\la_j}\big)(\la_j x)\big|\leq c  L^{2d} \e^{-\ga_0 L \d( x,\, \supp( f ) )} \|f\|_{\infty},
\eeqa
where we used
\beqa
\sum_{j=1}^{k-1} \la_j^{-2}=L^{-2} +(L^{-2})^2+\cdots+(L^{-2})^{k-1} \leq L^{-2}\fr{1}{1-L^{-2}}. 
\eeqa
{\ma Now the claim follows, recalling that} {\bl $\ga_0=c_1/ 4L$}. \qed

\section{Green functions with free boundary conditions} \label{Free-section}
\setcounter{equation}{0}

{\rcc The goal of this and the next section is to prove estimate~(\ref{images-estimate}) which was used in the proof of  Theorem~\ref{main-result}.}

\subsection{Laplacian on {\ma the infinite lattice}  $\xi\mathbb{Z}$} \label{free-subsection-one}

Consider the Hilbert space $\L^2(\xi \mathbb{Z})$ equipped with the scalar product
\beqa
\lan \v, \u\ran= \xi \sum_{\k\in \mathbb{Z}} \ov{\v}_\k \u_\k=\xi \sum_{x\in \xi\mathbb{Z} } \ov{\v(x)} \u(x). 
\eeqa
As before, we use the notation $\v_\k=\v(\xi \k)$, where $\k$ is an integer parameter corresponding to
the point $\xi \k$ of the lattice.

Now we define the discrete derivatives
\beqa
(\pa^{\xi}\v)_\k=\fr{1}{\xi} (\v_{\k+1} - \v_\k), \quad (\pa^{\xi,\da} \v)_\k=-\fr{1}{\xi} (\v_{\k} - \v_{\k-1})=- (\pa^{\xi}\v)_{\k-1}.
\eeqa
For a fixed $\xi$ these  are clearly  bounded operators. By Lemma~\ref{Leibniz-free} we obtain that 
$\pa^{\xi,\da}=(\pa^{\xi})^*$. It is also easy to see that $\pa^{\xi}$ is normal, see (\ref{De-def}) below.
By analogy with Lemma~\ref{Leibniz}, we  obtain the following:
 \bel\label{Leibniz-free} The Leibniz rule holds in the form
 \beqa
 (\pa^{\xi} (\v\u))_\k=(\pa^{\xi}\v)_\k \u_{\k+1} + \v_\k (\pa^{\xi}\u)_\k.  \label{Leibniz-x}
   \eeqa
 Integration by parts holds in the form
 \beqa
 \lan \v, \pa^{\xi} \u\ran = -\lan \pa^{\xi} \v,  \u_{\cdot+1}\ran = \lan \pa^{\xi,\da} \v,  \u\ran.
  \eeqa
  \eel
The one-dimensional Laplacian has the familiar form
\beqa
(\De^{\xi}_{d=1}\v)_\k=(-\pa^{\xi} \pa^{\xi,\da}\v)_\k=(-\pa^{\xi,\da} \pa^{\xi}\v)_\k=  \fr{\v_{\k+1} -2\v_\k+\v_{\k-1}}{\xi^2} \label{De-def}
\eeqa 
and is a bounded self-adjoint operator. We note the following:
\bel\label{projection-lemma} {\ma Let  $\k_*\in  \h \mathbb{Z}$ and  $P$ be the reflection operator}
\beqa
(P\v)_\k=\v_{2\k_*-\k}. \label{reflection-part}
\eeqa
Then  $P^2=1$ and $P=P^*$. Furthermore, 
\beqa
P \De^{\xi}_{d=1} P=\De^{\xi}_{d=1}.
\eeqa 
{\rcc (We note that $P$ depends on $\k_*$, although this is hidden in the notation).}
\eel
\proof $P^2=1$ is clear. To check $P=P^*$ we write
\beqa
\lan \v, P\u\ran=\xi \sum_{\k} \v_\k \u_{2\k_*-\k}=\xi \sum_{\k'} \v_{2\k_*-\k'} \u_{\k'}=\lan P\v, \u\ran.
\eeqa
Next, we note that
\beqa
(\pa^{\xi}P\v)_\k=\fr{1}{\xi} ((P\v)_{\k+1} - (P\v)_\k)= \fr{1}{\xi}(\v_{2\k_*-\k -1}-\v_{2\k_*-\k}).
\eeqa 
Hence
\beqa
(P\pa^{\xi}P\v)_\k= \fr{1}{\xi}(\v_{2\k_*-(2\k_*-\k) -1}-\v_{2\k_*-(2\k_*-\k)})=(\pa^{\xi,\da}\v)_\k.  
\eeqa
Therefore $P \De^{\xi}_{d=1} P=-(P\pa^{\xi}P) (P\pa^{\xi,\da}P)=-\pa^{\xi,\da} \pa^{\xi}=\De^{\xi}_{d=1}$.  \qed\\
 Let us comment on the relation between the lattice Laplacians with free and Neumann boundary conditions in one dimension.
\bed\label{D-Om-def-x} We say that a function $\v\in \L^2(\xi\mathbb{Z} )$ satisfies Neumann boundary conditions on $I$, if the following relations
hold on  the boundary
\beqa
\2 \2 (\pa^{\xi,\da} \v)(0)= 0, \label{first-relation-Neumann-x}\\ 
\2 \2 (\pa^{\xi} \v)(n-1)=0. 
\eeqa
We denote the subspace of such functions $D_{I}$. 
\eed
  Clearly, we have the following:
 \bel\label{identity-of-Laplacians-lemma-x}   For any $\v\in D_{I}$ 
\beqa
(\De^{\xi}_{I}\v_{I})(x)= (\De^{\xi}\v)(x), \quad x\in I,
\eeqa
where $\v_{I}\in \L^2(I)$ is the restriction of $\v$ to $I$.
  \eel
 Now we formulate a sufficient condition for $\v\in \L^2(\xi\mathbb{Z})$ to be
 an element of $D_{I}$. Let $P$ be a reflection with $\k_*=-1/2$ and $\ov{P}$ with
 $\k_*=n-1/2$ (cf. (\ref{reflection-part})). Thus we obtain  reflections w.r.t. lines  passing in the $(1/2)\xi$ - distance to the boundary of $I$
 and denote the actions on the lattice points by the same symbols. The  $d$-dimensional counterpart of Lemma~\ref{projection-lemma-2d-x} below (Lemma~\ref{projection-lemma-2d})  will be needed in the context of the method of images (see  Section~\ref{Images-section} and Lemma~\ref{method-of-images-lemma}).
 \bel\label{projection-lemma-2d-x} Let $\v\in \L^2(\xi\mathbb{Z})$ satisfy 
 \beqa
 (P\v)(x)=\v(x), \quad (\ov{P}\v)(x)=\v(x), 
 \eeqa 
 for $x$ at the ends of the interval $I$.  Then $\v\in D_{I}$. 
  \eel
 \begin{remark} This is analogous to the fact that a symmetric function has a vanishing derivative at zero.  
 \end{remark}
 \proof  Let us check (\ref{first-relation-Neumann-x})
 for demonstration. We write for $x=0 $
 \beqa
 \v_{0}=(P\v)_{0}=\v_{-1},
 \eeqa
 which amounts to  $(\pa^{\xi,\da} \v)(0)=0$. \qed

\subsection{Laplacian on {\ma the infinite lattice}  $\xi \mathbb{Z}^d$}
Now we work  on $\L^2( \xi \mathbb{Z}^{d} )=\L^2( \xi \mathbb{Z})^{\otimes d} $ and write $\k=(\k_0,\ldots, \k_{d-1})=(\k_{\mu})_{\mu=0,\ldots,d-1}$
for elements of $\mathbb{Z}^{d}$ parametrizing points $x:=\eta \k$ of $\xi \mathbb{Z}^d$. The scalar product in this space is given, accordingly, by
\beqa
\lan f,g\ran=\xi^d \sum_{\k \in \mathbb{Z}^{d}} \bar{f}_{\k} g_{\k}.
\eeqa

In the $d$-dimensional context the derivatives are denoted 
\beqa
\pa^{\xi}_\mu:=1\otimes \cdots \otimes \pa^{\xi}\otimes\cdots \otimes 1, \quad \pa^{\xi,\dagger}_\mu:=1\otimes \cdots \otimes \pa^{\xi,\dagger}\otimes\cdots \otimes 1.
\eeqa

Now the Laplacian is given by
\beqa
\De^{\xi}:= -\sum_{\mu=0}^{d-1}  (\pa^{\xi,\dagger}_\mu)(\pa^{\xi}_\mu).  
\eeqa
We have the following corollary of Lemma~\ref{projection-lemma}.
\bel Let $P$ be a reflection w.r.t. $\k_*\in \mathbb{Z}$ as in Lemma~\ref{projection-lemma}
and denote by 
\beqa
P_{\mu}=\underbrace{1\otimes \cdots \otimes P}_{\mu+1}\otimes \cdots \otimes 1,
\eeqa
the corresponding reflections w.r.t. the hyperplanes
\beqa
 \{(\underbrace{\k'_0,\ldots, \k_*}_{\mu+1},\ldots, \k'_{d-1})\,|\, \k'\in \mathbb{Z}^{d-1}\, \}.
\eeqa
The following relation holds:
\beqa
P_\mu\De^{\xi} P_\mu=\De^{\xi}, \quad \mu=0,1,\ldots,d-1. \label{P-mu-Delta}
\eeqa
{\rcc (We note that $P_{\mu}$ depend{\gr s} on $\k_*$, although this is hidden in the notation).}
\eel
We note as an aside that for any bounded Borel function $F: \real\to \complex$ we can define the operator $F(\De^{\xi})$ whose kernel satisfies
\beqa
F(\De^{\xi})(P_\mu x, P_\mu y)=F(\De^{\xi})(x,y), \quad \mu=0,1,\ldots d-1. \label{invariance-of-kernels}
\eeqa
Here we denoted the action of reflection on a lattice point $x$ also by $P_\mu$ so that $P_\mu\de^{\xi}_x=\de^{\xi}_{P_\mu x} $
and used (\ref{2d-kernel}), (\ref{P-mu-Delta}).

Like in Subsection~\ref{free-subsection-one}, we comment on the relation between the lattice Laplacians with free and Neumann boundary conditions in $d$ dimensions.
\bed\label{D-Om-def} We say that a function $\v\in \L^2(\xi\mathbb{Z}^d)$ satisfies Neumann boundary conditions on $\Om$, if the following relations
hold on the respective subsets of the boundary (\ref{boundary}):
\beqa
 (\pa^{\xi,\da}_\mu \v)(x)\2=\2 0  \quad \textrm{ for } \quad x\in  \pa \Om_\mu,  \label{first-relation-Neumann} \\
 (\pa^{\xi}_\mu \v)(x)\2=\20   \quad \textrm{ for } \quad x\in \pa \Om^{\mu},  
 \eeqa
$\mu=0,1,\ldots, d-1$. We denote the subspace of such functions $D_{\Om}$. 
\eed
  Clearly, we have the following:
 \bel\label{identity-of-Laplacians-lemma}   For any $\v\in D_{\Om}$ we have
\beqa
(\De^{\xi}_{\Om}\v_{\Om})(x)= (\De^{\xi}\v)(x), \quad x\in \Om,
\eeqa
where $\v_{\Om}\in \L^2(\Om)$ is the restriction of $\v\in \L^2(\eta \mathbb{Z}^d)$ to $\Om$.
  \eel
 Now we formulate a sufficient condition for $\v\in \L^2(\xi\mathbb{Z}^{d})$ to be
 an element of $D_{\Om}$. We fix $\mu$ and let $P$ be a reflection with $\k_*=-1/2$ and $\ov{P}$ with
 $\k_*=n-1/2$ (cf. Lemma \ref{projection-lemma}). Thus we obtain  reflections 
 w.r.t. hyperplanes  passing in the $(1/2)\xi$-distance to the boundary of $\Om$: 
  \beqa
P_{\mu}=\underbrace{1\otimes \cdots\otimes P}_{\mu+1} \otimes\cdots\otimes 1, \quad \ov{P}_{\mu}=\underbrace{1\otimes \cdots\otimes \ov{P}}_{\mu+1} \otimes\cdots\otimes 1 \label{mu-projections}
 \eeqa
and denote their actions on the lattice points by the same symbols. 
We have:
 \bel\label{projection-lemma-2d} Let $\v\in \L^2(\xi\mathbb{Z}^{d})$ satisfy 
 \beqa
(P_{\mu}\v)(x)\2=\2 \v(x), \quad x\in \pa\Om_{\mu}, \\
(\ov{P}_{\mu}\v)(x)\2=\2 \v(x), \quad x\in \pa\Om^{\mu},
 \eeqa 
  $\mu=0,1,\ldots, d-1$.  Then $\v\in D_{\Om}$. 
  \eel
 \proof  Let us check (\ref{first-relation-Neumann})  for demonstration. We fix $\mu$ and write for 
 $x=\xi \k \in  ( \underbrace{ I\times\cdots\times  \{0\}}_{\mu+1} \times\cdots \times I)$
  \beqa
\v_{(\k_0,\ldots, 0,\ldots, \k_{d-1})}=(P_{\mu}\v)_{(\k_0,\ldots, 0,\ldots \k_{d-1})}=\v_{\k_0,\ldots, -1,\ldots \k_{d-1}},
 \eeqa
 which amounts to  $(\pa^{\xi,\da}_\mu \v)(x)=0$. \qed

Finally, we define the lattice translation operators
\beqa
(T_{\k'} \v)_{\k}=\v_{\k-\k'}. \label{translation-def}
\eeqa
which form a unitary representation of $\mathbb{Z}^d$ acting on $\L^2(\xi\mathbb{Z}^d)$. We will also use the notation $T(x')$ for $x'=\xi \k'$. Since
\beqa
\lan T_{\k'} \u,\v\ran=\xi^d \sum_{\k\in  \mathbb{Z}^d} \ov{T_{\k'} \u}_{\k} \v_\k= \xi^d   \sum_{\k\in  \mathbb{Z}^d} \ov{\u}_{\k-\k'} \v_\k
=\xi^d \sum_{\k\in \mathbb{Z}^d} \ov{\u}_{\k} \v_{\k+\k'}= \lan  \u, T_{-\k'} \v\ran\eeqa
we also have $T_\k^*=T_{-\k}$. By obvious computations using the definitions of $\pa_{\mu}^{\xi}$ and $\pa_{\mu}^{\xi, \dagger}$ we  have
\beqa
T_{\k} \De^{\xi}  T_{\k}^*= \De^{\xi}. \label{translation-of-laplacian}
\eeqa
\subsection{Fourier transform}

We define the Fourier transform and inverse Fourier transform on $\L^2(\xi\mathbb{Z}^d)$ by
\beqa
(\mcF \v)(p):=\hat{\v}(p):=(2\pi)^{-d/2}\sum_{x\in \xi\mathbb{Z}^d} \xi^d \e^{-\i p\cdot x} \v(x), \quad 
\v(x)=(2\pi)^{-d/2} \int_{p\in \widehat{\xi\mathbb{Z}^d} } \e^{\i p\cdot x} \hat{\v}(p)\,dp, \label{Fourier-def}
\eeqa
where $\widehat{\xi\mathbb{Z}^d}$ is the  torus $[-\pi/\xi, \pi/\xi[^{\times d}$. {\bl We recall that the sum in (\ref{Fourier-def}) should not be taken literally:  $\mcF$ is defined first
on $\mcL^1(\eta\mathbb{Z}^d)$ and then extended  to $\mcL^2(\eta\mathbb{Z}^d)$ using the isometry
property checked below.}
\bel The Fourier transform (\ref{Fourier-def}) is a unitary  $\L^2(\xi\mathbb{Z}^d)\to \L^2(\widehat{\xi\mathbb{Z}^d})$.
\eel
\proof We check here only the isometry property to {\ma verify} normalization. We recall the formula
\beqa
\int _{[-\pi,\pi [^{\times d}} d\ti{p}\, \e^{\i \ti p \m}=(2\pi)^d\de_{\m,0}, \quad \m\in \mathbb{Z}^d.
\eeqa
Now we compute
\beqa
\lan \hat{\v}_1, \hat{\v}_2 \ran\2=\2 (2\pi)^{-d}\int_{[-\pi/\xi,\pi/\xi [^{\times d}} dp \sum_{x_1,x_2\in \xi\mathbb{Z}^d} \xi^{2d} \e^{\i p\cdot (x_1-x_2)} 
\ov{\v}_1(x_1) \v_2(x_2)\non\\
\2=\2  (2\pi)^{-d}\int_{[-\pi,\pi [^{\times d}} d\ti{p}\, {\ma \xi^{-d}} \sum_{\m_1,\m_2 \in \mathbb{Z}^d} \xi^{2d} \e^{\i \ti{p}\cdot (\m_1-\m_2)} 
\ov{\v}_1(\xi \m_1 ) \v_2(\xi \m_2)\non\\
\2=\2 \sum_{\m_1,\m_2 \in \mathbb{Z}^d} {\ma \xi^{d} } \de_{\m_1,\m_2}  \ov{\v}_1(\xi \m_1 ) \v_2(\xi \m_2)=\lan \v_1,\v_2\ran. \quad\quad\quad\quad  \textrm{\qed}
\eeqa

\bel The following relations hold
\beqa
\hat{\de}^{\xi}_y(p)\2=\2\fr{1}{(2\pi)^{d/2}}\e^{-\i p\cdot y}, \label{delta-Fourier}\\
\mcF \pa^{\xi}_{\mu}\mcF^{-1}\2=\2  \{ \fr{1}{\xi} (  \e^{\i \xi p_{\mu}}  - 1)\}_{p\in\widehat{\xi\mathbb{Z}^d} }, \label{Fourier-derivative}\\
\mcF \pa^{\xi,\dagger}_{\mu} \mcF^{-1}\2=\2 \{-\fr{1}{\xi} (  \e^{-\i \xi p_{\mu}}  - 1)\}_{p\in\widehat{\xi\mathbb{Z}^d} }, \label{Fourier-derivative-adjoint}\\
\mcF (-\De^{\xi}) \mcF^{-1}\2=\2  {\rcc \{ \fr{2}{\xi^{2}} \sum_{\mu=0}^{d-1} \big(1-\cos(p_{\mu}\eta) \big)\}_{p\in\widehat{\xi\mathbb{Z}^d} } } , \label{Fourier-Laplace}
\eeqa
where the r.h.s. in (\ref{Fourier-derivative})--(\ref{Fourier-Laplace}) denote multiplication operators on $\L^2(\widehat{\xi\mathbb{Z}^d})$.
\eel
\proof Relation~(\ref{delta-Fourier}) is clear. To prove (\ref{Fourier-derivative}), we get for $x=\xi \k$,
 $e_{\mu}=(\underbrace{0,\ldots,1}_{\mu+1},\ldots 0)$ 
\beqa
(\mcF \pa^{\xi}_\mu \v)(p)\2=\2(2\pi)^{-d/2}\sum_{\k \in \mathbb{Z}^d} \xi^d \e^{-\i \xi p\cdot \k}  \fr{1}{\xi} (\v_{\k+e_{\mu}} - \v_{\k})\non\\
  \2=\2 (2\pi)^{-d/2}\sum_{\k\in \mathbb{Z}^d} \xi^{d}  \fr{1}{\xi} (  \e^{-\i \xi p\cdot (\k-e_{\mu}  )}  \v_{\k} - \e^{-\i\xi p\cdot \k}\v_{\k})\non\\
\2=\2   \fr{1}{\xi} (  \e^{\i\xi p_\mu}  - 1) (\mcF \v)(p).
\eeqa
Relation~(\ref{Fourier-derivative-adjoint}) follows from~(\ref{Fourier-derivative}) by taking adjoints and (\ref{Fourier-Laplace})
follows from $\De^{\xi}=-\sum_{\mu=0}^{d-1}\pa_{\mu}^{\xi,\dagger}\pa^{\xi}_{\mu}$ via (\ref{Fourier-derivative}), (\ref{Fourier-derivative-adjoint}). \qed

\subsection{Averaging operators  on $\xi \mathbb{Z}^d$} \label{averaging-infinite-lattice}

The averaging operators on $\xi \mathbb{Z}^d$ are defined analogously to the averaging operators on a finite lattice. 
Thus we will denote them by the same symbol $Q_j$ and the discussion (\ref{averaging-start})--(\ref{QQ}) can be repeated \emph{mutatis mutandis}.
The operator $Q_j: \L^2(\xi \mathbb{Z}^d) \to \L^2(L^j \xi \mathbb{Z}^d)$ is given by the familiar formula
\beqa
(Q_j\v)(z)= \fr{1}{L^{jd}}  \sum_{   z_{\mud} \leq x'_{\mud}< z_{\mud}+L^j\xi  } \v(x') \label{Q-k-x-free}
\eeqa
{\gr and has norm equal to one, cf. Remark~\ref{remark-block-constant}.} Its adjoint has the form
\beqa
(Q^*_j \w)(x)=\w(y_x),  \label{y-x-free}
\eeqa
where $y_x$ is determined by  $x\in B_j(y_x)$. 

{\gr  As a preparation, let us  now write for $g\in \L^2(\mathbb{Z}^d)$
\beqa
f(x)=(Q_k^*g)(x)=g(y_x),
\eeqa
where $y_x$ was defined below (\ref{y-x}). The Fourier transform has the form
\beqa
\hat f(p)\2=\2(2\pi)^{-d/2} \sum_{x\in \xi \mathbb{Z}^d} \xi^d \e^{-\i p\cdot x} g([x_{\mud}])\non\\
            \2=\2 (2\pi)^{-d/2} \sum_{[x_{\mud}]\in \mathbb{Z} }\sum_{\ell_{\mud}=0}^{L^k-1}   \xi^d \e^{-\i p_{\nu}([x_{\nu}]+\xi\ell_{\nu})} g([x_{\mud}]) \non\\
            \2=\2 (2\pi)^{-d/2} u(p) \sum_{[x_{\mud}]\in \mathbb{Z}}  \e^{-\i p_{\nu}\cdot [x_{\nu}]} g([x_{\mud}]) =u(p) \hat{g}(p),     \label{hat-f}      
\eeqa
 where $u$, the Fourier kernel of $Q^*$, has the form
\beqa
u(p):=\xi^d \prod_{\mu=0}^{d-1}  \fr{1-\e^{-\i p_{\mu}}}{1-\e^{-\i p_{\mu}\xi}}. \label{u-def}
\eeqa 
We note that $\hat f$  is a function on $[-\pi/\xi, \pi/\xi[^{\times d}$ after extending $\hat{g}$  from $[-\pi, \pi[^{\times d}$ to $[-\pi/\xi, \pi/\xi[^{\times d}$ by periodicity.}

Furthermore,   $Q_j^*Q_j: \L^2(\xi\mathbb{Z}^d)\to  \L^2(\xi\mathbb{Z}^d)$ is given by
\beqa
(Q_j^*Q_j \v)(x) \2=\2\fr{1}{L^{jd}} \sum_{y_{x,\mud}\leq x'_{\mud}< y_{x,\mud}+L^j\xi} \v(x')= \fr{1}{L^{jd}} \sum_{  \xintj (L^j\xi) \leq x_{\mud}'<  \xintj (L^j\xi)+L^j\xi} \v(x'). \label{QQ-free}
\eeqa

We want to compute the Fourier transform of this expression, cf.   (\ref{Fourier-def}). 
In the following lemma it is used that $L$ is odd and $L^k\xi=1$.
\bel\label{Fourier-transform-lemma}  {\ma For $j=k$} the Fourier transform of {\ma (\ref{QQ-free})}  has the form
\beqa
(\wh{Q_k^*Q_k} \v)(p)\2=\2 u(p)  \sum_{\ell''_{\mud} =- \fr{L^k-1}{2}}^{  \fr{L^k-1}{2}}  
\ov{u(p+  2\pi\ell'')}  \hat{f}(p+2\pi\ell''),  \label{Fourier-transform-Q-Q}
\eeqa
where $u$ was defined in (\ref{u-def}).
\eel
\begin{remark} This agrees with formula~(2.45) of   \emph{\cite{Ba83}}. 
\end{remark}
\proof  See Appendix~\ref{App-Fourier-transform-lemma}. \qed 

\subsection{Green functions with free boundary conditions}\label{Green-free}

This subsection serves to decipher the discussion \cite[(2.44)-(2.51)]{Ba83}: The propagators  with free boundary conditions are operators on $\L^2(\xi\mathbb{Z}^d)$ given by
\beqa
G_k:=(-\Delta^{\xi, \bmu_k}+a_k Q^*_kQ_k)^{-1}. \label{G-k}
\eeqa
The existence of the inverse is visible in the Fourier space and will follow from the discussion below (see Lemma~\ref{G-k-lemma}). 
We consider the following equation for some $v, f\in \L^2(\xi\mathbb{Z}^d)$ 
\beqa
(-\Delta^{\xi, \bmu_k}+a_k Q^*_kQ_k)v=f. \label{D-Q}
\eeqa
We take the Fourier transform, using (\ref{Fourier-transform-Q-Q}) and (\ref{Fourier-Laplace})
\beqa
\De^{\xi}(p) \hat{v}(p) +a_k u(p)  \sum_{\ell''_{\mud} =- \fr{L^k-1}{2}}^{  \fr{L^k-1}{2}}  
\ov{u(p+  2\pi\ell'')}  \hat{v}(p+2\pi\ell'')=\hat{f}(p), \label{starting-equation}
\eeqa
where $ {\rcc \De^{\xi}(p):=\fr{2}{\xi^2}\sum_{\mu=0}^{d-1} (1-\cos(p_{\mu}\eta)) +\bmu_k}$,   
$u(p):=\xi^d \prod_{\mu=0}^{d-1}  \fr{1-\e^{-\i p_{\mu}}}{1-\e^{-\i p_{\mu}\xi}}$ {\gr (cf. (\ref{u-def}))} and we 
deliberately hide $\bmu_k$ in our notation as it will play a minor role in the following.

To solve  equation~(\ref{starting-equation}), we set
\beqa
\llan u, \hat v \rran(p):=\sum_{\ell''_{\mud} =- \fr{L^k-1}{2}}^{  \fr{L^k-1}{2}}  
\ov{u(p+  2\pi\ell'')}  \hat{v}(p+2\pi\ell''). \label{double-bracket}
\eeqa
We note that $\llan u, \hat v \rran$ depends on $k$ although this is hidden in the notation.
By Lemma~\ref{periodicity} below we have for any $\ell\in \mathbb{Z}^d$
\beqa
\llan u, \hat v \rran(p)= \llan u, \hat v \rran(p+2\pi \ell). \label{periodicity-bracket}
\eeqa
Now we obtain from~(\ref{starting-equation}) using notation (\ref{double-bracket}) and  $u_{\De}(p):= \fr{u(p)}{\De^{\xi}(p)}$, $\hat{f}_{\De}(p):= \fr{\hat{f}(p)}{\De^{\xi}(p)}$
\beqa
 \hat{v}(p) + a_k u_{\De}(p)  \llan u, \hat v \rran(p)  = \hat{f}_{\De}(p).  \label{main-equation-v}
\eeqa
Using (\ref{periodicity-bracket}), we obtain
\beqa
\2 \2 \llan u, \hat{v}\rran(p) + a_k \llan u, u_{\De}\rran(p)\,  \llan u, \hat v \rran(p)  = \llan u, \hat{f}_{\De}\rran (p), \\
\2 \2 \llan u, \hat{v}\rran(p)\big(1+a_k \llan u, u_{\De}\rran(p)\big)=  \llan u, \hat{f}_{\De}\rran (p), \\
\2 \2 \llan u, \hat{v}\rran(p)=\big(1+a_k \llan u, u_{\De}\rran(p)\big)^{-1} \llan u, \hat{f}_{\De}\rran (p). \label{double-bracket-intermediate-comp}
\eeqa
Now we substitute (\ref{double-bracket-intermediate-comp}) to (\ref{main-equation-v}), which gives a solution
\beqa
 \hat{v}(p)=\wh{(G_kf)}(p) \2 = \2 \hat{f}_{\De}(p) -   \fr{ a_k  u_{\De}(p)}{ 1+a_k \llan u, u_{\De}\rran(p) } \llan u, \hat{f}_{\De}\rran (p). 
  \label{hat-v}
 \eeqa
From this solution we can conclude {\rcc that $G_k$ is a bounded operator. (The norm may depend on parameters of the problem s.t. $\eta$, $\bmu_k$, $L$).}
\bel\label{G-k-lemma} $G_k$ is a bounded operator on $\L^2(\xi \mathbb{Z}^d)$ {\ma for any $\bmu_k\geq 0$}. 
\eel
\proof 
{\rcc Suppose first that $\bmu_k>0$. Then $ \De^{\eta}(p)\geq \bmu_k>0$ and the first term on the r.h.s. of
(\ref{hat-v}) satisfies $\|\hat{f}_{\De}\|_2\leq \bmu_k^{-1}\|\hat {f}\|_2$. As for the second term, we first note that 
$ \llan u, u_{\De}\rran(p)\geq 0$. Next, regarding that $u$ is
a bounded function (actually uniformly in $\eta$ for $p\in [-\pi/\eta,\pi/\eta[^{\times d}$), we note that
\beqa
\|\llan u, \hat{f}_{\De}\rran \|_2 \leq c \bmu_k^{-1} \|\hat f\|_2.
\eeqa 
This concludes the analysis of the case of $\bmu_k>0$.

Now suppose that $\bmu_k=0$. In this case the only possible obstruction to boundedness are non-square-integrable singularities in (\ref{hat-v}).}
{\rcc Let us justify that such singularities}  can only appear at $p=0$ for $p\in [-\pi/\eta, \pi/\eta[^{\times d}$.  
This is obvious for the  term  $\hat{f}_{\De}(p)$  on the r.h.s. of  (\ref{hat-v}). As for the second term, we first note that the denominator $1+a_k \llan u, u_{\De}\rran(p)$
is irrelevant, since $\llan u, u_{\De}\rran(p)\geq 0$. The remaining expression  $u_{\De}(p)\llan u, \hat{f}_{\De}\rran (p)$ might potentially also have {\rcc non-square-integrable}  singularities
for  $p +2\pi \ell''=0$, $- \fr{L^k-1}{2}\leq \ell''_{\mud}\leq \fr{L^k-1}{2}$, $\ell''\neq 0$, resulting from  {\rcc $\De^{\eta}(p)= \fr{1}{\eta^2}\sum_{\mu=0}^{d-1}|\e^{-\i p_{\mu}\eta} -1|^2$ }appearing in $ \hat{f}_{\De}$. These are, however,
cancelled by the factors $1-\e^{-\i p_{\mu}}=1-\e^{-\i (p_{\mu}+2\pi\ell''_{\mu})}$ appearing in $u_{\De}(p)$. 

{\rcc Let us now exclude a non-square-integrable singularity at $p=0$. For this purpose, } we reformulate (\ref{hat-v}) as follows
 \beqa
 \wh{(G_kf)}(p)\2= \2 \fr{ \hat{f}(p) }{\De^{\xi}(p)}  -   \fr{a_k u_{\De}(p) }{1 +a_k \llan u, u_{\De}\rran(p)} \llan u, \hat{f}_{\De}\rran_{\ell''=0} (p) \label{singular-G-k} \\
\2 \2 \phantom{444444444} -   \fr{a_k u_{\De}(p) }{1 +a_k \llan u, u_{\De}\rran(p)} \llan u, \hat{f}_{\De}\rran_{\ell''\neq 0} (p), \label{regular-G-k}
 \eeqa
 where the subscripts $\ell''=0,\ell''\neq 0$ pertain to the sum in (\ref{double-bracket}).  We note that (\ref{regular-G-k}) is square-integrable  
 in the interval $p\in [-\pi, \pi]^{\times d}$: {\rcc First, since $[-\pi, \pi]^{\times d}\ni  p\mapsto \De^{\eta}(p+2\pi \ell'')$ has no zeros  for $\ell''\neq 0$, we obtain 
\beqa
 \|\llan u, \hat{f}_{\De}\rran_{\ell''\neq 0} \|_2\leq C\| \hat{f} \|_2,
 \eeqa
 where $C$ may depend on $\eta$. Furthermore, we have
 \beqa
  \fr{| u_{\De}(p)| }{1 +a_k \llan u, u_{\De}\rran(p)}=\fr{ |u(p)|}{\De^{\eta}(p)+|u(p)|^2+a_k\llan u,u_{\De}  \rran_{\ell''\neq 0}(p) } \leq \fr{1}{|u(p)|},
 \eeqa
 where $|u(p)|\geq c>0$.}  Now we consider~(\ref{singular-G-k}). We rewrite it  as follows:
 \beqa
(\ref{singular-G-k})\2=\2 \bigg(1 - \fr{a_k \llan u, u_{\De}\rran_{\ell''=0}(p) }{ 1 +a_k  \llan u, u_{\De} \rran(p)}  \bigg)  \fr{  \hat{f}(p) }{\De^{\xi}(p) }\non\\
\2=\2 \bigg( \fr{1+a_k \llan u, u_{\De}\rran_{\ell''\neq 0}(p) }{1 +a_k  \llan u, u_{\De} \rran_{\ell''=0} (p) +  
a_k\llan u, u_{\De} \rran_{\ell''\neq 0}(p)}  \bigg)\fr{  \hat{f}(p) }{\De^{\xi}(p) } \non\\
\2=\2 \bigg( \fr{1+a_k \llan u, u_{\De}\rran_{\ell''\neq 0}(p) } {\De^{\xi}(p)  +a_k  |u(p)|^2 +  
a_k\llan u, u_{\De} \rran_{\ell''\neq 0}(p)   \De^{\xi}(p)  } \bigg) \hat{f}(p), \label{big-bracket}
\eeqa
 where the expression in bracket in (\ref{big-bracket}) is manifestly regular near $p=0$, since {\rcc $|u(p)|\geq c>0$}. \qed
 
Now we state a rough estimate for the decay of the kernel of $G_k$. Its weakness is
the dependence of the constants on $k$, hence on the lattice spacing $\xi=L^{-k}$. But it suffices to write formula (\ref{images-test-case-x}) below.  
\bel\label{rough-decay} The integral kernel of $G_k$ satisfies
\beqa
|G_k(x,y)|\leq c_k \e^{-c_{1,k}|x-y|}, \label{exponential-decay-claim}
\eeqa
where the constants $c_k$ and $c_{1,k}>0$ may depend on $k$.
\eel
\proof We have
\beqa
(G_k f)(x)\2=\2\fr{1}{(2\pi)^{d/2}} \int_{[-\pi/\xi,\pi/\xi[^{\times d} } dp\, \e^{\i p\cdot x}   \wh{(G_kf)}(p). 
\eeqa
Now we set $f(x)=\de^{\xi}_y(x)$,  hence   $\hat{f}(p)=(2\pi)^{-d/2} \e^{-\i p\cdot y}$ by (\ref{delta-Fourier}). Thus we can write, using (\ref{hat-v}),
\beqa
(G_k f)(x)\2=\2\fr{1}{(2\pi)} \int_{[-\pi/\xi,\pi/\xi[^{\times d} } dp\, \e^{\i p\cdot x}  
\bigg(\hat{f}_{\De}(p) -   \fr{a_ku_{\De}(p)}{ (1+a_k \llan u, u_{\De}\rran(p)\big) } \llan u, \hat{f}_{\De}\rran (p)\bigg).
\eeqa 
We have
\beqa
\hat{f}_{\De}(p)= \fr{1}{(2\pi)^{d/2}} \fr{\e^{-\i p\cdot y} }{\De^{\xi}(p) }, \quad \llan u, \hat{f}_{\De}\rran (p)=\sum_{\ell''_{\mud} =- \fr{L^k-1}{2}}^{  \fr{L^k-1}{2}}  
\ov{u(p+  2\pi\ell'')}  \fr{1}{(2\pi)^{d/2}} \fr{\e^{-\i(p+2\pi\ell''  )\cdot y} }{\De^{\xi}(p+2\pi {\bcc \ell''}) }.   
\eeqa
Consequently
\beqa
(G_k f)(x)\2=\2\fr{1}{(2\pi)^{d/2}} \int_{[-\pi/\xi,\pi/\xi[^{\times d} } dp\, \e^{\i p\cdot (x-y)} g(p),  
\eeqa
where $g$ is real analytic (cf. the proof of Lemma~\ref{G-k-lemma}) and periodic with period $2\pi/\xi$. Hence it remains periodic
in the real direction after analytic   continuation to  $S_{\infty, c^k_{\mrm{st}}}:=\{\,z=p+\i \etaq\in \complex^d \,|\, p\in \real,  |\etaq|< c^k_{\mrm{st}}\}$, where
$0<c^k_{\mrm{st}}\leq 1$. (Such $c^k_{\mrm{st}}$ exists, but may depend on $k$).  Thus, by the Cauchy theorem for a rectangular contour $C$, 
 we have  for a fixed $\mu$, 
 \beqa
 0\2=\2 \oint_C\, \e^{\i z_{\mu}(x-y)_{\mu}}g_{\mu}(z_{\mu})\,dz_{\mu}\non\\ 
 \2=\2 \int_{-\pi/\xi}^{\pi/\xi}\e^{\i p_{\mu}(x-y)_{\mu}}g_{\mu}(p_{\mu}) dp_{\mu}+\i\int_0^{  \etaq_{\mrm{st},\mu}  } \e^{\i (\pi/\xi+\i\etaq_{\mu})(x-y)_{\mu}}g_{\mu}(\pi/\xi+\i\etaq)d\etaq\non\\
  \2  \2+ \int_{\pi/\xi}^{-\pi/\xi}\e^{\i (p_{\mu}+ \i \etaq_{\mrm{st},\mu} )(x-y)_{\mu}} g_{\mu}(p_{\mu}+\i \etaq_{\mrm{st},\mu} ) dp_{\mu}+ \i\int_{  {\etaq_{\mrm{st},\mu}   }  }^{0} \e^{\i(-\pi/\xi+\i\etaq_{\mu})(x-y)_{\mu}}  
  g_{\mu}(-\pi/\xi+\i\etaq)d\etaq, \quad\quad\quad \label{Cauchy-theorem-one}
 \eeqa
 where $g_{\mu}(z_{\mu}):=g(z_0,\ldots, z_{\mu}, \ldots, z_{d-1})$  and $\etaq_{\mrm{st}}:=\h c^{k}_{\mrm{st}} \fr{(x-y)}{|x-y|}$. 
  We note that the second and fourth term on the r.h.s. of (\ref{Cauchy-theorem-one}) cancel  by periodicity and the remaining terms give  
\beqa
 \int_{-\pi/\xi}^{\pi/\xi}\e^{\i p_{\mu}(x-y)_{\mu}}g_{\mu}(p_{\mu}) dp_{\mu}=  \int_{-\pi/\xi}^{\pi/\xi}\e^{\i (p_{\mu}+ \i \etaq_{\mrm{st},\mu} )(x-y)_{\mu}} g_{\mu}(p_{\mu}+\i \etaq_{\mrm{st},\mu} ) dp_{\mu}. 
 \eeqa  
By iterating this argument, we obtain
\beqa
\int_{  [-\pi/\xi,\pi/\xi[^{\times d} }  \e^{\i p\cdot (x-y) }g(p) dp=  \int_{[-\pi/\xi,\pi/\xi[^{\times d} } \e^{ \i (p+ \i \etaq_{\mrm{st}} ) \cdot (x-y) } g(p+\i \etaq_{\mrm{st}} ) dp_{\mu}
\eeqa
which gives   (\ref{exponential-decay-claim}).  \qed

Now we will study the kernel of $G_kQ_k^*$.   Now we substitute (\ref{hat-f}) to (\ref{hat-v}):
\beqa
\wh{(G_k Q_k^*g)}(p) \2=\2 u_{\De}(p) \hat{g}(p) - a_k u_{\De}(p)  \big(1+a_k \llan u, u_{\De}\rran(p)\big)^{-1} \llan u, u_{\De}\hat{g}\rran (p) \non\\
\2=\2 \big(1  - \big(1+a_k \llan u, u_{\De}\rran(p)\big)^{-1} a_k\llan u, u_{\De}\rran (p)\big) u_{\De}(p) \hat{g}(p)\non\\
\2=\2 \fr{1}{1+a_k \llan u, u_{\De}\rran(p) }u_{\De}(p) \hat{g}(p),
\eeqa
where we used that $\hat{g}$ has period $2\pi$ to pull it out of the bracket $\llan\cdot\,,\cdot\rran$. Next, we write
\beqa
(G_k Q_k^*g)(x)=(2\pi)^{-d/2} \int_{[-\pi/\xi, \pi/\xi[^{\times d}} dp \,\e^{\i p\cdot x} \fr{1}{1+a_k \llan u, u_{\De}\rran(p) }u_{\De}(p) \hat{g}(p).
\eeqa
Now we set $g=\de^{1}_y$, which gives
\beqa
(G_k Q_k^*)(x,y)=(2\pi)^{-d} \int_{[-\pi/\xi, \pi/\xi[^{\times d}} dp \,\e^{\i p\cdot (x-y)} \fr{1}{1+a_k \llan u, u_{\De}\rran(p) }u_{\De}(p). 
\eeqa 
We note that $x\in \xi\mathbb{Z}^d$, whereas $y\in (L^k\xi)\mathbb{Z}^d=\mathbb{Z}^d$. Since $1/\xi=L^k$, we can write
\beqa
(G_k Q_k^*)(x,y)=(2\pi)^{-d} \int_{[-\pi, \pi[^{\times d}} dp\, \sum_{\ell'_{\mud}=-\fr{(L^k-1)}{2}}^{\fr{(L^k-1)}{2}}  \e^{\i (p+2\pi\ell')\cdot (x-y)} 
\fr{1}{1+a_k \llan u, u_{\De}\rran(p)}u_{\De}(p+2\pi \ell'),\,\,\, \label{G-Q-1}
\eeqa
where we used (\ref{periodicity-bracket}). The main result of this subsection is: 
\bel\label{exponential-decay} There exist numerical constants $c$ and $c_{\mrm{st}}>0$, s.t.
\beqa
|(G_k Q_k^*)(x,y)|\leq c \e^{-\h c_{\mrm{st}} |x-y|}. \label{exponential-decay-free}
\eeqa
\eel
\proof  The expression~(\ref{G-Q-1}) has the form
\beqa
(G_k Q_k^*)(x,y)\2=\2(2\pi)^{-d} \int_{[-\pi, \pi[^{\times d}}  f(p)\,dp, \\
 f(p)\2:=\2\sum_{\ell'_{\mud}=-\fr{(L^k-1)}{2}}^{\fr{(L^k-1)}{2}}  \e^{\i (p+2\pi\ell')\cdot (x-y)} 
\fr{1}{1+a_k \llan u, u_{\De}\rran(p)}u_{\De}(p+2\pi \ell').
\eeqa
We know from Lemma~\ref{bound-integrand} that $f$ has an analytic  continuation to  $S_{\real, c_{\mrm{st}}}:=\{\,z=p+\i\etaq\in \complex^d \,|\, \\ p\in \real,  |\etaq|< c_{\mrm{st}}\}$, where
$0<c_{\mrm{st}}\leq 1$, and is bounded in this region as in (\ref{statement-lemma}). Thus, by the Cauchy theorem for a rectangular contour $C$,  we have
for a fixed $\mu$
\beqa
 0= \oint_C  f_{\mu}(z_{\mu})\,dz_{\mu}\2 =\2 \int_{-\pi}^{\pi}f_{\mu}(p_{\mu}) dp_{\mu}+\i\int_0^{  \etaq_{\mrm{st},\mu}    }f_{\mu}(\pi+\i\etaq)d\etaq \non\\
  \2 + \2 \int_{\pi}^{-\pi}f_{\mu}(p_{\mu}+\i \etaq_{\mrm{st},\mu} ) dp_{\mu}+ i\int_{  {\etaq_{\mrm{st},\mu} }  }^{0} f(-\pi+\i\etaq)d\etaq, \label{Cauchy-formula} 
 \eeqa
 where $f_{\mu}(z_{\mu}):=f(z_0,\ldots, z_{\mu}, \ldots, z_{d-1})$ and $\etaq_{\mrm{st}}:=\h c_{\mrm{st}} \fr{(x-y)}{|x-y|}$. 
 We note that the second and fourth term on the r.h.s. of (\ref{Cauchy-formula}) cancel due to Lemma~\ref{periodicity}.
 Thus, by iteration, we obtain
\beqa
  \int_{[-\pi, \pi[^{\times d}}  f(p)\,dp= \int_{[-\pi, \pi[^{\times d}}  f(p+\i\etaq_{\mrm{st}})\,dp.
\eeqa 
Thus we can  write  
\beqa
\2 \2(G_k Q_k^*)(x,y)=(2\pi)^{-d} \int_{[-\pi, \pi[^{\times d} }f(p+\i   \etaq_{\mrm{st}}  ) dp \non\\
  \2 \2= (2\pi)^{-1} \int_{[-\pi, \pi[^{\times d}}\, \sum_{\ell'_{\mud}=-\fr{L^k-1}{2}}^{\fr{L^k-1}{2}}  \e^{\i (p+\i  \etaq_{\mrm{st}}  +2\pi\ell')\cdot (x-y)} 
\fr{1}{1+a_k \llan u, u_{\De}\rran(p+\i \etaq_{\mrm{st}}  )}u_{\De}(p+\i\etaq_{\mrm{st}}+2\pi \ell') dp.\quad\quad
  \eeqa
Now, making use of Lemma~\ref{bound-integrand}, we obtain (\ref{exponential-decay-free}). \qed\\
 \bel\label{periodicity} {\bl Suppose $f: \real^d\to \complex$ satisfies $f(p+2\pi L^k)=f(p)$,  $p\in\real$.   
    Then
  \beqa
  F(p):=\sum_{\ell'=-\fr{L^k-1}{2}}^{\fr{L^k-1}{2}} f(p+2\pi \ell')
  \eeqa
  satisfies  $F(p+2\pi)=F(p)$, $p\in\real$. If, in addition, $f$ is analytic in a strip $S_{\infty,c_{\mrm{st}}}:=\{\, z:=p+\i\etaq \in \complex \,|\,p\in \real^{d},  |\etaq|<c_{\mrm{st}}\,\}$
  then $f(z+2\pi L^k)=f(z)$ and $F(z+2\pi)=F(z)$ for all $z\in S_{\infty,c_{\mrm{st}}}$. }
    \eel
 \proof We note the following
 \beqa
 F({\bl p}+2\pi)=\sum_{\ell'=-\fr{L^k-1}{2}}^{\fr{L^k-1}{2}} f(p+2\pi (\ell' +1)),
  \eeqa
 thus  the shift amounts to relabelling the series, apart from the boundaries of summation.
Hence, we have
 \beqa
 F(p+2\pi)-F(p)= f(p+2\pi (\fr{(L^k-1)}{2}  +1))- f(p+2\pi (-\fr{(L^k-1)}{2}) )=0,
 \eeqa
 where the last equality follows from the periodicity of $f$ with period $2\pi L^k$. {\bl Regarding the last statement, it suffices to note that
 $z\mapsto f(z)$ and $z\mapsto f(z+2\pi L^k)$ are two analytic functions on $S_{\infty,c_{\mrm{st}}}$ which, by assumption, coincide on the real line. The same argument 
 applies to $F$.} \qed
\bel\label{bound-integrand} Let  $0< c \leq a_k \leq 1$.  We have for $-\fr{(L^k-1)}{2}\leq \ell'_{\mud}\leq \fr{(L^k -1)}{2}$, $p\in [-\pi,\pi[^{\times d}$,
\beqa
\bigg|\fr{1}{1+a_k \llan u, u_{\De}\rran(p)}u_{\De}(p+2\pi \ell')\bigg|\leq \fr{c}{\prod_{\mu=0}^{d-1}(1+|\ell'_{\mu}|)^{1+2/d} }. \label{statement-lemma}
\eeqa
 The bound remains true also for the analytic continuation of the function under the modulus on the l.h.s. of (\ref{statement-lemma}) to the strip $S_{\pi,c_{\mrm{st}}}:=\{\,z=p+\i\etaq\in \complex^d \,|\, p\in {\bcc ]-\pi,\pi[},  |\etaq|< c_{\mrm{st}}\}$, where
$0<c_{\mrm{st}}\leq 1$ is a  constant {\bl depending only on $d$}.  
\eel
\proof See Appendix~\ref{App-bound-integrand}. \qed\\
The most tedious part of the proof is to ensure that $c_{\mrm{st}}$ can be chosen uniformly in the lattice spacing (and other parameters). This aspect
is left to the reader in  \cite[(2.44)-(2.51)]{Ba83}. Some hints can be found in \cite{BOS89} in a different context. We give a lengthy but self-contained proof in Appendix~\ref{App-bound-integrand}. {\bcc Here we explain only the basic idea in the case $\bmu_0=0$. To start with, we write
\beqa
\fr{1}{1+a_k \llan u, u_{\De}\rran(p)}u_{\De}(p+2\pi \ell')=\fr{1}{\De^{\eta}(p)+a_k \llan u, u_{\De}\rran(p)\De^{\eta}(p) }  \fr{\De^{\eta}(p)}{\De^{\eta}(p+2\pi \ell')} u(p+2\pi \ell').
 \eeqa
This  regularizes $\De^{\gr \eta}(p+2\pi \ell')$ for $\ell'=0$ at $p=0$. More importantly, it facilitates the analysis of the first denominator, which we call $D$. We have
\beqa
D(p):=\De^{\eta}(p)+a_k \llan u, u_{\De}\rran(p)\De^{\eta}(p)=\De^{\eta}(p)+ a_k \llan u, u_{\De}\rran_{\ell''\neq 0}(p)\De^{\eta}(p)+  a_k|u(p)|^2,
\eeqa
where $ \llan u, u_{\De}\rran_{\ell''\neq 0}$ denotes the omission of $\ell''=0$ term. It appears explicitly as $ a_k|u(p)|^2$ and
we note that for $p\in [-\pi,\pi[$ we have the bound $a_k|u(p)|\geq c>0$. By positivity of the remaining terms, we 
conclude that $D(p)\geq a_k|u(p)|^2\geq c>0$. Thus, by the Taylor theorem
\beqa
|D(p+\i q)|=|D(p)+\i q\cdot \nabla D(p+\i q')|{\gr \geq }c-|q| | \nabla D(p+\i q')|.
\eeqa
So it suffices to show that $|\nabla D(z)|\leq c'$, for $z$ in a strip as in  the statement of Lemma~\ref{bound-integrand}, to ensure 
that the denominator does not vanish there. The dependence of the r.h.s. of (\ref{statement-lemma}) on $\ell'$ requires a more careful analysis.
 }

\section{The method of images}\label{Images-section}
\setcounter{equation}{0}

In this section we decode formula (2.43) of \cite{Ba83} and prove estimate (\ref{images-estimate}).  We  will relate the propagator $G_k(\Om)$ with Neumann boundary conditions    to  the propagator  $G_k$ with free boundary conditions
using the method of images. To this end, we need some information about the behaviour of the averaging operators
under projections: 
\bel \label{Q-reflection} The following properties hold
\beqa
P_{\mu} Q_{k}^*Q_{k} P_{\mu}= Q_{k}^*Q_{k}, \quad \ov{P}_{\mu} Q_{k}^*Q_{k} \ov{P}_{\mu} = Q_{k}^*Q_{k},\quad T(z) Q_{k}^*Q_{k} T(z)^*=Q_{k}^*Q_{k}, \quad z\in   \mathbb{Z},\eeqa
where the reflections $P_{\mu}, \ov{P}_{\mu}$, $\mu=1,\ldots,d-1$, are defined in  (\ref{reflection-part}), (\ref{mu-projections})
 and  translations $T(z)$ in  (\ref{translation-def}). Consequently,
\beqa
G_k(P_{\mu}x, P_{\mu}y)=G_k(\ov{P}_{\mu} x, \ov{P}_{\mu} y)=G_k(x+z, y+z)=G_k(x,y), \quad \mu=0,\ldots,d-1.
\eeqa
\eel
\proof We recall that $\Om$ is a union of unit boxes $\triangle_y$, $y\in \Om_k$, since we set $n-1=L^m$, $m\geq k$.  Thus $P_{\mu}$, $\ov{P}_{\mu}$, defined in  (\ref{reflection-part}), (\ref{mu-projections}),
preserve the pattern of unit boxes $\triangle_y$, $y\in \mathbb{Z}^{d}$. 
We recall from Remark~\ref{remark-block-constant} that $Q_{k}^*Q_{k}$ is the orthogonal projection on the subspace of functions in $\mcL^2(\eta\mathbb{Z}^d)$ which are
block-constant (i.e. constant on  blocks $\triangle_y$, $y\in \mathbb{Z}^{d}$). 
By assumption, the reflections $P_{\mu}$ and  $\ov{P}_{\mu}$ leave this subspace
invariant. Since they are self-adjoint, they also leave the orthogonal complement invariant. Thus we can write
\beqa
P_{\mu}  Q_{k}^*Q_{k} P_{\mu}f = P_{\mu}  Q_{k}^*Q_{k} P_{\mu}(f_{\mrm{bc}}+f_{\mrm{bc}}^{\bot}) =P_{\mu}P_{\mu} f_{\mrm{bc}}= f_{\mrm{bc}}=Q_{k}^*Q_{k}f,
\eeqa
where we decomposed $f$ into the block-constant part and its orthogonal complement.  The argument regarding $\ov{P}$ and $T(z)$
is analogous. \qed 

Define a set of image points  $\Img:=\{y_j\}_{j\in \nat}$,  by the following two 
requirements (cf. \cite{GJ}[Section 7.4])
\begin{itemize}
\item $y\in   \Img $,
\item The set $  \Img$ is invariant under the  reflections $P_{\mud}$ and $\ov{P}_{\mud}$ defined in (\ref{mu-projections}).
\end{itemize}
{\bcc This set is depicted in Figure~\ref{picture}}. The following relation between the Green functions with free and Neumann boundary conditions holds true:
\bel\label{method-of-images-lemma} For $x,y\in \Om$ the following identity holds
\beqa
G_k(\Om)(x,y)=\sum_{y_j\in \Img}G_k(x, y_j). \label{images-test-case-x}
\eeqa
\eel
\proof  Given Lemma~\ref{rough-decay} and the distribution of points $y_j$, as in the figure, we obtain that the sum in (\ref{images-test-case-x})  is convergent.  It suffices to check that 
\beqa
 (-\De_{\Om}^\xi+a_k Q_{\Om,k}^*Q_{\Om,k})   \textrm{(r.h.s. of  (\ref{images-test-case-x})) }=1.
\eeqa
Let $\ti I:=\xi [-1,0,\ldots,n-1, n]$ and $\ti \Om:=\ti I^{\times d}$ be a slightly larger box than $\Om$. We will
check that for each $y\in \Om$ and $x\in \xi\mathbb{Z}^d$ the expression
\beqa
F_y(x):=\chi_{\ti \Om}(x) \sum_{y_j\in \Img }G_k(x, y_j) \label{F-y-x}
\eeqa
is an element of $D_{\Om}\subset \L^2(\xi\mathbb{Z}^d)$ (cf. Definition~\ref{D-Om-def}). We will
use the criterion from Lemma~\ref{projection-lemma-2d}. Let $\ti{P}_{\mud}$ denote the reflections $P_{\mud}$ or 
$\ov{P}_{\mud}$. We consider 
$x\in \pa \Om$, so $\chi_{\ti \Om}(x)=\chi_{\ti \Om}(\ti{P}_{\mud}x)=1$. (We need $\chi_{\ti \Om}$ in (\ref{F-y-x}) only to ensure that $x\mapsto F_y(x)$ is in 
$\L^2(\xi\mathbb{Z}^d)$).  Then
 \beqa
 F_y(\ti{P}x){\ma =}\sum_{y_j\in \Img }G_k(\ti{P}x, y_j)=\sum_{y_j\in \Img}G_k(\ti{P}x, \ti{P}y_j)
 =\sum_{y_j\in\Img}G_k(x, y_j)=F_y(x), 
 \eeqa
where in the second step we used the invariance of the set $\Img$ under the reflections  and then we used $G_k(\ti{P}x, \ti{P}y)=G_k(x,y)$,
from Lemma~\ref{Q-reflection}.  Thus $F_y\in D_{\Om}$ and we obtain from 
Lemma~\ref{identity-of-Laplacians-lemma} and consistency of the free and Neumann averaging operators  that, for $x\in \Om$, $F_{y,\Om}:=F_y|_{\Om}$,
\beqa
\2 \2((-\De^{\xi}_{\Om}+a_k Q_{\Om,k}^*Q_{\Om,k} )F_{y, \Om})(x)= ((-\De^{\xi}+ a_k Q_{k}^*Q_{k} )F_{y})(x) \non\\
\2 \2=\sum_{y_j\in \Img, x'\in \xi\mathbb{Z}^d } \lan \de^{\xi}_x, (-\De^{\xi}+ a_k Q_{k}^*Q_{k} ) \de^{\xi}_{x'} \ran \lan \de^{\xi}_{x'}, \fr{1}{-\De^{\xi}+ a_k Q_{k}^*Q_{k}}  \de^{\xi}_{y_j} \ran
= \sum_{y_j\in \Img}  \lan\de^{\xi}_x, \de^{\xi}_{y_j} \ran =\de^{\eta}_{y}(x).
\eeqa
Here in the second step we used that for $x\in \Om$ we have $\lan \de^{\xi}_x, (-\De^{\xi}+ a_k Q_{k}^*Q_{k} ) \de^{\xi}_{x'} \ran=0$ unless $x'\in \ti \Om$ (so we could get rid of the function $\chi_{\ti \Om}$) and in the last step we made use of the fact that $y$ is the only element of $\Img$ inside $\Om$. Since $(y,x)\mapsto \de^{\eta}_{y}(x)$ is the
kernel of the identity operator on $\mcL^2(\Om)$, we conclude that $F_{y, \Om}(x)=G(\Om)(x,y)$. \qed\\
Now we can state and prove the main theorem of this section {\ma which confirms (\ref{images-estimate}).}
\bet\label{G-k-omega-Q-theorem} For $x\in \Om$ and $y\in \Om_{k}$ we have the following bound
\beqa
|(G_k(\Om) Q_{\Om,k}^*)(x,y)|\leq c \e^{-c_1|x-y|},
\eeqa
where $c,c_1$, $c_1>0$, are {\bl  constants depending only on $d$, in particular independent of the size of $\Om$}. 
\eet
\proof We compute the kernel we are interested in using Lemma~\ref{method-of-images-lemma}. We first note that, for any $g\in \L^2( \Om )$, 
\beqa
(G_k(\Om)g)(x)=\xi^d \sum_{j}\sum_{w\in \Om} G_k(x,w_j) g(w),  \label{main-th-first-formula}
\eeqa
where the first sum is over the image points and we remember that each $w_j$ is a function of $w$. Now suppose that $g=Q_{\Om,k}^*f$, $f\in \L^2(\Om_{k})$. That is
\beqa
g(w)=(Q_{\Om,k}^*f)(w)=f(z_w), \label{main-th-second-formula}
\eeqa
where $z_w\in \Om_{k}$ is the element of the coarse lattice defined by $w\in B_k(z_w)$ (cf. (\ref{y-x})).

Thus (\ref{main-th-first-formula}), (\ref{main-th-second-formula}) give
\beqa
(G_k(\Om)g)(x)=(G_k(\Om) Q_{\Om,k}^*f)(x)=\xi^d \sum_{j}\sum_{w\in \Om}  G_k(x,w_j) f(z_w).
\eeqa
Now the kernel has the form
\beqa
(G_k(\Om) Q_{\Om,k}^*)(x,y)\2 =\2(G_k(\Om) Q_{\Om,k}^* \de^{1}_y)(x) =\xi^d \sum_{j}\sum_{w\in \Om}  G_k(x, w_j) \de^{1}_y(z_w)  \non\\
   \2=\2 \xi^d\sum_{j}\sum_{w\in \Om}  G_k(x, w_j) \de_{z_w,y}=\xi^d \sum_{j}\sum_{w\in B_k(y)}  G_k(x, w_j),
   \eeqa
where $B_k(y)$ is the unit box with label $y$. We observe that, with $j=(j_0,\ldots,j_{d-1})\in \mathbb{Z}^d$, $w_{j ,\mu}=  (L^m\xi) (j_{\mu}+1)-w_{\mu}$ for $j_{\mu}$ odd and  $w_{j,\mu}=(L^m\xi) j_{\mu} +w_{\mu}$ for $j_{\mu}$ even, $\mu=0,1,\ldots, d-1$ (cf. figure and recall that $L^m\xi$ is the linear size of  $\Om$).
We can thus write, using Lemma~\ref{Q-reflection} 
\beqa
(G_k(\Om) Q_{\Om,k}^*)(x,y)=\xi^d\sum_{j}\sum_{w\in B_k(y)}  
G_k(( \ldots,  (L^m\xi) (j_{\mu}+1)-x_{\mu},\ldots, x_{\nu}- (L^m\xi) j_{\nu},\ldots ), w),
\label{expression-to-decay}
\eeqa
where $j_{\mu}$ is odd and $j_{\nu}$ is even in (\ref{expression-to-decay}).

Now we compute the kernel
\beqa
(G_k Q_{k}^*)(x,y)\2=\2 \xi^d\sum_{w\in \xi\mathbb{Z}^d} G_k(x,w) Q_k^*\de^1_y(w) \non\\
\2=\2 \xi^d\sum_{w\in \xi\mathbb{Z}^d} G_k(x,w) \de_{z_w, y} =\xi^d \sum_{w\in B_k(y)} G_k(x,w),
\eeqa
{\rcc whose exponential decay we know from Lemma~\ref{exponential-decay}.}  (In the last step we used $x\in \Om, y\in \Om_{k}$).  Hence, (\ref{expression-to-decay}) gives
\beqa
(G_k(\Om) Q_{\Om,k}^*)(x,y)\2=\2 \sum_{j}  (G_k Q_{k}^*)
(( \ldots,  (L^m\xi) (j_{\mu}+1)-x_{\mu},\ldots, x_{\nu}- (L^m\xi) j_{\nu},\ldots )  , y). \label{GQ-star-formula}
\eeqa
From this formula and Lemma~\ref{exponential-decay} the statement of the theorem is relatively clear.
Thus we postpone further details to Appendix~\ref{G-k-omega-Q-App}. \qed

\appendix 
\section{Proof of Lemma~\ref{Combes-Thomas}} \label{Combes-Thomas-App}
\setcounter{equation}{0}

In the proof we will write $\pa^{\xi}_{\mu}:=\pa^{\xi}_{\mu,\Om}$, $\De^{\xi,\bmu_k}:=\De^{\xi,\bmu_k}_{\Om}$ and $Q_{k}:=Q_{k,\Om}$ for
brevity. Let us start with some preparations. Define an operator $\e_{q}$ by its action on functions
$f \in \mcL^{2}(\Omega)$ as 
\beqa
(\e_{q} f) (x) = \e^{q \cdot x} f(x),
\eeqa
where $q \in \mathbb{R}^{2}$ is a vector s.t. $|q| \leq 1$. Then 
we compute, using Lemma~\ref{Leibniz},
\beqa
((\pa^{\xi}_{\mu})_qf)(x):= \e_{-q}\{ \pa^{\xi}_{\mu} ( f \e_q )\}(x)
\2=\2 \e_{-q}(x) (\pa^{\xi}_{\mu}f)(x) \e_q(x+\xi e_{\mu}) + \e_{-q}(x) (\pa^{\xi}_{\mu}\e_q)(x) f(x)  \non\\
\2=\2 \fr{\e^{q_{\mu}\xi} -1 }{\xi} f(x)+   \e^{\xi q_{\mu}} (\pa_{\mu}^{\xi}f)(x)  \non\\
\2=\2 q_{\mu} E_{q_{\mu}}  f(x)+ \e^{\xi q_{\mu}}(\pa^{\xi}_{\mu}f)(x), \label{derivative-computation}
\eeqa
where $E_{q_{\mu}}:=\int_0^1ds\, \e^{sq_{\mu}\xi }$ is independent of $x$. Since $|q|\leq 1$, we have $|E_{q_{\mu}}|\leq \e$. (We note that it is 
helpful for estimate~(\ref{E-q-estimate}) below to have the shift  $(\,\cdot\,+\xi e_{\mu})$ in $\e_q$ and not in $f$).
Recall that for $y \in \Om_{k}$
\beqa
(Q_{k} f) (y) =\fr{1}{L^{kd}}\sum_{x\in B_k(y)} f(x).  
\eeqa
Define
\beqa
(Q_{k, q} f) (y) :=(\e_{-q} Q_k \e_{q}f)(y) = \fr{1}{L^{kd}}\sum_{x\in B_k(y)} \e^{q \cdot (x-y)} f(x). \label{Q-k-q}
\eeqa
We will also need that $(Q_{k}^{*} f)(x) = f(y)$  if  $x\in B_k(y)$, hence
\beqa
(Q_{k, -q}^{*} f)(x) \2=\2 (\e_{-q} Q_k^* \e_q f)(x)= \e^{q\cdot (y-x)} f(y),  \label{Q-k-q-star}
\eeqa
where in the second line we use (\ref{Q-k-q}).

After this preparation we move on to the proof of Lemma~\ref{Combes-Thomas}. Let 
\beqa
\mathrm{D}_{q} :=\e_{-q}  \big[-\Delta^{\xi,\bmu_k} + a_{k}Q_{k}^{*}Q_{k} \big] \e_{q}.
\eeqa
We claim that there exists a constant $c'$ such that for $f \in \mcL^{2}(\Omega)$
\beqa
\label{D-shift}
|\lan f, [\mathrm{D}_{q} - \mathrm{D}_{0}] f \ran | \leq c' |q| \lan f, (-\Delta^{\xi} + 1) f \ran. 
\eeqa
First consider the term $\lan f, [\e_{-q} Q_{k}^{*}Q_{k} \e_{q} - Q_{k}^{*}Q_{k}] f \ran$.
From  (\ref{Q-k-q}) and (\ref{Q-k-q-star}) this can be written as $\lan f, [Q_{k, -q}^{*}Q_{k, q} - Q_{k}^{*}Q_{k}] f \ran$.
We estimate
\beqa
\|(Q_{k, q} - Q_{k}) f\|_2^2\2=\2\sum_{y\in \Om_{k}}  |\fr{1}{L^{kd}} \sum_{x\in B_k(y)} [\e^{q \cdot (x-y)} -1] f(x)|^2\non\\
\2\leq \2  cq^2  \sum_{y\in \Om_{k}}  \fr{1}{L^{kd}} \sum_{x\in B_k(y)}  \fr{1}{L^{kd}} \sum_{x'\in B_k(y)}     (f(x)^2+f(x')^2) \non\\
\2 \leq\2   cq^2  \fr{1}{L^{kd}} \sum_{x\in \Om } f(x)^2\leq cq^2 \|f\|_2^2, \label{Q-k-q-shift}
\eeqa
where the sum over $y$ and the sums over the boxes combined to the sum over the entire lattice. We used that $|x-y|\leq \sqrt{d}$, if
$x$ in the unit box with label $y$. Estimate~(\ref{Q-k-q-shift})  gives
\beqa
|\lan f, [\e_{-q} Q_{k}^{*}Q_{k} \e_{q} - Q_{k}^{*}Q_{k}] f \ran| \leq  c|q|  \| f \|_{2}^{2}, \label{Q-shift}
\eeqa
{\rcc where we applied the Cauchy-Schwarz inequality.}

The remaining part is estimated as follows 
\beqa
\2 \2|\lan f, [\e_{-q}(-\Delta^{\xi,\bmu_k})\e_{q} - (-\Delta^{\xi,\bmu_k})]  f \ran | \non\\
\2 \2\phantom{4444444}= | \lan \e_q \pa^{\xi}_{\mu}  (\e_{-q} f), \e_{-q}\pa^{\xi}_{\mu} (\e_{q} f) \ran - \lan \pa^{\xi}_{\mu} f, \pa^{\xi}_{\mu} f\ran| \non\\
\2 \2\phantom{4444444}= | \lan  (-q_{\mu}) E_{( -q_{\mu})}  f+ \e^{\gr -\xi q_{\mu}} (\pa^{\xi}_{\mu}f),  q_{\mu} E_{q_{\mu}}  f+ \e^{\xi q_{\mu}} (\pa^{\xi}_{\mu}f)  \ran - \lan \pa_{\mu}^{\xi}  f, \pa_{\mu}^{\xi}  f\ran |\non\\
\2 \2\phantom{4444444} \leq c |q| \lan f, (-\De^{\xi}+1) f\ran|, \label{E-q-estimate}
\eeqa
where {\gr summation over $\mu$ is understood.  We used here (\ref{integration-by-parts}),  (\ref{derivative-computation}), $|q|^2\leq |q|$ and 
$|\lan \pa^{\xi}_{\mu}f,f\ran|\leq \h \lan f, (-\De^{\xi}+1) f\ran|$ }. Using (\ref{E-q-estimate}) and (\ref{Q-shift}) we check (\ref{D-shift}).

 Recall from Lemma~\ref{inverse-lemma} that $\lan f, \mathrm{D}_{0} f \ran \geq c_{0}  \lan f, (-\Delta^{\xi} + 1) f \ran$, where
 $\mathrm{D}_0:=(-\De^{\xi,\bmu_k}+a_k Q_k^*Q_k)$.
Thus we can write, choosing $q$  s.t. $c'|q|\leq c_0/2$, {\ma where $c'$ appeared in (\ref{D-shift}), }
\beqa
|\lan f, \mathrm{D}_{q} f \ran| \2\geq\2   \lan f, \mathrm{D}_{0} f \ran
- |\lan f, (\mathrm{D}_{q} - \mathrm{D}_{0}) f \ran| \non \\
\2\geq\2  c_{0} \,\, |\lan f, (-\Delta^{\xi} + 1) f \ran| -  \frac{c_{0}}{2}| \lan f, (-\Delta^{\xi} + 1) f \ran| \non\\
\2=\2  \fr{1}{2} c_{0} | \lan f, (-\Delta^{\xi} + 1) f \ran| \geq \fr{1}{2} c_{0}  \|f\|^{2}_{2}. \label{D-q-est}
\eeqa
We note that $\mathrm{D}_q$ is invertible as a composition of invertible mappings. Substituting $f = \mathrm{D}_{q}^{-1} h$ to (\ref{D-q-est}) we get
\beqa
\frac{1}{2} c_{0}   \|\mathrm{D}_{q}^{-1} h \|^{2}_{2} \leq \lan  \mathrm{D}_{q}^{-1} h, h\ran
\leq \|h\|_{2} \|\mathrm{D}_{q}^{-1} h \|_{2} \quad  \Rightarrow \quad
\|\mathrm{D}_{q}^{-1} h\|_{2} \leq 2 c_{0}^{-1} \|h\|_{2}.
\eeqa
Since $\mathrm{D}_{q}^{-1} = \e_{-q} G_{k}(\Omega) \e_{q}$ this reads
\beqa
\|\e_{-q} G_{k}(\Om) \e_{q} h\|_{2} \leq c \|h \|_{2}.
\eeqa
Now let $c_1 := \mrm{min} \{\frac{1}{2} (c')^{-1} c_{0}, 1\}$. Then, {\rcc for $|q| \leq  c_0/(2c')$ as specified above (\ref{D-q-est})},
and  $\supp(f) \subset \triangle_{y}$,  $\supp(f^{\prime}) \subset \triangle_{y^{\prime}}$
\beqa
|\lan f, G_{k}(\Om) f^{\prime} \ran| \2=\2 |\langle \e_{q} f, (\e_{-q}G_{k}(\Om)\e_{q}) \e_{-q} f^{\prime} \ran| \non\\
\2\leq\2 c \|\e_{q} f \|_2 \, \|\e_{-q} f^{\prime}\|_2 \non\\
\2\leq\2  {\rcc c''}  \e^{q \cdot (y-y^{\prime})} \| f \|_{2} \|f^{\prime}\|_{2}.
\eeqa
Taking $q = c_1 \frac{-(y-y^{\prime})}{|y-y^{\prime}|}$ gives the result. 

\section{Proof of Lemma \ref{ren-iteration-x} }\label{ren-iteration}
 \setcounter{equation}{0}

  The computation is analogous as in the proof of \cite[Lemma 2]{Di13}. {\gr The statement follows from the fact that a convolution of two Gaussian functions is again a Gaussian.} It suffices to check $T^{L^j\xi}_{a,L} T^{\xi}_{a_j, L^j}=T^{\xi}_{a_{j+1}, L^{j+1}} $ as (\ref{T-product}) follows
by iteration.  We write
\beqa
(T^{L^j\xi}_{a,L} T^{\xi}_{a_j, L^j})[\Om, \rho](\psi')\2=\2 \bigg(  \fr{b_1^{L^j\xi} }{2\pi}  \bigg)^{ \fr{|\Om_{j+1}|}{2} }  \int d\psi\,   \e^{-\h b_1^{L^j\xi}  \sum_{y'\in \Om_{j+1}} |\psi'(y')-(Q_{\Om_j}\psi)(y')|^2    }   (T^{\xi}_{a_j, L^j} \rho)(\psi)\non\\
\2=\2  \bigg(  \fr{b_1^{L^j\xi} }{2\pi}  \bigg)^{ \fr{|\Om_{j+1}|}{2} }  \int d\psi\,   \e^{-\h b_1^{L^j\xi}  \sum_{y'\in \Om_{j+1}} |\psi'(y')-(Q_{\Om_j}\psi)(y')|^2    } \non\\
\2 \2\times  \bigg(  \fr{b_j^{\xi} }{2\pi}  \bigg)^{ \fr{|\Om_{j}|}{2} } \int d\phi\,   \e^{-\h b_j^{\xi}  \sum_{y\in \Om_{j}} |\psi(y)-(Q_{\Om,j}\phi)(y)|^2    }    \rho(\phi).
\eeqa
The problem boils down to computing the integral
\beqa
\int d\psi\,  \e^{-\h b_1^{L^j\xi}  \sum_{y'\in \Om_{j+1}} |\psi'(y')-(Q_{\Om_j}\psi)(y')|^2    } \e^{-\h b_j^{\xi}  \sum_{y\in \Om_{j}} |\psi(y)-(Q_{\Om,j}\phi)(y)|^2  }. \label{integral-to-compute}
\eeqa
We denote the {\rcc expression} in the exponential by
\beqa
F(\psi):=\h b_1^{L^j\xi}  \sum_{y'\in \Om_{j+1}} |\psi'(y')-(Q_{\Om_j}\psi)(y')|^2+\h b_j^{\xi}  \sum_{y\in \Om_{j}} |\psi(y)-(Q_{\Om,j}\phi)(y)|^2.
\eeqa
We will find the minimum $\psi_0$ and then expand $\psi=\psi_0+\ti\psi$. {\ma Since the linear terms vanish, we get}
\beqa
F(\psi)=F(\psi_0)+\h \sum_{\ti y, \ti y'} F''(\psi_0)(\ti y, \ti y') \ti\psi(\ti y) \ti\psi(\ti y'), \quad {\rcc F''(\psi_0)(\ti y, \ti y'):=\pa_{\psi(\ti y)} \pa_{\psi(\ti y')} F(\psi) |_{\psi=\psi_0}}.
\eeqa
Then we will obtain, {\ma referring to (\ref{Gaussian})},
\beqa
(\ref{integral-to-compute})=\int d\psi\, \e^{-F(\psi)}\2=\2\e^{-F(\psi_0)} \int d\ti{\psi}\, \e^{- \h \sum_{\ti y, \ti y'} F''(\psi_0)({\rcc \ti y, \ti y'}) \ti\psi(\ti y) \ti\psi(\ti y') }
= \e^{-F(\psi_0)}  \sqrt{\fr{(2\pi)^{|\Om_j|}}{\det(F'')}}, \label{gaussian-formula}
\eeqa
{\rcc provided that $F''(\psi_0)$ is invertible. (The latter property can can be read off from (\ref{invertibility-ddF}) below). } 

We  compute the derivative of $F$
\beqa
\fr{\pa}{\pa \psi(\ti y) } (Q_{\Om_j}\psi)(y')\2 =\2 \fr{\pa}{\pa \psi(\ti y) } \fr{1}{L^d} \sum_{ y'_{\mud} \leq z_{\mud}< y'_{\mud}+L (L^j\xi) } \psi(z)
= \fr{1}{L^d} \sum_{ y'_{\mud} \leq z_{\mud}< y'_{\mud}+L (L^j\xi) } \de_{z,\ti y}  = \fr{1}{L^d} \one_{B_{j+1}(y') }( \ti y) , \quad
\eeqa
where $\one_{B_{j+1}(y') }$ is the characteristic function of the $L$-box $B_{j+1}(y')$ in the lattice $\Om_j$. Hence,
\beqa
\fr{\pa}{\pa \psi(\ti y) } F(\psi)\2 =\2 -b_1^{L^j\xi}  \sum_{y'\in \Om_{j+1}} \big(\psi'(y')-(Q_{\Om_j}\psi)(y')\big)\cdot \fr{1}{L^d} \one_{B_{j+1}(y') }( \ti y)  + b_j^{\xi}  \sum_{y\in \Om_{j}} \big(\psi(y)-(Q_{\Om,j}\phi)(y) \big)\cdot \de_{y,\ti y} \non\\
\2=\2- \fr{b_1^{L^j\xi}}{L^d}  \big(\psi'(y'_{\ti y})-(Q_{\Om_j}\psi)(y'_{\ti y})\big)+  b_j^{\xi}\big(\psi(\ti y)-(Q_{\Om,j}\phi)( \ti y) \big). \label{Q-derivative}
\eeqa
Thus at the minimum $\psi_0$ we have, using (\ref{y-x}),
\beqa
\2 \2- \fr{b_1^{L^j\xi} (b_j^{\xi})^{-1}  }{L^d}  \big(Q^*\psi'- Q^*Q\psi_0 \big)+  \psi_0-Q_{j}\phi=0,\label{invertibility-ddF} \\
\2 \2 \big(1+ \fr{b_1^{L^j\xi} (b_j^{\xi})^{-1}  }{L^d} Q^*Q \big)\psi_0=\fr{b_1^{L^j\xi} (b_j^{\xi})^{-1}  }{L^d}Q^*\psi'+ Q_{j}\phi.
\eeqa
We note that for any projection $P$ we have, for $|b|<1$,
\beqa
\fr{1}{1+bP}=((1+b)^{-1} -1)P+1=-\fr{b}{1+b} P+1,
\eeqa
cf.  (\ref{projection-computation}) below.  (In our case $b:= \fr{b_1^{L^j\xi} (b_j^{\xi})^{-1}  }{L^d}<1$ holds, because $L>1$). Thus we get 
\beqa
\psi_0\2=\2  \big(-\fr{b^2}{1+b} + b\big) Q^*\psi'-\fr{b}{1+b} Q^*Q_{j+1}\phi+Q_{j}\phi\non\\
\2=\2 \fr{b}{1+b} Q^*\psi'-\fr{b}{1+b} Q^*Q_{j+1}\phi+Q_{j}\phi.
\eeqa
We compute
\beqa
 \|\psi'- Q\psi_0\|^2_{2,\Om_{j+1}}\2=\2\|\psi'- \fr{b}{1+b} \psi' +\fr{b}{1+b} Q_{j+1}\phi - Q_{j+1}\phi \|^2_{2,\Om_{j+1}} \non\\
\2=\2 \fr{1}{(1+b)^2} \|\psi'-Q_{j+1}\phi\|^2_{2,\Om_{j+1}},\label{first-norm}\\
 \|\psi_0-Q_{j}\phi \|_{2,\Om_j}^2\2=\2\fr{b^2}{(1+b)^2}\| Q^*\psi'- Q^*Q_{j+1}\phi\|_{2,\Om_j}^2=\fr{b^2}{(1+b)^2}\| \psi'- Q_{j+1}\phi\|_{2,\Om_{j+1}}^2. \label{second-norm}
\eeqa
Now we write, using (\ref{first-norm}), (\ref{second-norm}), 
\beqa
F(\psi_0)\2=\2\h b_1^{L^j\xi} (L^{j+1}\xi)^{-d}  \|\psi'- Q\psi_0\|^2_{2,\Om_{j+1}}+\h b_j^{\xi}  (L^j\xi)^{-d}  \|\psi_0-Q_{j}\phi \|_{2,\Om_j}^2\non\\
\2=\2 \h \fr{1}{(1+b)^2} (L^j\xi)^{-d} \big( b_1^{L^j\xi} L^{-d} + b_j^{\xi} b^2 \big)\| \psi'- Q_{j+1}\phi\|_{2,\Om_{j+1}}^2\non\\
\2=\2 \h \fr{1}{(1+b)^2} \big( b_1^{L^j\xi}  + L^d b_j^{\xi} b^2 \big)  \sum_{y'\in \Om_{j+1}} | \psi'(y')- Q_{j+1}\phi(y') |^2. \label{F-formula}
 \eeqa
We have that, with $b:= \fr{b_1^{L^j\xi} (b_j^{\xi})^{-1}  }{L^d}$, $b_1^{L^j\xi}:=a_1(L^{j+1}\xi)^{d-2}$, $b_j^{\xi}:=a_j (L^j\xi)^{d-2}$
\beqa
\h \fr{1}{(1+b)^2} \big( b_1^{L^j\xi}  + L^d b_j^{\xi} b^2 \big)\2=\2 \h\fr{1}{(1+b)^2}  b_1^{L^j\xi} (1+b) =\h\fr{1}{b_j^{\xi}+  \fr{b_1^{L^j\xi}}{L^d} }  b_j^{\xi} b_1^{L^j\xi} \non\\
\2=\2\h\fr{a_1 a_j (L^{j+1}\xi)^{d-2} (L^j\xi)^{d-2}}{a_j (L^j\xi)^{d-2} + a_1(L^{j+1}\xi)^{d-2}/L^d } \non\\
\2=\2  \h(L^{j+1}\xi)^{d-2} \fr{a_1a_j}{a_j+a_1/L^2} =\h (L^{j+1}\xi)^{d-2} a_{j+1}.
\eeqa
This, together with (\ref{F-formula}), (\ref{gaussian-formula}) concludes the proof.

\section{Proof of Lemma~\ref{G-lemma}.}\label{G-appendix}
\setcounter{equation}{0}

We write
\beqa
\rho_j(\psi)\2=\2T^{\xi}_{a_j,L^j}[\Om,\rho_0](\psi)\non\\
\2=\2\bigg(  \fr{b_j^{\xi} }{2\pi}  \bigg)^{ \fr{|\Om_{j}|}{2} } \int d\phi\,   \e^{-\h b_j^{\xi}  \sum_{y\in \Om_{j}} |\psi(y)-(Q_{\Om,j}\phi)(y)|^2    }    \rho_0(\phi)\non\\
\2=\2\bigg(  \fr{b_j^{\xi} }{2\pi}  \bigg)^{ \fr{|\Om_{j}|}{2} }   \int d\phi\,   \e^{-\h b_j^{\xi}  \sum_{y\in \Om_{j}} |\psi(y)-(Q_{\Om,j}\phi)(y)|^2 } 
  \e^{-\h\lan \phi, \De^{(0), \xi}(\Om) \phi\ran}{\bcc \e^{\i \lan J, \phi\ran}}.
\eeqa
We define the function
\beqa
F(\phi):=\h b_j^{\xi}  \sum_{y\in \Om_{j}} |\psi(y)-(Q_{\Om,j}\phi)(y)|^2+\h \xi^d\sum_{x\in \Om}  \phi(x) (\De^{(0), \xi}(\Om) \phi)(x) 
\eeqa
and compute {\bcc the derivatives} (cf. (\ref{Q-derivative})) 
\beqa
\fr{\pa}{\pa \phi(\ti x)} F(\phi)\2 =\2-\fr{b_j^{\xi}}{L^{jd}}((Q_{\Om,j}^*\psi)(\ti x) -(Q_{\Om,j}^*Q_{\Om,j}\phi)(\ti x)  )+ \xi^d  (\De^{(0), \xi}(\Om)\phi)(\ti x),\\  
{\bcc \fr{\pa}{\pa \phi(\ti x)} \fr{\pa}{\pa \phi(\ti x')}F(\phi)} \2 =\2 {\bcc[ \fr{b_j^{\xi}}{L^{jd}} {\rcc \eta^d} Q_{\Om,j}^*Q_{\Om,j} + \xi^{{\rcc 2}d}  \De^{(0), \xi}(\Om)]{\ma(\ti x,\ti x')}=
\eta^{{\rcc 2} d} [(G_j^{\eta}(\Om))^{-1}}]{\ma (\ti x, \ti x')},
\eeqa
where we made use of (\ref{Laplacian-s-a}). Thus the {\ma first} derivative of $F$ vanishes at
\beqa
\phi_0= \fr{b_j^{\xi}}{(L^{j}\xi)^d  }   G_j^{\xi}(\Om) Q_{\Om,j}^*\psi=\fr{a_j}{(L^j\xi)^2 } G_j^{\xi}(\Om) Q_{\Om,j}^*\psi=:{\bcc \hil_j^{\eta}(\Om)\psi}.
\eeqa
We can write $\phi=\phi_0+\ti\phi$, which gives
\beqa
F(\phi)=F(\phi_0)+\h \lan \ti\phi,  (G_j^{\eta}(\Om))^{-1} \ti\phi\ran
\eeqa
so that, {\ma referring to (\ref{Gaussian})}, 
\beqa
\int d\phi\, \e^{-F(\phi)} {\bcc \e^{\i \lan J, \phi\ran}} \2=\2\e^{-F(\phi_0)} {\bcc \e^{\i \lan J,\phi_0 \ran}} \int d\ti{\phi}\, \e^{- \h  \lan \ti\phi,  (G_j^{\eta}(\Om))^{-1} \ti\phi\ran   }  {\bcc \e^{\i \lan J, \ti{\phi}\ran}}\non\\
\2=\2 \e^{-F(\phi_0)}  {\bcc \e^{\i \lan J,\phi_0 \ran}} \sqrt{(2\pi)^{|\Om|}\det(G_j^{\eta}(\Om))} {\bcc \e^{-\h \lan J,  G_j^{\eta}(\Om)  J\ran}}.
\eeqa

Now to determine  $\De^{(j), L^j\xi}(\Om)$ it suffices to compute for $\phi_0=\fr{a_j}{(L^j\xi)^2 } G_j^{\xi}(\Om) Q_{\Om,j}^*\psi $
\beqa
F(\phi_0)\2=\2 \h a_j (L^j\xi)^{-2}  \|\psi- Q_{\Om,j}\phi_0 \|_{2,\Om_j}^2+\h\lan \phi_0, \De^{(0), \xi}(\Om) \phi_0\ran_{\Om}\non\\
\2=\2 \h a_j (L^j\xi)^{-2}  \| \psi   - \fr{a_j}{(L^j\xi)^2 } Q_{\Om,j}G_j^{\xi}(\Om) Q_{\Om,j}^*\psi    \|_{2,\Om_j}^2\non\\
\2 \2+\h  \fr{a_j^2}{(L^j\xi)^4 }  \lan  G_j^{\xi}(\Om) Q_{\Om,j}^*\psi    , \De^{(0), \xi}(\Om)    G_j^{\xi}(\Om) Q_{\Om,j}^*\psi  \ran_{\Om}\non\\
\2=\2 \h a_j (L^j\xi)^{-2}  \| \psi \|^2_{2,\Om_j}  - \bigg(\fr{a_j}{(L^j\xi)^2 }\bigg)^2 \lan \psi,  Q_{\Om,j}G_j^{\xi}(\Om) Q_{\Om,j}^*\psi  \ran_{\Om_j}\non\\
\2 \2+\h\big(a_j (L^j\xi)^{-2} \big)^3 \|Q_{\Om,j}G_j^{\xi}(\Om) Q_{\Om,j}^*\psi    \|^2_{2,\Om_j}\non\\
\2 \2+\h \bigg( \fr{a_j}{(L^j\xi)^2 } \bigg)^2  \lan  G_j^{\xi}(\Om) Q_{\Om,j}^*\psi    , \De^{(0), \xi}(\Om)    G_j^{\xi}(\Om) Q_{\Om,j}^*\psi  \ran_{\Om},
\eeqa
{\rcc where we refer to (\ref{De+bmu}) and (\ref{G-k-xi-sec4}) for definitions of  $\De^{(0), \xi}(\Om)$ and $G_j^{\xi}(\Om)$.}

Now in the last expression we write $\De^{(0), \xi}(\Om)=G_j^{\xi}(\Om)^{-1} -a_j (L^j\xi)^{-2} Q_{\Om,j}^* Q_{\Om,j}$ which leads to
\beqa
F(\phi_0)=\h a_j (L^j\xi)^{-2}  \| \psi \|^2_{2,\Om_j}-\h\bigg(\fr{a_j}{(L^j\xi)^2 }\bigg)^2 \lan \psi,  Q_{\Om,j}G_j^{\xi}(\Om) Q_{\Om,j}^*\psi  \ran_{\Om_j}
=\h \lan\psi,\De^{(j),L^j\xi}(\Om)\psi \ran_{\Om_j},
\eeqa
which concludes the proof. 

\section{A formula for $C^{(j),L^j\xi}(\Om)$} \label{App-C-Dimock}
\setcounter{equation}{0}

We repeat here the discussion from \cite[Appendix C]{Di13} slightly adapted to our situation. {\ma We start by recalling
the standard Gaussian integral formula
\beqa
\int_{\real^M} d\phi\, \e^{-\h \lan \phi, C \phi\ran +\lan J,\phi\ran }=\sqrt{\fr{(2\pi)^{M} }{\det C} } \e^{\h \lan \ov{J}, C^{-1} J\ran } \label{Gaussian} 
\eeqa
valid for any positive definite $M\times M$  matrix $C$ and a complex-valued vector $J$. }
{\bcc We use the short-hand notations valid only for this appendix  
\beqa
G_j^{\xi}:=G^{\xi}_{j}(\Om), \quad C^{\ma (j)}:=C^{(j), L^j\xi}(\Om), \quad Q_j:=Q_{\Om,j}, \quad \De^{\ma (j)}:=\De^{(\kk), L^\kk \xi}(\Om),\quad   \ti{a}_{j,\ell}:=a_j(L^\ell\xi)^{-2}. \label{short-hand-notation}
\eeqa
We recall that
\beqa
\De^{\ma (j)}:=[\ti a_{j,j}-\ti{a}_{j,j}^2 Q_j G^{\xi}_j Q_j^*]_{\Om_j}, \quad  G_j^{\xi}:=[-\De^{\xi}+\bmu_k+ \ti{a}_{j,j}Q_j^*Q_j]_{\Om}^{-1}.
\eeqa
}
Lemma~\ref{App-C-lemma} below  gives a formula for
\beqa
C^{\ma (j)}:=[\De^{\ma (j)}+\fr{\ti{a}_{1,j} }{L^2}Q^*Q]_{\Om_j}^{-1}=\big[\ti{a}_{j,j}-\ti{a}_{j,j}^2 Q_j G_jQ_j^*  +\fr{ \ti{a}_{1,j} }{L^2}Q^*Q\big]_{\Om_j}^{-1} \label{C-k-r}
\eeqa
which is used in the proof of Theorem~\ref{Key-formula-thm}.
\bel\label{App-C-lemma} We have
\beqa
C^{\ma (j)}=A_{j}+\ti{a}_{j,j}^2 A_{j}Q_j G^{\xi}_{j+1} Q_j^* A_{j}, \label{C-vs-A-lemma}
\eeqa
where 
\beqa
A_{j}:=[\ti{a}_{j,j}+\ti{a}_{1,j} L^{-2} Q^* Q]_{\Om_j}^{-1}, \quad
G^{\xi}_{j+1}:=\big[ -\De^{\xi}+ \bmu_k+ \fr{\ti{a}_{j+1,j}}{L^2} Q_{j+1}^* Q_{j+1}\big]_{\Om}^{-1}. \label{A-notation-app}
\eeqa
\eel
\proof Using definition (\ref{C-k-r}) and the Gaussian integration formula (\ref{Gaussian}), we obtain for a suitable normalization constant ${\ma Z_j}$:
\beqa
\exp\bigg( \h \lan f, C^{\ma (j)} f\ran \bigg)={\ma Z_j} \int d\Phi\, \exp\bigg(\lan \Phi, f\ran-\fr{\ti{a}_{1,j}}{2L^2} \|Q\Phi\|^2_2-\h \lan \Phi, \De^{\ma (j)}\Phi\ran     \bigg). \label{first-exponent}
\eeqa
Apart from this, we checked in Lemma~\ref{G-lemma} that
\beqa
\exp(-\h \lan \Phi, \De^{\ma (j)} \Phi\ran) ={\ma Z_j} \int d\phi \exp\bigg(  -\fr{\ti{a}_{j,j}}{2}\|\Phi-Q_j\phi\|^2_2 - \h \lan \phi, (-\De^{\xi}_{\Om}+\bmu_k)\phi \ran    \bigg).
\eeqa
{\rcc (We note that this expression  is proportional to $\rho_j^{J=0}(\Phi)$).}  Inserting this in (\ref{first-exponent}), we obtain
\beqa
\2 \2\exp\bigg( \h \lan f, C^{\ma (j)} f\ran \bigg)\non\\
\2 \2={\ma Z_j} \int d\phi \, d\Phi\, \exp\bigg(\lan \Phi, f\ran-\fr{\ti{a}_{1,j}}{2L^2} \|Q\Phi\|^2_2
-\fr{\ti{a}_{j,j}}{2}\|\Phi-Q_j\phi\|^2_2 - \h \lan \phi, (-\De^{\xi}_{\Om}+\bmu_k)\phi \ran\bigg)\non\\
\2 \2={\ma Z_j} \int d\phi\,d\Phi\, \exp\bigg(  \lan \Phi, f+\ti{a}_{j,j}Q_j\phi\ran-\h \lan \Phi, \big( \ti{a}_{j,j}+\ti{a}_{1,j}L^{-2} Q^*Q \big)\Phi\ran \non\\
\2 \2\phantom{444444444444444444444444444444444444444}-\fr{ \ti{a}_{j,j}}{2} \|Q_j\phi\|^2_2 -  \h \lan \phi, (-\De^{\xi}_{\Om}+\bmu_k)\phi \ran  \bigg)\non\\
\2 \2= {\ma Z_j'} \int d\phi \exp\bigg( \h\lan ( f+\ti{a}_{j,j}Q_j\phi ), A_{j} ( f+\ti{a}_{j,j} Q_j\phi ) \ran  -\fr{\ti{a}_{j,j}}{2} \|Q_j\phi\|^2_2 -  \h \lan \phi, (-\De^{\xi}_{\Om}+\bmu_k)\phi \ran \bigg),
\eeqa
{\ma where $Z_j'$ is a new normalization constant}. Let us now  compute $A_{j}:=(\ti{a}_{j,j}+\ti{a}_{1,j}L^{-2} Q^* Q)^{-1}$.  We note that if $P$ is a projection, $a,b>0$, $a^{-1}b<1$, then
\beqa
(a+bP)^{-1}\2=\2a^{-1}(1+a^{-1}bP)^{-1}=a^{-1} \sum_{\ell=0}^{\infty} (a^{-1}b)^{\ell} P^{\ell}\non\\ 
\2=\2 a^{-1} I+a^{-1} \bigg( \fr{1}{1+ a^{-1}b} -1\bigg) P = a^{-1}(I-P)+\fr{1}{a+b} P. \label{projection-computation}
\eeqa
For $a=\ti{a}_{j,j}$, $b=\ti{a}_{1,j}L^{-2}$ we have  $a^{-1}b=\ti{a}_{j,j}^{-1} \ti{a}_{1,j}L^{-2}<1$ by (\ref{short-hand-notation}), (\ref{a-j}). Thus (\ref{projection-computation}) gives
\beqa
A_j=\fr{1}{\ti{a}_{j,j}} + \bigg(\fr{1}{\ti{a}_{j,j}+\ti{a}_{1,j}L^{-2}} - \fr{1}{\ti{a}_{j,j}} \bigg)Q^*Q = \fr{1}{\ti{a}_{j,j}} -\fr{\ti{a}_{j+1,j} L^{-2} }{\ti{a}_{j,j}^2} Q^*Q. 
\label{A-k-formula-new}
\eeqa
Consequently
\beqa
& &\exp\bigg( \h \lan f, C^{\ma (j)} f \ran \bigg)\non\\
& & =  {\ma Z_j'}\int d\phi\, \exp\bigg( \h\lan ( f+\ti{a}_{j,j}Q_j\phi ), A_{j} ( f+\ti{a}_{j,j} Q_j\phi ) \ran  
-\fr{\ti{a}_{j,j}}{2} \|Q_j\phi\|_2^2 -  \h \lan \phi, (-\De^{\xi}_{\Om}+\bmu_k)\phi \ran \bigg)\non\\
& &={\ma Z_j'}  \exp{\bigg(\h \lan f, A_{j} f\ran \bigg)} \int d\phi\, \exp{\bigg(\h \lan \ti{a}_{j,j}   Q_j A_{j}f,\phi \ran \bigg)} \exp{\bigg(-  \h \lan \phi, (-\De^{\xi}_{\Om}+\bmu_k + 
\fr{\ti{a}_{j+1,j}}{L^2}  Q_{j+1}^{*}Q_{j+1} )\phi \ran\bigg)} \non\\
& & = {\ma Z_j''}\exp{\bigg(\h \lan f, A_{j} f\ran  + \fr{\ti{a}_{j,j}^2}{2}\lan f, A_{j} Q_j G^{\xi}_{j+1} Q_j^* A_{j} f\ran \bigg)}, \label{C-vs-A}
\eeqa
where  we noted that, given  (\ref{A-k-formula-new}), 
\beqa
\big( -\De^{\xi}_{\Om}+ \bmu_k+\ti{a}_{j,j} Q_j^*Q_j  - \ti{a}_{j,j}^2 Q_j^* A_{j} Q_j\big)
 =\big( -\De^{\xi}_{\Om}+ \bmu_k+ \fr{\ti{a}_{j+1,j}}{L^2} Q_{j+1}^* Q_{j+1}\big)  = (G^{\xi}_{j+1})^{-1}.
\eeqa
{\rcc Setting $f=0$ in (\ref{C-vs-A}), we obtain ${\ma Z_j''}=1$. Then  (\ref{C-vs-A-lemma}) follows.}  \qed 
\section{Proof of Lemma~\ref{Fourier-transform-lemma}}  \label{App-Fourier-transform-lemma}
 \setcounter{equation}{0}

{\bl We  compute the Fourier transform of (\ref{QQ}) first for $f\in \mcL^1(\eta\mathbb{Z}^d)$:}
\beqa
(\wh{Q_k^*Q_k} f)(p)\2=\2 \fr{1}{(2\pi)^{d/2} L^{kd} } \sum_{x\in \xi\mathbb{Z}^d}\xi^d \e^{-\i p\cdot x}  \sum_{[x_{\mud}] \leq x_{\mud}'< [x_{\mud}]  +1} f(x')\non\\
\2=\2 \fr{1}{(2\pi)^{d/2}  L^{kd} } \sum_{\ell_{\mud}=0}^{L^k-1} \sum_{x\in \xi\mathbb{Z}^d }\xi^d \e^{-\i p\cdot x} f([x_{\mud}]  +\ell_{\mud} \xi) \non\\
\2=\2 \fr{1}{ (2\pi)^{d/2} L^{kd} } \sum_{\ell_{\mud}=0}^{L^k-1}  \sum_{\ell'_{\mud}=0}^{L^k-1} \sum_{ [x_{\mud}]  \in \mathbb{Z} }\xi^d \e^{-\i p_{\nu} ( [x_{\nu}]  +\ell'_{\nu}\xi)} f( [x_{\mud}]   +\ell_{\mud}\xi)\non\\
\2=\2 \fr{1}{ (2\pi)^{d/2} L^{kd} } \sum_{\ell_{\mud}=0}^{L^k-1}  \sum_{\ell'_{\mud}=0}^{L^k-1} \sum_{\m \in \mathbb{Z}^d } \xi^d \e^{-\i p_{\nu} (\m_{\nu}  +\ell'_{\nu}\xi)} 
f(\m+\ell \xi ) \non\\
\2=\2 \fr{1}{ (2\pi)^{d/2} L^{kd} } \sum_{\ell_{\mud}=0}^{L^k-1}  \sum_{\ell'_{\mud}=0}^{L^k-1}  \e^{-\i p_{\nu}\ell'_{\nu}\xi } \sum_{\m \in \mathbb{Z}^d }\xi^d  \e^{-\i p_{\nu} \m_{\nu} } f(\m +\ell \xi)\non\\
\2=\2 \fr{\xi^d}{ (2\pi)^{d/2} L^{kd} } \bigg(\prod_{\nu=0}^{d-1} \fr{1-\e^{-\i p_{\nu}}}{1-\e^{-\i p_{\nu}\xi}}\bigg) \sum_{\ell_{\mud}=0}^{L^k-1} \sum_{\m \in \mathbb{Z}^d } 
\e^{-\i p_{\nu} \m_{\nu}} f(\m+\ell \xi).
\eeqa
{\bl Thus, by Lemma~\ref{delta-lemma} below, we obtain}
\beqa
(\wh{Q_k^*Q_k} f)(p)\2=\2 \fr{1}{L^{2kd} }\bigg(\prod_{\mu=0}^{d-1}  \fr{1-\e^{-\i p_{\mu}}}{1-\e^{-\i p_{\mu}\xi}} \bigg)  \sum_{\ell''_{\mud} =- \fr{L^k-1}{2}}^{  \fr{L^k-1}{2}}  
\bigg( \prod_{\nu=0}^{d-1} \fr{1-\e^{\i (p_{\nu}+2\pi \ell''_{\nu} )} }{1-\e^{\i (p_{\nu}+2\pi\ell''_{\nu}) \xi }} \bigg)    \hat{f}(p+2\pi\ell''). \label{Fourier-Q-star-Q}
\eeqa
{\bl Finally,  this formula is extended to all $f\in \mcL^2(\eta \mathbb{Z}^d)$ using unitarity of the Fourier transform,  boundedness of  $Q_k^*Q_k$,
 and boundedness of the multiplication operators and shifts by $2\pi\ell''$ on the r.h.s. of (\ref{Fourier-Q-star-Q}). 
\bel\label{delta-lemma} The following relation holds for $f\in \mcL^1(\eta\mathbb{Z}^d)$
\beqa
(2\pi)^{-d/2}\sum_{\ell_{\mud}=0}^{L^k-1} \sum_{\m \in \mathbb{Z}^d } \e^{-\i p\cdot \m} f(\m+\ell \xi)= 
\sum_{\ell''_{\mud} =- \fr{L^k-1}{2}}^{  \fr{L^k-1}{2}}  
\bigg( \prod_{\nu=0}^{d-1} \fr{1-\e^{\i (p_{\nu}+2\pi \ell''_{\nu} )} }{1-\e^{\i (p_{\nu}+2\pi\ell''_{\nu}) \xi }} \bigg)    \hat{f}(p+2\pi\ell''),
\eeqa
where we extended $\hat{f}$ from $[-\pi/\xi, \pi/\xi[^{\times d}$ to a function on $\real^d$ by periodicity.
 \eel
\proof  We note that for any $x\in \eta \mathbb{Z}^d$,
\beqa
\fr{1}{L^{kd}} \sum_{\ell''_{\mud} =- \fr{L^k-1}{2}}^{  \fr{L^k-1}{2}} \e^{-\i 2\pi \ell'' x}
\2=\2 \e^{-\i \pi (L^k-1)x }\fr{1}{L^{kd}}  \sum_{\ell''_{\mud}=0}^{L^k-1} \e^{-\i 2\pi \ell''   x}= \left\{ \begin{array}{ll}  1 & \textrm{if $x\in \mathbb{Z}^d,$}\\  0 & \textrm{otherwise,}\end{array} \right.
\eeqa
where we used that $L$ is odd.  Thus we can write
\beqa
(2\pi)^{-d/2}\sum_{\m \in \mathbb{Z}^d } \e^{-\i p\cdot \m} f(\m+\ell \xi)
\2=\2 \fr{1}{L^{kd}}\sum_{\ell''_{\mud} =- \fr{L^k-1}{2}}^{  \fr{L^k-1}{2}} (2\pi)^{-d/2}\sum_{x \in \eta\mathbb{Z}^d } 
\e^{-\i (p +2\pi \ell'')\cdot x}   f(x+\ell \xi) \non\\
\2=\2 \fr{1}{L^{kd}}\sum_{\ell''_{\mud} =- \fr{L^k-1}{2}}^{  \fr{L^k-1}{2}} (2\pi)^{-d/2}\sum_{x \in \eta\mathbb{Z}^d } 
\e^{-\i (p +2\pi \ell'')\cdot (x-\ell \xi) }   f(x)\non\\
\2=\2  \sum_{\ell''_{\mud} =- \fr{L^k-1}{2}}^{  \fr{L^k-1}{2}} \e^{\i (p +2\pi \ell'')\cdot \ell \xi } 
\hat{f}(p +2\pi \ell''). 
\eeqa
Now the summation over $\ell$ gives the claim. \qed }

\section{Proof of Lemma~\ref{bound-integrand} } \label{App-bound-integrand}

\setcounter{equation}{0}

 Recall that
\beqa
\De^{\xi}(p)\2:=\2 \fr{1}{\xi^{2}} \sum_{\mu=0}^{d-1} |1-\e^{-\i p_{\mu}\xi} |^2+\bmu_k=\fr{2}{\xi^{2}} \sum_{\mu=0}^{d-1}  (1-\cos( p_{\mu}\xi))+\bmu_k=
\fr{4}{\xi^{2}} \ovsum_{\mu=0}^{d-1}\sin^2\big(\fr{ p_{\mu}\xi}{2}\big), 
\eeqa
where the star over the sum means addition of $\fr{\xi^2}{4}\bmu_k= \fr{1}{4} \bmu_0$. 
Now we list other relevant definitions:
\beqa
  u(p)\2:=\2 \xi^d \prod_{\nu=0}^{d-1}\bigg( \fr{1-\e^{-\i p_{\nu}}}{1-\e^{-\i p_{\nu}\xi}} \bigg)  =\xi^d\prod_{\nu=0}^{d-1}\bigg(  \fr{\e^{-\i \fr{p_{\nu}}{2}} }{  \e^{-\i\fr{p_{\nu} \xi }{2}} }   \fr{\sin(\fr{p_{\nu}}{2} )}{\sin(\fr{p_{\nu}\xi}{2} ) }\bigg), \\
u_{\De}(p)\2:=\2 \fr{ \xi^{d+2} }{4} \bigg(\prod_{\al=0}^{d-1} \fr{\e^{-\i\fr{p_{\al}}{2}} }{  \e^{-\i\fr{p_{\al} \xi }{2}} } \bigg) \fr{1}{ \ovsum_{\mu=0}^{d-1}\sin^2(\fr{ p_{\mu}\xi}{2})   }   \bigg(\prod_{\nu=0}^{d-1}\fr{\sin(\fr{p_{\nu}}{2} )}{\sin(\fr{p_{\nu}\xi}{2} ) }\bigg),\\
u_{\De}(p+2\pi \ell') 
\2:=\2     \fr{ \xi^{d+2} }{4} \bigg(\prod_{\al=0}^{d-1} \fr{\e^{-\i\fr{p_{\al}}{2}} }{  \e^{-\i\fr{(p_{\al}+2\pi \ell'_{\al}  )\xi }{2}} } \bigg) 
\fr{1}{ \ovsum_{\mu=0}^{d-1}\sin^2\big(\fr{ p_{\mu} \xi }{2}   +  \pi \ell'_{\mu}\xi     \big)   }   
\bigg(\prod_{\nu=0}^{d-1}\fr{\sin(\fr{p_{\nu}}{2} )}{\sin\big(\fr{p_{\nu} \xi}{2}+\pi \ell'_{\nu}\xi \big) }\bigg),  \label{u-De}\\  
\llan u,  u_{\De} \rran(p)\2:=\2 \sum_{\ell''_{\mud}  =- \fr{L^k-1}{2}}^{  \fr{L^k-1}{2}}  \fr{|u(p+  2\pi\ell'')|^2}{\De^{\xi}(p+2\pi\ell'')} \non\\
\2=\2 \fr{\xi^{2d+2}}{4}  \sum_{\ell''_{\mud} =- \fr{L^k-1}{2}}^{  \fr{L^k-1}{2}} \fr{1}{ \ovsum_{\mu=0}^{d-1}\sin^2\big(\fr{ p_{\mu} \xi }{2}   +  \pi \ell''_{\mu}\xi     \big)   }   \prod_{\nu=0}^{d-1} \fr{\sin^2(\fr{p_{\nu}}{2} )}{\sin^2( \fr{p_{\nu}\xi}{2} + \pi \ell'' \xi )  },
\eeqa
where we used in (\ref{u-De}) that $p_{\al} \mapsto \e^{-\i\fr{p_{\al}}{2}}\sin(\fr{p_{\al}}{2} )$ has period $ 2\pi$  and in the last line  that $p_{\al} \mapsto \sin^2({\ma p_{\al}/2} )$ has  period ${\ma 2}\pi$. 

The function appearing under the modulus in (\ref{statement-lemma}) 
has the form (where we set $a:=a_k$ for brevity)
\beqa
H(p)\2:=\2 \fr{1}{1+a \llan u, u_{\De}\rran(p)} u_{\De}(p+2\pi \ell') \label{before-continuation-m} \non\\
\2=\2  \fr{ \xi^{d+2} }{4} \bigg(\prod_{\al=0}^{d-1} \fr{\e^{-\i\fr{p_{\al}}{2}} }{  \e^{-\i\fr{(p_{\al}+2\pi \ell'_{\al}  )\xi }{2}} } \bigg) \label{exponential-functions-part-m} \\
\2\times \2 \fr{1}{ 1+a\fr{\xi^{2d+2}}{4}  \sum_{\ell''_{\mud} =- \fr{L^k-1}{2}}^{  \fr{L^k-1}{2}} 
\fr{ 1 }{ \ovsum_{\mu=0}^{d-1}\sin^2\big(\fr{ p_{\mu} \xi }{2}   +  \pi \ell''_{\mu}\xi     \big)   } 
  \prod_{\nu=0}^{d-1} \fr{\sin^2(\fr{p_{\nu}}{2} )}{\sin^2( \fr{p_{\nu}\xi}{2} + \pi \ell''_{\nu} \xi )  }  } 
\label{big-denominator-m} \\
\2\times \2 \fr{1 }{ \ovsum_{\mu=0}^{d-1}\sin^2\big(\fr{ p_{\mu} \xi }{2}   +  \pi \ell'_{\mu}\xi     \big)   }   
\bigg(\prod_{\nu=0}^{d-1}\fr{\sin(\fr{p_{\nu}}{2} )}{\sin\big(\fr{p_{\nu} \xi}{2}+\pi \ell'_{\nu}\xi \big) }\bigg). \label{last-factor-m}
\eeqa
{\rcc In the following two subsections we will study the case of large mass and of small mass, respectively. 
For this purpose we fix a {\bl constant $c_*>0$ depending only on $d$} and establish Lemma~\ref{bound-integrand} first under the assumption $\fr{1}{4}\bmu_0\geq c_{*}\eta^2$
and then under the complementary assumption $\fr{1}{4}\bmu_0<c_{*}\eta^2 $. }
\subsection{The case of large mass: $\fr{1}{4}\bmu_0\geq c_{*}\eta^2$, $c_{*}>0$}
The function in the bracket in (\ref{exponential-functions-part-m}) is clearly entire analytic, thus bounded in any compact set. By its definition, for any such set the bound can be chosen uniformly in $\eta \leq 1/2$.  In other words, the function
\beqa
H_1(z):=\fr{\xi^{d+2}}{4} \bigg(\prod_{\al=0}^{d-1} \fr{\e^{-\i\fr{z_{\al}}{2}} }{  \e^{-\i\fr{(z_{\al}+2\pi \ell'_{\al}  )\xi }{2}} } \bigg) \label{H-1-treatment}
\eeqa
satisfies $|H_1(z)|\leq c \xi^{d+2}$ for $z$ in the closure of $S_{\pi,1}:=\{\,p+\i\etaq\in \complex^d\,|\, p\in ]-\pi,\pi[,  |q|<1\,\}$. Let us now analyse the
function $\ti{H}_3$ defined by (\ref{last-factor-m}):
\beqa
\ti{H}_3(z):=\fr{1 }{ \sum_{\mu=0}^{d-1}\sin^2\big(\fr{ z_{\mu} \xi }{2}   +  \pi \ell'_{\mu}\xi     \big)+ \fr{1}{4}\bmu_0   }   
\bigg(\prod_{\nu=0}^{d-1}\fr{\sin(\fr{z_{\nu}}{2} )}{\sin\big(\fr{z_{\nu} \xi}{2}+\pi \ell'_{\nu}\xi \big) }\bigg).
\eeqa
First, we note that for $\ell'=0$ we have by Lemmas~\ref{sin-z}, \ref{added-lemma} and the bound $\fr{1}{4}\bmu_0\geq c_{*}\eta^2$ 
\beqa
|\ti{H}_3(z)|\leq \fr{c}{\eta^{2+d}c_*}. \label{tilde-H-3-zero}
\eeqa
Now let us assume $\ell'\neq 0$.  We have, by Lemmas~\ref{added-lemma}, \ref{sin-z}, \ref{simple-analyticity}, 
\beqa
|\ti{H}_3(z)| \2\leq\2 \fr{c}{\eta^{2+d}} \fr{ 1 }{ \sum_{\mu=0}^{d-1} | z_{\mu}   + 2 \pi \ell'_{\mu} | ^2 }\fr{1}{\prod_{\nu=0}^{d-1}(1+|\ell'_{\nu}| )} \non\\
\2 \leq \2 \fr{c}{\eta^{2+d}} \fr{1}{\sum_{\mu, \ell'_{\mu}\neq 0 }(1+\ell'_{\mu})^2 }   \fr{1}{\prod_{\nu=0}^{d-1}(1+|\ell'_{\nu}| )} \non\\
\2 \leq \2 \fr{c}{\eta^{2+d}}    \fr{1}{\prod_{\nu=0}^{d-1}(1+|\ell'_{\nu}| )^{1+2/d} },  \label{tilde-H-3-estimate}
\eeqa
where in the last step we used that the geometric mean is smaller than the arithmetic mean.
Considering (\ref{tilde-H-3-zero}), we see that (\ref{tilde-H-3-estimate}) holds for all $\ell'\in \mathbb{Z}^d$.

The function in the  middle factor (\ref{big-denominator-m}) will be called $\ti{H}_2=1/\ti{F}$.  We will find a   strip $S_{\pi,c_{\mrm{st}}}=\{\, p+\i\etaq\,|\,p\in ]-\pi,\pi[^{\times d},\, |\etaq|<c_{\mrm{st}}\}$ s.t. $\ti{F}$ 
has no zeros there. We have
\beqa
\ti{F}(z)\2:=\2 1 +a\fr{\xi^{2d+2}}{4}  \sum_{\ell''_{\mud} =- \fr{L^k-1}{2}}^{  \fr{L^k-1}{2}}
 \fr{ 1 }{ \ovsum_{\mu=0}^{d-1}\sin^2\big(\fr{ z_{\mu} \xi }{2}   +  \pi \ell''_{\mu}\xi     \big)   }   
 \prod_{\nu=0}^{d-1} \fr{\sin^2(\fr{z_{\nu}}{2} )}{\sin^2( \fr{z_{\nu}\xi}{2} + \pi \ell''_{\nu} \xi )}.  \label{full-F-m} 
\eeqa
 By the Taylor theorem, we can write
 \beqa
 \ti{F}(p+\i\etaq)=\ti{F}(p)+\i\etaq\cdot (\nabla \ti{F})(p+\i s\etaq), 
 \eeqa
 for some $0\leq s \leq 1$. {\rcc (Due to this restriction, possible dependence of  $s$ on parameters of the problem does not cause any complications).}  Using Lemma~\ref{denominator-lemma-m} below, we obtain
 \beqa
 |\ti{F}(p+\i\etaq)|\geq |\ti{F}(p)|-|\etaq|  |(\nabla \ti{F})(p+\i s\etaq)| \geq 1 -|\etaq| c a
 \eeqa 
 for some $c \geq 0$. Thus there exists a numerical constant $c_{  \mrm{st} }>0$ s.t. the function $H$ satisfies in the strip $p\in [-\pi,\pi[^{\times d}$, 
 $|\etaq|<c_{  \mrm{st} }$ the bound 
 \beqa
 |H(z)|\leq \fr{c}{\prod_{\mu=0}^{d-1}(1+|\ell'_{\mu}| )^{1+2/d} }.
 \eeqa
Thus we have proven Lemma~\ref{bound-integrand} under the assumption $\fr{1}{4} \bmu_0\geq c_* \eta^2$, $c_*>0$. In the next subsection we will treat the case 
$\fr{1}{4} \bmu_0< c_* \eta^2 $.
\bel\label{denominator-lemma-m} For $\ti{F}$ defined in (\ref{full-F-m}), there holds, for $p\in [-\pi, \pi]^{\times d}$, $|\etaq_{\mud}|\leq 1/d$, 
\beqa
\2 \2 \ti{F}(p)\geq 1, \label{Lower-bound-F-m} \\ 
\2 \2|\nabla \ti{F}(p+\i\etaq)|\leq c  a. \label{F-j-estimates-m}
\eeqa
\eel
\proof Inequality~(\ref{Lower-bound-F-m}) is obvious from (\ref{full-F-m}).   Now we write
\beqa
\ti{F}(z)\2:=\2 1 +a\fr{\xi^{2d+2}}{4}  \sum_{\ell''_{\mud} =- \fr{L^k-1}{2}}^{  \fr{L^k-1}{2}} 
 \underbrace{ \fr{ 1 }{ \ovsum_{\mu=0}^{d-1}\sin^2\big(\fr{ z_{\mu} \xi }{2}   +  \pi \ell''_{\mu}\xi     \big)   } }_{F_{\ell''}^{(1)}(z) }  
  \underbrace{ \prod_{\nu=0}^{d-1} \fr{\sin^2(\fr{z_{\nu}}{2} )}{\sin^2( \fr{z_{\nu}\xi}{2} + \pi \ell''_{\nu} \xi )} }_{F_{\ell''}^{(2)}(z) }.  \label{full-F-m-x} 
 \eeqa
To estimate $\nabla \ti F$,  we  first consider $\ell''=0$. We have, {\rcc by Lemma~\ref{sin-z} and the assumption $\fr{1}{4} \bmu_0\geq c_* \eta^2$,}
\beqa
|F^{(1)}_{\ell''=0}(z)| \2=\2 \bigg|\fr{ 1 }{ \ovsum_{\mu=0}^{d-1}\sin^2\big(\fr{ z_{\mu} \xi }{2}   \big)   } \bigg|\leq \fr{1}{c_*\eta ^2}, \\
| \pa_{z_{\al}} F^{(1)}_{\ell''=0}(z) | \2=\2 \bigg|\fr{ \fr{\xi}{2} \sin\big( z_{\al} \xi     \big) }{(\ovsum_{\mu=0}^{d-1}\sin^2\big(\fr{ z_{\mu} \xi }{2}   \big))^2 }\bigg|\leq \fr{c\xi^2}{c_*^2 \xi^4}.
 \eeqa
Furthermore,  Lemma~\ref{complex-der-lemma} gives, 
\beqa
|F_{\ell''=0}^{(2)}(z)|, |\pa_{z_{\al}}F_{\ell''=0}^{(2)}(z)|\leq \fr{c}{\eta^{2d}}.
\eeqa
Hence, 
\beqa
|\pa_{z_{\al}}  (F^{(1)}_{\ell''=0} F_{\ell''=0}^{(2)})(z)| \leq  \fr{c}{\eta^{2d+2}}. \label{ell=0-part}
\eeqa
Now let us now  assume $\ell''\neq 0$.  We have, by Lemma~\ref{added-lemma}, 
\beqa
\big|  F_{\ell''}^{(1)}(z)  \big| \2  \leq  \2  \bigg| \fr{ 1 }{ \ovsum_{\mu=0}^{d-1}\sin^2\big(\fr{z_{\mu} \xi }{2}   +  \pi \ell''_{\mu}\xi     \big) } \bigg| \leq 
  \fr{ c }{ \sum_{\mu=0}^{d-1} |\fr{z_{\mu} \xi }{2}   +  \pi \ell''_{\mu}\xi |^2 } \non\\
\2 \leq \2 \fr{c}{\eta^2} \fr{1}{\sum_{\mu, \ell''_{\mu}\neq 0}(1+|\ell''_{\mu}|)^2} \leq \fr{c}{\eta^2} \fr{1}{\prod_{\mu=0}^{d-1} (1+|\ell''_{\mu}|)^{  2/d }  },
\eeqa
where in the third step we used Lemma~\ref{non-zero-ell-bound} and in the last step the fact that the geometric mean is smaller than the arithmetic mean.
Regarding the derivative, we write
\beqa
\big| \pa_{z_{\al}} F_{\ell''}^{(1)}(z)  \big| \2=\2
\bigg|    \fr{ \fr{\xi}{2} \sin\big( z_{\al} \xi   +  2\pi \ell''_{\al}\xi     \big) }{    \big(\ovsum_{\mu=0}^{d-1}\sin^2\big(\fr{ z_{\mu} \xi }{2}   +  \pi \ell''_{\mu}\xi     \big) \big)^2  } \bigg| \leq c\fr{ \fr{\xi}{2} |z_{\al} \xi   +  2\pi \ell''_{\al}\xi | }{  \big(\sum_{\mu=0}^{d-1} |\fr{ z_{\mu} \xi }{2}   +  \pi \ell''_{\mu}\xi  \big|^2\big)^2  } \non\\
\2 \leq \2 c\fr{\eta^2(1+|\ell''_{\al}|) }{\eta^4 \big(\sum_{\mu, \ell''_{\mu}\neq 0}(1+|\ell''_{\mu}| )^2\big)^2 } 
 \leq  \fr{c}{\eta^2} \fr{\prod_{\al=0}^{d-1} (1+ |\ell''_{\al}| ) }{\prod_{\mu=0}^{d-1} (1+ |\ell''_{\mu}| )^{4/d}  } \non\\
\2\leq\2 \fr{c}{\eta^2} \bigg( \prod_{\al'=0}^{d-1} (1+ |\ell''_{\al'}| ) \bigg)^{1-4/d},
\eeqa 
where in the second step we used Lemmas~\ref{added-lemma}, \ref{sin-z}, in the third step $|z_{\al}|\leq c$ and  Lemma~\ref{non-zero-ell-bound} and in the fourth
step the fact that the geometric mean is smaller than the arithmetic mean.

On the other hand, Lemma~\ref{complex-der-lemma} gives
\beqa
|  F_{\ell''}^{(2)}(z) |,   |\pa_{z_{\al}}  F_{\ell''}^{(2)}(z)| \leq \fr{c}{\xi^{2d}} \fr{1}{ \prod_{\nu=0}^{d-1}(1+|\ell''_{\nu}|)^2 }.
\eeqa
Thus, altogether,
\beqa
|\pa_{z_{\al}}  (F^{(1)}_{\ell''} F_{\ell''}^{(2)})(z)| \leq  \fr{c}{\eta^{2d+2}}  \fr{1}{ \prod_{\nu=0}^{d-1}(1+|\ell''_{\nu}|)^{1+4/d}}, \label{all-ell-bound}
\eeqa
which holds for all $\ell''\in \mathbb{Z}^d$ by (\ref{ell=0-part}). Substituting (\ref{all-ell-bound}) to (\ref{full-F-m-x}) and extending the region of summation
in $\ell''$ to whole $\mathbb{Z}^d$ we obtain (\ref{F-j-estimates-m}). \qed
\subsection{The case of small mass: ${\rcc \fr{1}{4}}\bmu_0< c_*\eta^2$}
We obtain from~(\ref{before-continuation-m})--(\ref{last-factor-m})
\beqa
H(p)\2:=\2 \fr{1}{1+a \llan u, u_{\De}\rran(p)} u_{\De}(p+2\pi \ell') \label{before-continuation} \non\\
\2=\2  \fr{ \xi^{d+2} }{4} \bigg(\prod_{\al=0}^{d-1} \fr{\e^{-\i\fr{p_{\al}}{2}} }{  \e^{-\i\fr{(p_{\al}+2\pi \ell'_{\al}  )\xi }{2}} } \bigg) \label{exponential-functions-part} \\
\2\times \2 \fr{1}{{\bcc \ovsum_{\mu'=0}^{d-1}\sin^2\big(\fr{ p_{\mu'} \xi }{2}\big)}+a\fr{\xi^{2d+2}}{4}  \sum_{\ell''_{\mud} =- \fr{L^k-1}{2}}^{  \fr{L^k-1}{2}} 
\fr{\bcc  \ovsum_{\mu'=0}^{d-1}\sin^2\big(\fr{ p_{\mu'} \xi }{2}\big) }{ \ovsum_{\mu=0}^{d-1}\sin^2\big(\fr{ p_{\mu} \xi }{2}   +  \pi \ell''_{\mu}\xi     \big)   } 
  \prod_{\nu=0}^{d-1} \fr{\sin^2(\fr{p_{\nu}}{2} )}{\sin^2( \fr{p_{\nu}\xi}{2} + \pi \ell''_{\nu} \xi )  }  } 
\label{big-denominator} \\
\2\times \2 \fr{\bcc \ovsum_{\mu'=0}^{d-1}\sin^2\big(\fr{ p_{\mu'} \xi }{2}\big) }{ \ovsum_{\mu=0}^{d-1}\sin^2\big(\fr{ p_{\mu} \xi }{2}   +  \pi \ell'_{\mu}\xi     \big)   }   
\bigg(\prod_{\nu=0}^{d-1}\fr{\sin(\fr{p_{\nu}}{2} )}{\sin\big(\fr{p_{\nu} \xi}{2}+\pi \ell'_{\nu}\xi \big) }\bigg), \label{last-factor}
\eeqa
where we divided and multiplied by  ${\bcc \ovsum_{\mu'=0}^{d-1}\sin^2\big(\fr{ p_{\mu'} \xi }{2}\big) }$. 
The function appearing in (\ref{exponential-functions-part}) we treated already in (\ref{H-1-treatment}). 
Now we consider the function appearing in (\ref{last-factor}). We will show that it is bounded in the closure of $S_{\pi,1}$, and thus (by inspection) analytic in $S_{\pi,1}$. We rewrite it as follows: 
\beqa
H_3(z)\2:=\2 \fr{ \sum_{\mu'=0}^{d-1}\sin^2\big(\fr{ z_{\mu'} \xi }{2} \big) + \bmu_0/4  }{ \sum_{\mu=0}^{d-1}\sin^2\big(\fr{ z_{\mu} \xi }{2}   +  \pi \ell'_{\mu}\xi     \big) + \bmu_0/4  }  
\bigg(\prod_{\nu=0}^{d-1}\fr{\sin(\fr{z_{\nu}}{2} )   }{\sin\big(\fr{z_{\nu} \xi}{2}+\pi \ell'_{\nu}\xi \big)  }\bigg).
\eeqa
For $\ell'\neq 0$, we have, by Lemmas~\ref{sin-z}, \ref{added-lemma}  
\beqa
| H_3(p)|\2\leq\2 c \fr{\sum_{\mu'=0}^{d-1} |\fr{z_{\mu'} \xi }{2}|^2 +c_* \eta^2 }{ \sum_{\mu=0}^{d-1}|\fr{z_{\mu} \xi }{2}   +  \pi \ell'_{\mu}\xi|^2   } 
\bigg(\prod_{\nu=0}^{d-1}\fr{  | \fr{z_{\nu}}{2}|  }{ |\fr{z_{\nu} \xi}{2}+\pi \ell'_{\nu}\xi | }\bigg)\non\\
\2\leq\2 \fr{c}{\eta^{d}} \fr{ 1 }{ \sum_{\mu=0}^{d-1}|\fr{z_{\mu}  }{2}   +  \pi \ell'_{\mu}|^2   }  \bigg(\prod_{\nu=0}^{d-1}\fr{  | \fr{z_{\nu}}{2}|  }{ |\fr{z_{\nu}}{2}+\pi \ell'_{\nu} | }\bigg)
\non\\
\2\leq\2  \fr{c}{\eta^{d}} \fr{ 1 }{ \sum_{\mu, \ell'_{\mu}\neq 0 } (1+\ell'_{\mu})^2   }  \bigg(\prod_{\nu=0}^{d-1} \fr{1}{(1+|\ell'_{\nu}|) }  \bigg)\non\\
\2\leq \2 \fr{c}{\xi^{d}} \fr{ 1}
{ (1   +  |\ell'_{0}|)^{1+2/d} \ldots  ( 1   +  |\ell'_{d-1}|)^{1+2/d}    },  \label{H-3-bound}
\eeqa
where in the second step we used the fact that $z_{\mu'}$ belong to compact sets, in the third step we applied Lemmas~\ref{non-zero-ell-bound}, \ref{simple-analyticity} 
and in the last step we use that the geometric mean is always smaller that the arithmetic mean.  By Lemma~\ref{sin-z},
the bound remains true for $\ell'=0$.

The function in the  middle factor (\ref{big-denominator}) will be called $H_2=1/F$.  We will find a   strip $S_{\pi,c_{\mrm{st}}}=\{\, p+\i\etaq\,|\,p\in ]-\pi,\pi[^{\times d},\, |\etaq|<c_{\mrm{st}}\}$ s.t. $F$ 
has no zeros there. We have
\beqa
F(z)\2:=\2 {\ma \ovsum_{\mu'=0}^{d-1}} \sin^2(\fr{z_{\mu'} \xi}{2} )   +a\fr{\xi^{2d+2}}{4}  \sum_{\ell''_{\mud} =- \fr{L^k-1}{2}}^{  \fr{L^k-1}{2}}
 \fr{  {\ma \ovsum_{\mu'=0}^{d-1}}\sin^2\big(\fr{ z_{\mu'} \xi }{2} \big)  }{ \ovsum_{\mu=0}^{d-1}\sin^2\big(\fr{ z_{\mu} \xi }{2}   +  \pi \ell''_{\mu}\xi     \big)   }   
 \prod_{\nu=0}^{d-1} \fr{\sin^2(\fr{z_{\nu}}{2} )}{\sin^2( \fr{z_{\nu}\xi}{2} + \pi \ell''_{\nu} \xi )} 
\label{full-F} \non\\
\2=\2 a\fr{\xi^{2d+2}}{4}     \prod_{\nu=0}^{d-1} \fr{\sin^2(\fr{z_{\nu}}{2} )}{\sin^2( \fr{z_{\nu}\xi}{2} )}\label{F-1} \\
\2 \2 +a\fr{\xi^{2d+2}}{4}  \sum_{{\bcc \ell''\neq 0},\, \ell''_{\mud} =- \fr{L^k-1}{2}}^{  \fr{L^k-1}{2}} \fr{\ovsum_{\mu'=0}^{d-1}\sin^2\big(\fr{ z_{\mu'} \xi }{2} \big)  }{ \ovsum_{\mu=0}^{d-1}\sin^2\big(\fr{ z_{\mu} \xi }{2}   +  \pi \ell''_{\mu}\xi     \big)   }   \prod_{\nu=0}^{d-1} \fr{\sin^2(\fr{z_{\nu}}{2} )}{\sin^2( \fr{z_{\nu}\xi}{2} + \pi \ell''_{\nu} \xi )  }  \label{F-2}\\
\2 \2+ {\ma \ovsum_{\mu'=0}^{d-1}} \sin^2(\fr{z_{\mu'}\xi}{2} ),  \label{F-3}
\eeqa
{\bcc where (\ref{F-1}) is the  $\ell''=0$ term}. By the Taylor theorem, we can write
 \beqa
 F(p+\i\etaq)=F(p)+\i\etaq\cdot (\nabla F)(p+\i s\etaq), 
 \eeqa
 for some $0\leq s \leq 1$.  Using Lemma~\ref{denominator-lemma} below, we obtain
 \beqa
 |F(p+\i \etaq)|\geq |F(p)|-|\etaq|  |(\nabla F)(p+\i s\etaq)| \geq c_1 a \xi^2 -|\etaq| c_2 a\xi^2 \label{F-lower-bound}
 \eeqa 
 for some $c_1, c_2>0$. 
 
 Thus there exists a numerical constant $c_{  \mrm{st} }>0$ s.t. the function $H$ satisfies in the strip $p\in [-\pi,\pi[^{\times d}$, 
 $|\etaq|<c_{  \mrm{st} }$ the bound 
 \beqa
 |H(z)|\leq \fr{c}{\prod_{\mu=0}^{d-1}(1+|\ell'_{\mu}| )^{1+2/d} }.
 \eeqa
We used that $\eta^{-d-2}$ coming from (\ref{F-lower-bound}), (\ref{H-3-bound}) is cancelled by $\eta^{d+2}$ appearing  in (\ref{exponential-functions-part}). This completes
the proof of Lemma~\ref{bound-integrand}.
\bel\label{denominator-lemma} For $F$ defined in (\ref{full-F}), there holds, for $p\in [-\pi, \pi]^{\times d}$, $|\etaq_{\mud}|\leq 1/d$, 
\beqa
\2 \2 F(p)\geq ca\xi^2  >0, \label{Lower-bound-F} \\ 
\2 \2|\nabla F(p+\i\etaq)|\leq c  a \xi^2. \label{F-j-estimates}
\eeqa
\eel

\proof We denote the three terms in (\ref{F-1}), (\ref{F-2}), (\ref{F-3})  by $F_1, F_2, F_3$, respectively. We recall the notation $\sinc(z):=\fr{\sin(z)}{z}$. We start with the lower bound in (\ref{Lower-bound-F}): 
We have, by  Lemma~\ref{sin-z},
\beqa
F_1(p)=a\fr{\xi^{2d+2}}{4}  \prod_{\nu=0}^{d-1} \fr{\sin^2(\fr{p_{\nu}}{2} )}{\sin^2( \fr{p_{\nu}\xi}{2} )}
=a\fr{\xi^{2d+2}}{4\xi^{2d} } \prod_{\nu=0}^{d-1} \fr{\sinc^2(\fr{p_{\nu}}{2} )    }{ \sinc^2( \fr{p_{\nu}\xi}{2})   }\geq ac_+\xi^2,
\eeqa
where $c_+>0$. Since $F_2(p), F_3(p)\geq 0$, we obtain (\ref{Lower-bound-F}).

Now we compute and estimate the derivatives: By Lemma~\ref{complex-der-lemma},
\beqa
|\pa_{z_{\mu}}F_1(z)|\2 =\2 \fr{a \xi^{2d+2}}{4}\big| \pa_{z_{\mu}} \prod_{\nu=0}^{d-1} \bigg( \fr{\sin(\fr{z_{\nu}}{2} )}{ \sin( \fr{z_{\nu}\xi}{2} )  } \bigg)^2\big|
\leq c a \xi^2.
\eeqa
Now we consider, for $\ell''\neq 0$ 
\beqa
F_{2,\ell''}(z):= a\fr{\xi^{2d+2}}{4}  \underbrace{\fr{\ovsum_{\mu'=0}^{d-1}\sin^2\big(\fr{ z_{\mu'} \xi }{2} \big)  }{ \ovsum_{\mu=0}^{d-1}\sin^2\big(\fr{ z_{\mu} \xi }{2}   +  \pi \ell''_{\mu}\xi     \big)   } }_{F^{(1)}_{2,\ell''}(z)} \underbrace{  \prod_{\nu=0}^{d-1} \fr{\sin^2(\fr{z_{\nu}}{2} )}{\sin^2( \fr{z_{\nu}\xi}{2} + \pi \ell''_{\nu} \xi )  } }_{F^{(2)}_{2,\ell''}(z)}.
\eeqa
We first estimate the auxiliary functions $F^{(1)}_{2,\ell''}, F^{(2)}_{2,\ell''}$ and their derivatives. We have by Lemma~\ref{added-lemma}
\beqa
|F^{(1)}_{2,\ell''}(z)|
\leq c \fr{\sum_{\mu'=0}^{d-1} |z_{\mu'}|^2 +c_*}{ \sum_{\mu=0}^{d-1}  |z_{\mu}   +  2\pi \ell''_{\mu}|^2 }  \leq
\fr{c}{ \sum_{\mu, \ell''_{\mu}\neq 0 }  (1   +   |\ell''_{\mu}| )^2 }\leq  c\fr{1}{\prod_{\mu=0}^{d-1}(1+|\ell''_{\mu}|)^{2/d}  },
\eeqa
where we made use in the second step of $\sum_{\mu'=0}^{d-1} |z_{\mu'}|^2\leq c$, of Lemma~\ref{non-zero-ell-bound} and in the last step of the fact that the geometric mean is smaller than the arithmetic mean. Furthermore, setting $z^{\nu}:=z_{\nu}+2\pi\ell''_{\nu}$ and applying Lemma~\ref{complex-der-lemma} we get
\beqa
\big|F^{(2)}_{2,\ell''}(z)\big|
\leq \fr{c}{\xi^{2d}}  \fr{1}{ \prod_{\nu=0}^{d-1}(1+|\ell''_{\nu}|)^2  }.
\eeqa

Next,  we consider the derivatives.  We compute:
\beqa
\2 \2 |\pa_{z_{\al}}F^{(1)}_{2,\ell''}(z)|=\bigg|\pa_{z_{\al}} \fr{ \ovsum_{\mu'=0}^{d-1}\sin^2\big(\fr{ z_{\mu'} \xi }{2} \big)  }{ \ovsum_{\mu=0}^{d-1}\sin^2\big(\fr{ z^{\mu} \xi }{2}     \big)   }  \bigg|\non\\
\2=\2 \xi \bigg| \fr{  \sin\big(\fr{ z_{\al} \xi }{2} \big) \cos(\fr{ z_{\al} \xi }{2} \big) \ovsum_{\mu=0}^{d-1}\sin^2\big(\fr{ z^{\mu} \xi }{2}      \big)-\big( \ovsum_{\mu'=0}^{d-1}\sin^2\big(\fr{ z_{\mu'} \xi }{2} \big)\big) \sin\big(\fr{ z^{\al} \xi }{2}      \big)\cos(\fr{ z^{\al} \xi }{2}    )  }
{ \big({\ma \ovsum_{\mu=0}^{d-1}}\sin^2\big(\fr{ z^{\mu} \xi }{2}    \big)  \big)^2 }  \bigg| \non\\
\2\leq \2 c \fr{  |z_{\al}|  (\sum_{\mu=0}^{d-1} |z^{\mu}|^2+c_*)+\big( \sum_{\mu'=0}^{d-1}  |z_{\mu'}|^2 +c_* \big)  |z^{\al}|   }
{ \big(\sum_{\mu=0}^{d-1}  |z^{\mu}|^2   \big)^2 } 
\leq  c\fr{1}{\big(\sum_{\mu=0}^{d-1}  |z^{\mu}|^2   \big)} +c\fr{|z^{\al}|}{\big(\sum_{\mu=0}^{d-1}  |z^{\mu}|^2   \big)^2 }\non\\
\2 \leq\2 c\fr{1}{\sum_{\mu, \ell_{\mu}''\neq 0}  (1+|\ell''_{\mu}| )^2 }+c \fr{(1+|\ell''_{\al}| )}{\big(\sum_{\mu, \ell_{\mu}''\neq 0}  (1+|\ell''_{\mu}| )^2   \big)^2 }
\leq c\fr{1}{\sum_{\mu, \ell''_{\mu}\neq 0}  (1+|\ell''_{\mu}| )^2 } \leq c\fr{1}{\prod_{\mu=0}^{d-1}  (1+|\ell''_{\mu}| )^{2/d} },
 \eeqa
where, in the next to the last line we used Lemmas~\ref{sin-z}, \ref{added-lemma} and in the last line  Lemma~\ref{non-zero-ell-bound} and the fact that the geometric mean is
smaller that the arithmetic mean.
Now we move on to the derivative of $F^{(2)}_{2,\ell''}$   and  use Lemma~\ref{complex-der-lemma}    
\beqa
\big|\pa_{z_{\al}} F^{(2)}_{2,\ell''}(z) \big|\leq \fr{c}{\xi^{2d}} \fr{1}{ \prod_{\nu=0}^{d-1}(1+|\ell''_{\nu}|)^2}.
\eeqa
Thus, altogether, we get
\beqa
|\pa_{z_{\al}} F_{2,\ell''}(z)|\leq ca \xi^{2}  \fr{1}{ \prod_{\nu=0}^{d-1}(1+|\ell_{\nu}|)^{2+2/d} }.
\eeqa
Similarly for
\beqa
 F_{2}(z):= \sum_{{\bcc \ell''\neq 0},\, \ell''_{\mud} =- \fr{L^k-1}{2}}^{  \fr{L^k-1}{2}} F_{2,\ell''}(z) \quad\textrm{ we have }\quad  |\pa_{z_{\al}} F_{2}(z)|\leq ca \xi^{2}
\eeqa
by extending the region of summation over $\ell''$ to $\mathbb{Z}^d$.

Finally, we note the obvious relation regarding $F_3(z)=\sum_{\mu'=0}^{d-1} \sin^2(\fr{z_{\mu'}\xi}{2} )$:
\beqa
|\pa_{z_{\al}}F_3(z)|= |\xi  (\fr{z_{\al}\xi}{2} ) \sinc(\fr{z_{\al}\xi}{2} )  \cos(\fr{z_{\al}\xi}{2} ) |\leq c\xi^2,
\eeqa
where we used Lemma~\ref{sin-z}. \qed
\subsection{Technical lemmas}
\bel \label{complex-der-lemma}  For $z=p+\i q$, $p\in [-\pi, \pi]^{\times d}$, $|\etaq|\leq 1$
\beqa
\bigg|\pa^{\al}_{z_{\mu}} \bigg(\prod_{\nu=0}^{d-1} \fr{\sin^2(\fr{z_{\nu}}{2} )}{\sin^2( \fr{z_{\nu}\xi}{2} +  \pi\ell''_{\nu}\eta)  } \bigg)\bigg| \leq 
\fr{c}{\xi^{2d}} \fr{1}{ \prod_{\nu=0}^{d-1}(1+|\ell_{\nu}|)^2 }, \quad |\al|=0,1.
\eeqa
\eel
\proof  We set $z^{\nu}:=z_{\nu}+2\pi\ell''_{\nu}$ {\ma and recall that $|\ell''_{\nu}|\leq \fr{(L^k-1)}{2}$, so $|\mrm{Re} (\fr{ z^{\nu}\eta}{2}) |\leq \fr{\pi}{2}$}. We compute
\beqa
\bigg|\prod_{\nu=0}^{d-1} \fr{\sin^2(\fr{z_{\nu}}{2} )}{\sin^2( \fr{z^{\nu}\xi}{2}  )  }\bigg|=\fr{1}{\xi^{2d}}  
\bigg|\prod_{\nu=0}^{d-1} \fr{\sinc^2(\fr{z_{\nu}}{2} ) z^2_{\nu}   }{\sinc^2( \fr{z^{\nu}\xi}{2}  ) (z^{\nu})^2  }\bigg|\leq \fr{c}{\xi^{2d}} \prod_{\nu=0}^{d-1} \bigg|\fr{z_{\nu}   }{ z^{\nu}  }\bigg|^2
\leq \fr{c}{\xi^{2d}}  \fr{1}{ \prod_{\nu=0}^{d-1}(1+|\ell''_{\nu}|)^2  },
\eeqa
where we applied Lemmas~\ref{sin-z},\ref{simple-analyticity}. Now we move on to the derivative:
\beqa
\bigg|\pa_{z_{\mu}} \bigg(\prod_{\nu=0}^{d-1} \fr{\sin^2(\fr{z_{\nu}}{2} )}{\sin^2( \fr{z^{\nu}\xi}{2} )  } \bigg)\bigg|  
\2=\2  2\bigg| \bigg( \fr{\sin(\fr{z_{\mu}}{2} )}{ \sin( \fr{z^{\mu}\xi}{2} )  } \bigg) \pa_{z_{\mu}}\bigg( \fr{\sin(\fr{z_{\mu}}{2} )}{ \sin( \fr{z^{\mu}\xi}{2} )  } \bigg)\prod_{\nu\neq \mu} \bigg( \fr{\sin(\fr{z_{\nu}}{2} )}{ \sin( \fr{z^{\nu}\xi}{2} )  } \bigg)^2\bigg|\non\\
\2=\2  2 \bigg|\bigg( \fr{\sinc(\fr{z_{\mu}}{2} )}{ \sinc( \fr{z^{\mu}\xi}{2} )  }\fr{z_{\mu}}{\xi z^{\mu}} \bigg) \pa_{z_{\mu}}\bigg( \fr{\sin(\fr{z_{\mu}}{2} )}{ \sin( \fr{z^{\mu}\xi}{2} )} \bigg)
 \prod_{\nu\neq \mu} \bigg( \fr{\sinc(\fr{z_{\nu}}{2} )}{ \sinc( \fr{z^{\nu}\xi}{2} )  } \fr{z_{\nu}}{z^{\nu}\xi } \bigg)^2  \bigg|\non\\
\2\leq \2  \fr{c}{\xi^{2d-1}} \fr{1}{(1+|\ell''_{\mu}|)\prod_{\nu\neq \mu}(1+|\ell''_{\nu}|)^2  }  \bigg|    \fr{\h  \cos(\fr{z_{\mu}}{2})\sin(\fr{z^{\mu}\xi}{2}) -\fr{\xi}{2} \cos(\fr{z^{\mu}\xi}{2})\sin(\fr{z_{\mu}}{2}) }
{\sin^2(\fr{z^{\mu}\xi}{2} ) }  \bigg|.\non\\
\2= \2  \fr{c}{\xi^{2d-1}} \fr{1}{(1+|\ell''_{\mu}|)\prod_{\nu\neq \mu}(1+|\ell''_{\nu}|)^2  } 
 \bigg|    \fr{\h  \cos(\fr{z_{\mu}}{2})\sinc(\fr{z^{\mu}\xi}{2}) \fr{z^{\mu}\xi}{2} -\fr{\xi}{2} \cos(\fr{z^{\mu}\xi}{2}) \sin(\fr{z_{\mu}}{2}) }
{ \sinc^2(\fr{z^{\mu}\xi}{2} )( \fr{z^{\mu}\xi}{2})^2  }  \bigg|\non\\
\2\leq \2  \fr{c}{\xi^{2d}} \fr{1}{ \prod_{\nu=0}^{d-1}(1+|\ell''_{\nu}|)^2 },
\eeqa
where in the third step  we used  Lemmas~\ref{sin-z}, \ref{simple-analyticity} {\rcc and in the last step Lemmas~\ref{sin-z}, \ref{non-zero-ell-bound}, provided that $\ell_{\mu}''\neq 0$ as required in this latter lemma. In the case $\ell_{\mu}''=0$ we have, trivially, $(1+|\ell''_{\mu}|)=(1+|\ell''_{\mu}|)^2$ and the numerator under the
absolute value above can be rewritten and estimated as follows
\beqa
\big| \fr{z_{\mu}\xi}{4} \big|  \big|  \cos(\fr{z_{\mu}}{2})\sinc(\fr{z_{\mu}\xi}{2})  - \cos(\fr{z_{\mu}\xi}{2}) \sinc(\fr{z_{\mu}}{2})\big|\leq c |z_{\mu}|^3\eta
\eeqa
by inspection of Taylor expansions of $\cos$ and $\sinc$. This compensates $(\fr{z^{\mu}\xi}{2})^2$ in the denominator (up to one inverse power of $\eta$ which is
taken into account) and concludes the proof.} \qed
\bel\label{added-lemma}  The following bounds hold true for $\ell\neq 0$, $z_{\mud}=p_{\mud}+\i q_{\mud}$, $p_{\mud}\in [-\pi, \pi]$,   $|q_{\mud}|\leq 1/d$ and any $\de \geq 0$
\beqa
\sum_{\mu=0}^{d-1} \big|\fr{z_{\mu}\eta}{2}+  \pi\ell_{\mu}\eta \big|^2 \2\leq\2 c \big| \sum_{\mu=0}^{d-1} \sin^2\big(\fr{z_{\mu}\eta}{2}+  \pi\ell_{\mu}\eta \big)+\de\big|, \label{length-vs-sin}\\
\de\2 \leq\2 c \big| \sum_{\mu=0}^{d-1} \sin^2\big(\fr{z_{\mu}\eta}{2}+  \pi\ell_{\mu}\eta \big)+\de\big|. \label{delta-lower-bound}
\eeqa
\eel
\proof Let us compute for arbitrary $w=x+\i y\in \complex$ 
\beqa
\sin^2\big(\fr{w}{2} \big)
\2=\2 \fr{1}{2}\big(1 -  \cos(x+\i y) \big) \non\\
\2=\2 \fr{1}{2}\big(1 -  \cos(x)\cos(\i y)+\sin(x)\sin(\i y)  \big) \non\\  
\2=\2  \fr{1}{2}\big(1 - \cos(x)\ch(y)+ \i\sin(x)\sh(y)  \big) \non\\
\2=\2  \fr{1}{2}\big(1-\ch(y)+\ch(y)  - \cos(x)\ch(y)+ \i\sin(x)\sh(y)  \big) \non\\
\2=\2 \fr{1}{2}\big(1-\ch(y)\big) + \fr{1}{2}\ch(y)(1 - \cos(x))  +\fr{\i}{2} \sh(y)\sin(x) \non\\
\2=\2 - \sh^2\big(\fr{y}{2}\big)  + \ch(y)\sin^2\big(\fr{x}{2}\big) + \fr{\i}{2} \sh(y)\sin(x).
\eeqa
Hence, setting $z^{\mu}:=z_{\mu}+ 2\pi\ell_{\mu}$
\beqa
 \bigg| \sum_{\mu=0}^{d-1} \sin^2\big(\fr{z^{\mu}\eta}{2} \big) +\de \bigg| \2\geq\2   -\sum_{\mu=0}^{d-1} \sh^2(\fr{q_{\mu} \eta}{2} )  
 + \sum_{\mu=0}^{d-1} \ch(q_{\mu}\eta )\sin^2(\fr{p^{\mu}\eta }{2} ) +\de \non\\ \2 \geq \2  -\sum_{\mu=0}^{d-1} \sh^2(\fr{q_{\mu} \eta}{2} )  + \sum_{\mu=0}^{d-1}\sin^2(\fr{p^{\mu}\eta }{2})+\de. \label{sum-denominator}
\eeqa
{\rcc By assumption, there is an index} $\hmu$  for which $\ell_{\hmu}\neq 0$ and suppose, {\rcc for definiteness}, $\ell_{\hmu}>0$.
For $p^{\hmu}:= p_{\hmu}+2\pi \ell_{\hmu}$ and $p_{\hmu}\in [-\pi, \pi]$ we have
\beqa
\fr{p^{\hmu}\eta }{2} \geq  \fr{\pi \eta }{2}\quad \Rightarrow\quad \sin^2(\fr{p^{\mu}\eta }{2})\geq \sin^2(\fr{\pi \eta }{2}). \label{sin-estimate}
\eeqa
In fact, since $k\in \nat$, $L>1$, we have $\xi\leq 1/2$. Thus $\fr{p_{\hmu}\xi}{2} \in [-\pi \xi/2, \pi \xi /2 ]$. By shifting this interval by 
$\pi \eta\leq   \pi \ell_{\hmu}\eta \leq \fr{\pi}{2}$, we stay away of zeros of $\sin^2(\,\cdot\,)$. {\rcc (We used here that $\ell_{\hmu}\leq (L^k-1)/2$ and $\eta=L^{-k}$).}
Now for $x\in [0,1/2]$ we can write
\beqa
\sh(x/d)=\fr{\sh(x/d) }{\sin(\pi x)} \sin(\pi x)=\fr{1}{\pi d} \fr{\mrm{shc}(x/d)}{\sinc(\pi x) }  \sin(\pi x)\leq \fr{3}{\pi d} {\sin(\pi x) }, \label{shc}
\eeqa
where  $\mrm{shc}(x):=\fr{\sh(x)}{x}, \sinc(x):=\fr{\sin(x)}{x}$ and we used that $\sinc(\pi x)\geq 1/2$ and $\mrm{shc}(x/d)\leq 3/2$ for $x\in [0,1/2]$ as can be read off from their graphs.
(We also stress for future reference that $3/\pi<1$). Hence, using  (\ref{sin-estimate}),  (\ref{shc})
\beqa
\sh^2\big( \fr{(q_{\mu} d)\eta}{2d} \big) \leq \fr{1}{\rcc d^2}\bigg( \fr{3}{\pi }\bigg)^2 \sin^2\big(\pi \fr{(q_{\mu} d) \eta}{2} \big) \leq \fr{1}{\rcc d^2} \bigg( \fr{3}{\pi}\bigg)^2 \sin^2(\pi \fr{\eta}{2} )\leq  
\fr{1}{\rcc d^2}\bigg( \fr{3}{\pi}\bigg)^2 \sin^2(\fr{p^{\hmu} \eta }{2}).
\eeqa
Consequently,
\beqa
\sum_{\mu=0}^{d-1}\sh^2\big( \fr{(q_{\mu} d)\eta}{2d} \big) \leq  \fr{1}{\rcc d}\bigg( \fr{3}{\pi}\bigg)^2 \sin^2(\fr{p^{\hmu} \eta }{2}).
\eeqa
Thus coming back to (\ref{sum-denominator}), we can write for some {\bl constant $c>0$ depending only on $d$}
\beqa
 \bigg| \sum_{\mu=0}^{d-1} \sin^2\big(\fr{z^{\mu}\eta}{2} \big)+\de \bigg| \2\geq\2 c\sum_{\mu=0}^{d-1}\sin^2(\fr{p^{\mu}\eta }{2})+\de \geq c \sum_{\mu=0}^{d-1} \bigg|\fr{p^{\mu}\eta }{2}\bigg|^2+\de, \label{sinx-vs-x}
  \eeqa
which proves (\ref{delta-lower-bound}).
Finally, arguing like  in (\ref{sin-estimate}) and using $|q_{\mu}| \leq 1/d$,
\beqa
\bigg| \fr{(p_{\hmu}+2\pi\ell_{\hmu}) \eta }{2} \bigg|^2\geq  \bigg( \fr{\pi \eta}{2} \bigg)^2 \geq   \sum_{\mu=0}^{d-1} \bigg(\fr{ q_{\mu} \eta}{2} \bigg)^2,
\eeqa
hence
\beqa
 \bigg| \fr{(p_{\hmu}+2\pi\ell_{\hmu}) \eta }{2} \bigg|^2\geq \h \bigg| \fr{(p_{\hmu}+2\pi\ell_{\hmu}) \eta }{2} \bigg|^2   + \h \sum_{\mu=0}^{d-1} \bigg|\fr{ q_{\mu}  \eta }{2} \bigg|^2.
\eeqa
which, {\rcc combined with (\ref{sinx-vs-x}), gives (\ref{length-vs-sin})}. \qed
\bel\label{sin-z} We have, for $\mrm{Re}(z)\in [-\pi/2, \pi/2]$,  $|\mrm{Im}(z)| \leq 1$
\beqa
\2 \2 0<c_-\leq |\fr{\sin(z)}{z}|\leq c_+, \label{sinc} \\
\2 \2 \quad\quad\quad  | \pa_{z} \big(\fr{\sin(z)}{z}\big)| \leq c_0, 
\eeqa
where $c_0,c_-, c_+$ are independent of $z$ within the above restrictions.
\eel
\proof Except for the lower bound in  (\ref{sinc}), all the bounds simply follow from the fact that entire analytic functions
are bounded on compact sets. Regarding the lower bound,  we compute
\beqa
\bigg|\fr{\sin(z)}{z} \bigg|^2\2=\2\fr{ \big|\e^{\i p} \e^{-\etaq} -  \e^{-\i p} e^{\etaq} \big|^2  }{p^2+\etaq^2} =\fr{ \e^{-2\etaq} - \e^{2\i p} - \e^{-2\i p}    + \e^{2\etaq}   }{p^2+\etaq^2}\non\\
\2=\24\fr{ (\fr{\e^{\etaq}-\e^{-\etaq}}{2})^2  + ( \fr{\e^{\i p} - \e^{-\i p}}{2\i} )^2      }{p^2+\etaq^2}=4\fr{\sin^2(p)+\sinh^2(\etaq)}{p^2+\etaq^2} \non\\
\2=\24\fr{\fr{\sin^2(p)}{p^2}p^2 +\fr{\sinh^2(\etaq)}{\etaq^2} \etaq^2 }{p^2+\etaq^2}.
\eeqa
The last expression can be estimated by
\beqa
0<c_1\leq 4\fr{ \fr{\sin^2(p)}{p^2}p^2 +\fr{\sinh^2(\etaq)}{\etaq^2} \etaq^2 }{p^2+\etaq^2}\leq c_2,
\eeqa
where we used that $0<c\leq  \fr{\sin^2(p)}{p^2}, \fr{\sinh^2(\etaq)}{\etaq^2}$ in the relevant region.   \qed
\bel\label{non-zero-ell-bound} The following bound holds true for $z\in [-\pi, \pi] +\i [-1,1]$ and $\ell\in \mathbb{Z}$, $\ell\neq 0$,
\beqa
|z-2\pi \ell|\geq c_+(1+|\ell|)
\eeqa 
for some $c_+>0$.
\eel
\proof We write
\beqa
|z-2\pi \ell|\geq |p-2\pi\ell|\geq 2\pi|\ell| -\pi\geq \fr{\pi}{2}(1+|\ell|). \quad \textrm{\qed}
\eeqa
\bel\label{simple-analyticity}  The following bound holds true for $z\in [-\pi, \pi[ +\i [-1,1[$ and $\ell\in \mathbb{Z}$
\beqa
\bigg| 1+\fr{2\pi \ell}{z}  \bigg|\geq \fr{1+|\ell|}{6}.
\eeqa
\eel
\proof We write
\beqa
\bigg| 1+\fr{2\pi \ell}{z}  \bigg| \geq \left\{ \begin{array}{ll}
1 & \textrm{for $\ell=0$,}\\
\fr{2\pi |\ell|}{|z|}-1  & \textrm{for $\ell\neq 0$.}
\end{array} \right.
\eeqa
It is clear that the claimed bound holds for $\ell=0$. As for the remaining  case, we note that $|z|\leq \pi+1\leq \fr{3}{2}\pi$ hence
\beqa
\fr{2\pi |\ell|}{|z|}-1\geq  \fr{4}{3} |\ell|-1 \geq  \fr{1+|\ell|}{6},
\eeqa
which concludes the proof. \qed
\section{Completion of the proof of Theorem~\ref{G-k-omega-Q-theorem}} \label{G-k-omega-Q-App}
\setcounter{equation}{0}

Making use of Lemma~\ref{exponential-decay}, we obtain from (\ref{GQ-star-formula})
\beqa
|(G_k(\Om) Q_{\Om,k}^*)(x,y)|\2\leq \2 c \sum_{j} \bigg(\prod_{j_{\mu}\,\,\mrm{even} } 
\e^{-\h c_{\mrm{st} } | x_{\mu}-y_{\mu}- (L^m\xi) j_{\mu}| } \bigg)  {\rcc \bigg( }\prod_{j_{\nu} \,\, \mrm{odd}} \e^{-\h c_{\mrm{st}}|(L^m\xi)  (j_{\nu}+1)-y_{\nu} - x_{\nu}| }\bigg), \\
\2=\2 \sum_{\si_0,\ldots \si_{d-1}\in \{\mrm{even} / \mrm{odd} \} }  c\prod_{\al=0}^{d-1} \bigg(\sum_{j_\al\in \mathbb{Z}_{\si_\al}} h_{\si_\al,j_\al}(x_\al,y_\al)\bigg), \label{odd-even-product}\\
h_{\mrm{even},j_{\mu}}(x_{\mu},y_{\mu})\2:=\2\e^{-\h c_{\mrm{st} } | x_{\mu}-y_{\mu}- (L^m\xi) j_{\mu}| }, \quad
h_{\mrm{odd},j_{\nu}}(x_{\nu},y_{\nu}):=\e^{-\h c_{\mrm{st} } | (L^m\xi)  (j_{\nu}+1)-y_{\nu} - x_{\nu}   | }
\eeqa
and $\mathbb{Z}_{\si_{\al}}$ denotes the set of even or odd integers, respectively. Since the sum over $\si$ has $2^d$ terms, it suffices to control the sums over $j_{\al}$.\\ \vspace{0.2cm}

\nin$\bullet$ Let us first analyse the sum  over  $j_{\mu}$ even. To simplify notation in the analysis below we write $x:=x_{\mu}, y:=y_{\mu}$ and $j:=j_{\mu}$. Suppose first that $j\geq 2$. Then
\beqa
| x-y- (L^m\xi) j|=(L^m\xi)(j-2)+ 2(L^m\xi) - (x-y) \geq (L^m\xi)(j-2) +|x-y|.
\eeqa
Here the last inequality boils down for $x\geq y$ (resp. $x\leq y$) to $L^m\xi\geq (x-y)$ (resp.   $L^m\xi\geq 0$) which hold true since $L^m\xi$ is the linear size of $\Om$. 
For $j=0$ we obviously have
\beqa
| x-y- (L^m\xi) j|=|x-y|.
\eeqa
Now for $j\leq -2$ we can write
\beqa
|x-y- (L^m\xi) j|= (L^m\xi)(- j)+x-y \2=\2 (L^m\xi)(- j-2)+ 2(L^m\xi) + x-y\non\\
 \2\geq\2  (L^m\xi)(- j-2)+|x-y|.
\eeqa
Here the last inequality boils down to for $x\geq y$ (resp. $x\leq y$) to $L^m\xi\geq 0$
(resp. $L^m\xi\geq y-x$).  

Coming back to (\ref{odd-even-product}), we have
\beqa
 \sum_{j\,\, \mrm{even}} c \e^{-\h c_{\mrm{st} } | x-y- (L^m\xi) j| } 
\leq    c   \e^{-\h c_{\mrm{st}} |x-y| }\2+\2
 \sum_{j\geq 2,\,\,j \,\mrm{even}} c \e^{-\h c_{\mrm{st}}((L^m\xi)(j-2)+| x-y|) } \non\\
\2 \2+  \sum_{j\leq -2,\,\,j\, \mrm{even}}  c \e^{-\h c_{\mrm{st}}((L^m\xi)(-j-2)+| x-y|) } \leq  c \e^{-\h c_{\mrm{st}} |x-y|},
\eeqa
where we used that $L^m\xi\geq 1$  (since $\xi=L^{-k}$, $m\geq k$) to avoid any dependence of constants on $\eta$ {\rcc and then summed up the geometric series}. 

\vspace{0.2cm}

$\bullet$ Let us now analyse the  sum in  (\ref{odd-even-product}) over odd $j_{\nu}$. To simplify notation in the analysis below we write $x:=x_{\nu}$, $y:=y_{\nu}$ and $j:=j_{\nu}$. 
Suppose first that  $j \geq 3$. Then 
\beqa
|(L^m\xi)  (j+1)-y-x| \2=\2   (L^m\xi)  (j+1)-y-x \non\\
 \2=\2 (L^m\xi)  (j-1)+ 2L^m\xi -y-x \geq (L^m\xi)  (j-1)+|y-x|,
\eeqa
where the last equality boils down for $y\geq x$ (resp. $y\leq x$) to $L^m\xi\geq y$ (resp. $L^m\xi\geq x$).
Let us now consider $j=1$. We have
\beqa
|2(L^m\xi)-y-x| = 2(L^m\xi)-y-x \geq |y-x|.
\eeqa
In fact, for $y\geq x$ (resp. $y\leq x$) this inequality boils down to
$y\leq L^m\xi$ (resp. $x\leq  L^m\xi$).

Next, we take $j=-1$. Then
\beqa
|(L^m\xi)  (j+1)-y-x|=|-y-x|=y+x\geq |y-x|,
\eeqa
where we used that $x, y\geq 0$. Now suppose $j\leq -3$. Then 
\beqa
|(L^m\xi)  (j+1)-y-x|\2=\2- (L^m\xi)  (j+1)+y+x= - (L^m\xi)  (j+3) +2(L^m\xi) +y+x \non\\
                             \2\geq\2   (L^m\xi)  |j+3| +|y-x|. \label{negative-j}
\eeqa
where we used again that $x,y\geq 0$.

Coming back to (\ref{odd-even-product}), we can write
\beqa
\sum_{j>0,\,\,j\,\, \mrm{odd}} c \e^{-\h c_{\mrm{st}}|(L^m\xi)  (j+1)-y - x| } \2\leq\2  \sum_{j>0,\,\,j \,\,\mrm{odd}} 
c \e^{-\h c_{\mrm{st}}( (L^m\xi)  (j-1)+|y-x| ) }\non\\
\2=\2c \sum_{j>0,\,\,j\,\, \mrm{odd}}   \e^{-\h c_{\mrm{st}}(L^m\xi)(j-1)  }   \e^{-\h c_{\mrm{st}} |y-x|  } \non\\
\2\leq \2 c  \sum_{j>0,\,\,j\,\, \mrm{odd}}   \e^{-\h c_{\mrm{st}}(j-1)  }  \e^{-\h c_{\mrm{st}} |y-x|  } \non\\
\2 = \2 c' \e^{-\h c_{\mrm{st}} |y-x|  },
\eeqa
where we noted that $L^m\xi\geq 1$. The sum of $j<0$ is treated analogously, using (\ref{negative-j}).


\end{document}